\documentclass[12pt]{article}
\pdfoutput=1
\usepackage{epsf,amsfonts,amssymb,epsfig,amsmath,graphics,cite}
\usepackage[dvipsnames]{xcolor}
\usepackage{graphicx,subfig}
\usepackage{booktabs}
\usepackage{cancel}
\usepackage{soul}
\usepackage{float}
\usepackage{verbatim}
\usepackage{mathrsfs}
\usepackage{empheq}
\usepackage{enumerate}
\usepackage{bbm}
\usepackage{slashed}

\usepackage[colorlinks=true
,urlcolor=blue
,anchorcolor=blue
,citecolor=blue
,filecolor=blue
,linkcolor=blue
,menucolor=blue
,linktocpage=true
]{hyperref}

\usepackage{tikz}
\usetikzlibrary{cd}
 
\makeatletter

\newcommand{\Rmnum}[1]{\expandafter\@slowromancap\romannumeral #1@}
\makeatother

\newcommand{\fixme}[1]{{\color{black}#1}}

\newcommand{\nn}{\notag }
\def\be{\begin{equation}}
\def\ee{\end{equation}}

\newcommand{\ii}{\mathrm{i}}
\newcommand{\ex}{\mathrm{e}}
\newcommand{\diff}{\mathrm{d}}
\newcommand{\dd}{\mathrm{d}}
\newcommand{\R}{\mathbb{R}}
\newcommand{\Z}{\mathbb{Z}}
\newcommand{\vol}{\mathrm{vol}}

\newcommand{\cutoff}{\delta}

\newcommand{\xinew}{\zeta}

\newcommand{\identity}{\mathbbm{1}}

\newcommand{\mf}[1]{\mathfrak{#1}}

\newcommand{\g}{\mf{g}}

\newcommand{\cA}{\mathcal{A}}
\newcommand{\cB}{\mathcal{B}}
\newcommand{\cC}{\mathcal{C}}

\newcommand{\cF}{\mathcal{F}}
\newcommand{\cG}{\mathcal{G}}

\newcommand{\cI}{\mathcal{I}}
\newcommand{\cJ}{\mathcal{J}}
\newcommand{\cK}{\mathcal{K}}
\newcommand{\cL}{\mathcal{L}}

\newcommand{\cN}{\mathcal{N}}

\newcommand{\cQ}{\mathcal Q}
\newcommand{\cR}{\mathcal{R}}
\newcommand{\cS}{\mathcal{S}}

\newcommand{\cU}{\mathcal{U}}
\newcommand{\cV}{\mathcal{V}}
\newcommand{\cW}{\mathcal{W}}

\renewcommand{\tilde}{\widetilde}
\newcommand{\tz}{\tilde{z}}

\newcommand{\Itot}{I^{\mathrm{Total}}}
\newcommand{\Rup}{R^{\rm up}}

\newcommand{\tinyspace}{\mskip1mu}
\newcommand{\smallspace}{\mskip2mu}

\newcommand{\hook}{\mathbin{\rule[.2ex]{.4em}{.03em}\rule[.2ex]{.03em}{.9ex}}}

\newcommand{\Pold}{\mathbb{P}}

\newcommand{\rd}{{\rm d}}
\newcommand{\ph}[1]{\phantom{#1}}
\newcommand{\KKVec}{\ell}
\newcommand{\KKForm}{\alpha}
\newcommand{\SUSYVec}{\cK}
\newcommand{\FourdSUSYVec}{\xi}
\newcommand{\FourdSUSYForm}{\eta}
\newcommand{\e}{{\rm e}}

\newcommand{\rsigma}{c_J}
\newcommand{\JS}{c_R}

\newcommand{\ztilde}{\tilde{z}}

\newcommand{\tepsilon}{\tilde{\epsilon}}
\newcommand{\tX}{\tilde{X}}
\newcommand{\tW}{\tilde{W}}
\newcommand{\tL}{\tilde{L}}

\newcommand{\nvxi}{a}

\numberwithin{equation}{section}       

\begin{document}

\begin{titlepage}

\vskip 1cm

\begin{center}

{\Large \bf Equivariant localization for $D=5$ \\ 
\vskip 0.2cm
 gauged supergravity}

\vskip 1cm
{Pietro Benetti Genolini$^{\mathrm{a}}$, Jerome P. Gauntlett$^{\mathrm{b}}$,
Yusheng Jiao$^{\mathrm{b}}$,\\
\vskip 0.1cm
Jaeha Park$^{\mathrm{b}}$  and James Sparks$^{\mathrm{c}}$}

\vskip 1cm

${}^{\mathrm{a}}$\textit{D\'epartement de Physique Th\'eorique, Universit\'e de Gen\`eve,\\
24 quai Ernest-Ansermet, 1211 Gen\`eve, Suisse\\}

\vskip 0.2cm

${}^{\mathrm{b}}$\textit{Blackett Laboratory, Imperial College, \\
Prince Consort Rd., London, SW7 2AZ, U.K.\\}

\vskip 0.2cm

${}^{\mathrm{c}}$\textit{Mathematical Institute, University of Oxford,\\
Andrew Wiles Building, Radcliffe Observatory Quarter,\\
Woodstock Road, Oxford, OX2 6GG, U.K.\\}

\vskip 0.2 cm

\end{center}

\vskip 0.5 cm

\begin{abstract}
\noindent  

We consider supersymmetric solutions of $D=5$ Euclidean gauged supergravity coupled to
an arbitrary number of vector multiplets. We consider solutions that admit both 
the R-symmetry Killing vector, $\cK$, constructed as a bilinear in the 
Killing spinor, as well as an additional Killing vector $\ell$.
Using $\ell$ to perform a dimensional reduction to $D=4$, $\mathcal{N}=2$ gauged
supergravity, we show how the $D=5$ on-shell action can be computed using equivariant localization. We illustrate the formalism with some examples, computing the supersymmetric Casimir energy and the supersymmetric index of the dual SCFT without using the explicit supergravity solutions.

\end{abstract}

\end{titlepage}

\tableofcontents
\newpage

\section{Introduction}\label{app:WCP}

It has recently been appreciated that supersymmetric solutions of supergravity
theories with an R-symmetry generically have a set of equivariantly closed forms which
can be constructed from the supergravity fields and Killing spinor bilinears.
Furthermore, these forms can be used to compute various physical observables via localization i.e. 
the Berline--Vergne--Atiyah--Bott (BVAB) fixed point formula \cite{BV:1982,Atiyah:1984px} can be utilized 
to express various integrals in terms of quantities evaluated at the fixed point set of the R-symmetry Killing vector \cite{BenettiGenolini:2023kxp}.

In \cite{BenettiGenolini:2023kxp,BenettiGenolini:2024xeo,BenettiGenolini:2024hyd,BenettiGenolini:2024lbj} this approach was used to compute the on-shell action of $D=4$, $\mathcal{N}=2$ Euclidean
gauged supergravity. In the case of minimal gauged supergravity the results of \cite{BenettiGenolini:2019jdz}, which were obtained by a direct computation, were recovered using this new perspective and further generalized. In addition, the formalism was extended to $D=4$, $\mathcal{N}=2$ Euclidean gauged supergravity coupled to an arbitrary number of vector multiplets. 
Using boundary counterterms that are associated with a supersymmetric renormalization scheme, remarkably, it was shown
that the $D=4$ on-shell action can be expressed entirely in terms of the fixed point set of the supersymmetric Killing vector in the bulk of the solution. In particular, the contribution from the boundary counterterms precisely cancel a boundary contribution that arises after 
using the BVAB theorem for the bulk action.

The goal of this paper is to extend the results of \cite{BenettiGenolini:2023kxp,BenettiGenolini:2024xeo,BenettiGenolini:2024hyd,BenettiGenolini:2024lbj} to $D=5$
Euclidean
gauged supergravity, coupled to $n$ vector multiplets. 
We focus on supergravity solutions on a manifold $M_{(5)}$ which have, in addition to the $D=5$ R-symmetry Killing vector, $\cK$, a second Killing vector $\ell$.
We assume that $\ell$ is nowhere vanishing and generates a circle fibration of the $D=5$ spacetime with 
base space $M_{(4)}$, which may have orbifold singularities.\footnote{One could also consider more general kinds of singularities, such as those arising in the $D=5$ Kaluza--Klein monopole \cite{Sorkin:1983ns,Gross:1983hb}, but that will not be the focus here.} Furthermore, $M_{(4)}$ will have an asymptotic
boundary that is associated with $M_{(5)}$ asymptotically approaching Euclidean $AdS_5$.
We also assume that the push-forward of $\cK$ is not identically zero, and
hence descends to give a Killing vector $\xi$ on $M_{(4)}$. With these ingredients, we carry out a dimensional
reduction of the $D=5$ supergravity theory to obtain a $D=4$, $\mathcal{N}=2$ Euclidean
gauged supergravity on $M_{(4)}$ coupled to $n+1$ vector multiplets. Furthermore, we find that
$\xi$ is the $D=4$ R-symmetry Killing vector which can be constructed as a bilinear of the $D=4$
Killing spinors. We then use the $D=4$ results of \cite{BenettiGenolini:2023kxp,BenettiGenolini:2024xeo,BenettiGenolini:2024hyd,BenettiGenolini:2024lbj} to compute the $D=5$ on-shell action.

While our approach is conceptually straightforward, there are several technical issues which make the analysis 
of $D=5$ supergravity significantly more challenging than that of $D=4$ supergravity. The first is that
identifying the holographic boundary terms which are required for a
supersymmetric renormalization scheme is more involved
 than for $D=4$. 
 For the class of solutions with boundary $S^1\times M_3$, such counterterms have been identified
in minimal gauged supergravity \cite{BenettiGenolini:2016qwm, BenettiGenolini:2016tsn}
and can be used to compute the supersymmetric Casimir energy of the dual SCFT,
an intrinsic quantity in the dual SCFT \cite{Assel:2015nca}. For other classes of solutions
one can take a pragmatic approach of removing divergences by suitably employing a  heuristic background subtraction procedure.
Here we will illustrate how the localization procedure can be used for both classes.

A second issue which arises is that after reducing on $\ell$ and
using the BVAB theorem on the base space $M_{(4)}$, 
we obtain a contribution to the action from the fixed point set of the $D=4$ R-symmetry Killing vector $\xi$, 
some boundary terms (as in \cite{BenettiGenolini:2023kxp,BenettiGenolini:2024xeo,BenettiGenolini:2024hyd,BenettiGenolini:2024lbj}), as well
as an additional term which is the integral of a closed four-form $\Lambda_4$. Locally, we can write 
$\Lambda_4=\rd\Lambda_3$, but in general $\Lambda_3$ is not globally defined, and moreover, can depend
on the choice of gauge for the gauge fields. A third, and related, issue is that in the examples we consider,
for certain choices of Kaluza--Klein (KK) vector $\ell$, the final result for the $D=5$ on-shell action can be expressed purely
in terms of the fixed point data of the $D=4$ supersymmetric Killing vector $\xi$ on $M_{(4)}$, but
for other choices of $\ell$ one also gets contributions from the boundary as well as from $\Lambda_4$.

Nevertheless, we show how the localization formalism can be used to obtain the $D=5$ on-shell action 
for two classes of solutions, without using the explicit solutions. The first is Euclidean $AdS_5$ with $S^1\times S^3$ boundary
and using the counterterm action of \cite{BenettiGenolini:2016qwm, BenettiGenolini:2016tsn}, we recover the known result for the supersymmetric Casimir energy. 
The second is the complex locus of supersymmetric Euclidean black hole solutions of \cite{Cabo-Bizet:2018ehj}. Using background subtraction we obtain a result for the $D=5$ action in minimal gauged supergravity
that is dual to the supersymmetric index of the dual SCFT. We also extend this to arbitrary numbers of vector multiplets, confirming the conjecture of \cite{Hosseini:2018dob}.

Before discussing the plan of the paper, we briefly comment on other approaches to localization in $D=5$ supergravity that have appeared.
An analysis of $D=5$ ungauged supergravity coupled to vector multiplets was carried out in \cite{Cassani:2024kjn}. 
For a class of solutions with toric symmetry, an equivariantly closed extension of the on-shell action was constructed and used to compute the on-shell action. A connection to $D=4$ solutions was also made.   
In addition, inspired by the generalization of the BVAB theorem to
odd dimensions given in \cite{Goertsches:2015vga}, 
a computation of the on-shell action
for $D=5$ minimal gauged supergravity was presented in \cite{Colombo:2025ihp}, which 
is applicable to 
Euclidean solutions that can be obtained from a Wick rotation of Lorentzian supergravity solutions.

These works, like ours, require the existence of a second Killing vector in $D=5$, in addition to 
the supersymmetric Killing vector. In contrast to \cite{Cassani:2024kjn}, we use the additional isometry, which is assumed to act locally freely, to reduce to $D=4$ from the start, and  then localize in $D=4$  using the supersymmetric Killing vector. Instead 
\cite{Colombo:2025ihp} uses the mathematical result in  \cite{Goertsches:2015vga} which essentially applies a transverse BVAB theorem to the additional 
isometry, having effectively reduced on the supersymmetric Killing vector direction.
The reduction/localization are effectively then carried out after exchanging the roles of the two Killing vectors, when comparing the approach here to that of \cite{Colombo:2025ihp}. 
Thus, we expect that the issues that we highlighted above, and discuss further in the text, will also arise in the approach of \cite{Colombo:2025ihp} (where background subtraction was utilized to regulate the $D=5$ action). 
We also remark that our formalism is applicable to $D=5$ Euclidean supergravity, which includes Euclidean solutions that can be obtained from a Wick rotation of Lorentzian supergravity solutions, as studied in \cite{Cassani:2024kjn,Colombo:2025ihp}, but also solutions that cannot be obtained in this way. 
Finally, we note that another recent approach to localization in supergravity has
emphasized the significance of the equivariant volume in even dimensions \cite{Martelli:2023oqk, Colombo:2023fhu}.

The plan of the paper is as follows. In section \ref{bigsec:dimred}
we present the $D=5$ Euclidean supergravity theory of interest and carry out the dimensional reduction to $D=4$.
We also discuss supersymmetric counterterms as well as the background subtraction procedure.
In section \ref{examplesads5}
we discuss the example of Euclidean $AdS_5$ with $S^1\times S^3$ boundary, while
in section \ref{examplesblackhole}
we discuss supersymmetric black holes.
We briefly conclude in section \ref{sec:fincomms}.
We have five appendices: appendix \ref{appconvs}
includes our conventions for $D=5$ and $D=4$ Clifford algebras; 
appendix \ref{app:minsugra}
discusses how to recover minimal gauged supergravity from the general set-up with an arbitrary number of vector multiplets;
appendix \ref{app:5d_Bilinears} summarizes
some relations for $D=5$ bilinears and also includes some equivariantly closed polyforms in $D=5$; 
appendix \ref{sec:usefulformulaed4} presents some useful results for $D=4$ Euclidean gauged supergravity; appendix \ref{appdimredkse}
includes some details of the KK reduction of the $D=5$ Killing spinor equations to $D=4$.

\section{The dimensional reduction of \texorpdfstring{$D=5$}{D=5} supergravity}\label{bigsec:dimred}

\subsection{\texorpdfstring{$D=5$}{D=5} supergravity theory: Lorentzian}
\label{subsec:5d_Lorentzian}

We consider $D=5$ gauged supergravity coupled to $n$ Abelian vector multiplets. We first discuss the Lorentzian theory.
The theory is constructed using
a very special real manifold of real dimension $n$, specified by a symmetric tensor $C_{IJK}$, with $I,J,K = 1, \dots, n+1$, with real coordinates $Y^I$ satisfying the cubic constraint
	\begin{equation}\label{cubicconst}
		\cV_{(5)}(Y) \equiv \frac{1}{6} C_{IJK} Y^I Y^J Y^K = 1\,.
	\end{equation}
This can be parametrized by $n$ real scalars $\phi^i$, $i=1,\dots,n$.
There are $n+1$ gauge fields $\cA^I$ with curvatures $\cF^I \equiv \rd \cA^I$.
In addition, we also have a set of $n+1$ real Fayet--Iliopoulos (FI) parameters $\zeta_I$. 

The bosonic action has the form 
\begin{align}
\label{eq:Lorentzian5dAction}
	S_{(5)} &= \frac{1}{16\pi G_{(5)}} \int_{M_{(5)}} \Big[ \left( R - g_{ij}(\phi) *_{(5)} \dd\phi^i\wedge \dd\phi^j - \g^2 V_{(5)} \right) \vol_{(5)} \nn\\
	& \qquad \qquad \qquad \qquad  -  G_{IJ}(\phi) *_{(5)} \cF^I \wedge \cF^J - \frac{1}{6} C_{IJK} \cF^I \wedge \cF^J \wedge \cA^K \Big]\,,
\end{align}
where $G_{(5)}$ is the $D=5$ Newton constant, 
\begin{equation}
\label{eq:VerySpecialGeometry_1}
g_{ij}(\phi) = \partial_i Y^I \partial_j Y^J G_{IJ} \Big |_{\cV_{(5)}=1} 	 \, , \qquad 
	 G_{IJ}(\phi) = - \frac{1}{2} \frac{\partial^2 \ }{\partial Y^I \partial Y^J} \log \cV_{(5)} \Big|_{\cV_{(5)}=1}\, ,
\end{equation}
and the scalar potential is
\begin{equation}
\label{V5}
	V_{(5)}= \zeta_I \zeta_J \left( g^{ij} \partial_i Y^I \partial_j Y^J - \frac{4}{3} Y^I Y^J \right) \, .
\end{equation} 
For some further details, see e.g. \cite{Hosseini:2017mds,Benini:2020gjh, BenettiGenolini:2024kyy}.
We will assume that there is an $AdS_5$ vacuum and
  \begin{align}
 L^{-1}= \g \,,
 \end{align}
is the inverse of the radius of the $AdS_5$.
We also have
\begin{equation}\label{g5ayrel}
	\frac{1}{\g^3 G_{(5)}} = \frac{8a}{\pi} \, ,
\end{equation}
where $a$ is the central charge of the dual $d=4$ SCFT in the large $N$ limit (assuming it exists\footnote{We return to this point in the discussion section.}); for $\cN=4$ super-Yang-Mills $a=N^2/4$. 

A solution is supersymmetric if there is a non-trivial solution 
$\chi\fixme{_L}$ to the Killing spinor and gaugino equations
\begin{align}
      &0=\left[\nabla_m-\frac{\ii}{2} \g \zeta_I {\cA}^I_m+\frac{1}{6} \g W_{(5)} {\Gamma}_m +\frac{\ii}{8}Y_I{\cF}_{np}^I({\Gamma}_{m}{}^{np}-4\delta_m^n{\Gamma}^p)\right]\chi\fixme{_L}\,, \nonumber \\
\label{L5dKSE}
      &0=\left[-\frac{\ii}{2}g_{ij}\partial_m\phi^j {\Gamma}^m+\frac{\ii}{2} \g \partial_i W_{(5)} +\frac{3}{8}\partial_i Y_I{\cF}_{mn}^I {\Gamma}^{mn}\right]\chi\fixme{_L} \, .
\end{align}  
Here ${\Gamma}_m$ are the $D=5$ gamma matrices and we have defined the $D=5$ superpotential
\begin{align}\label{W5d}
W_{(5)} \equiv \zeta_I Y^I\, .
\end{align}
Notice that we may then write the potential in \eqref{V5} as
\begin{align}\label{P5d}
V_{(5)} = g^{ij}\partial_i W_{(5)} \partial_j W_{(5)} - \frac{4}{3} W_{(5)}^2\, . 
\end{align}
We also have $Y_I \equiv \frac{2}{3} G_{IJ} Y^J$ and $Y^IY_I =1$\fixme{, and we will assume in the following that the scalars parametrize a symmetric space, which guarantees symmetry properties of the tensor $C_{IJK}$ that will be useful in the reduction to four dimensions}.
From \eqref{L5dKSE}, the equations satisfied by the charge conjugate spinor
(in the conventions of
appendix \ref{appconvs})
are given by
\begin{align}
      &0=\left[\nabla_m + \frac{\ii}{2}\g \zeta_I {\cA}^I_m - \frac{1}{6} \g W_{(5)} {\Gamma}_m + \frac{\ii}{8}Y_I{\cF}_{np}^I({\Gamma}_{m}{}^{np}-4\delta_m^n{\Gamma}^p)\right]\chi^c_{{L}}\,, \nonumber \\
\label{L5dKSEc}
      &0=\left[-\frac{\ii}{2}g_{ij}\partial_m\phi^j {\Gamma}^m - \frac{\ii}{2} \g\partial_i W_{(5)} + \frac{3}{8}\partial_i Y_I{\cF}_{mn}^I {\Gamma}^{mn}\right]\chi^c_{{L}} \, .
\end{align}  
The special case of $D=5$ minimal gauged supergravity is discussed in appendix \ref{app:minsugra}.

\subsection{\texorpdfstring{$D=5$}{D=5} supergravity theory: Euclidean}
To construct the $D=5$ Euclidean theory, which we study in the remainder of the paper, we follow the approach of \cite{Freedman:2013ryh}. We allow all fields in the Lorentzian action \eqref{eq:Lorentzian5dAction} to become complex, keeping the FI parameters $\zeta_I$ 
and the tensor $C_{IJK}$ real, 
and perform a Wick rotation to find the Euclidean theory with action given by
\begin{align}
\label{eq:Euclidean5dAction}
	I_{(5)} &= - \frac{1}{16\pi G_{(5)}} \int_{M_{(5)}} \Big[ \left( R - g_{ij} *_{(5)} \dd\phi^i\wedge \dd\phi^j - \g^2 V_{(5)}\right) \vol_{(5)} \nn \\
	& \qquad \qquad \qquad \qquad  -  G_{IJ}*_{(5)} \cF^I \wedge \cF^J - \frac{\ii}{6} C_{IJK} \cF^I \wedge \cF^J \wedge \cA^K \Big]\,.
\end{align}
We also double the spinors, leading to two Dirac spinors $\chi$ and $\tilde{\chi}$ that satisfy the equations derived from \eqref{L5dKSE} and \eqref{L5dKSEc}:
\begin{align}
\label{eq:E5dKSEepsilon}
      &0=\left[\nabla_m-\frac{\ii}{2}\g \zeta_I {\cA}^I_m+\frac{1}{6}\g W_{(5)} {\gamma}_m +\frac{\ii}{8}Y_I{\cF}_{np}^I({\gamma}_{m}{}^{np}-4\delta_m^n{\gamma}^p)\right]\chi\,, \nn\\
      &0=\left[-\frac{\ii}{2}g_{ij}\partial_m\phi^j {\gamma}^m+\frac{\ii}{2} \g \partial_i W_{(5)} +\frac{3}{8}\partial_i Y_I{\cF}_{mn}^I {\gamma}^{mn}\right]\chi\, \, , 
\end{align}
and
\begin{align}
\label{eq:E5dKSEepsilontilde}
	      &0=\left[\nabla_m + \frac{\ii}{2} \g \zeta_I {\cA}^I_m - \frac{1}{6} \g W_{(5)} {\gamma}_m + \frac{\ii}{8}Y_I{\cF}_{np}^I({\gamma}_{m}{}^{np}-4\delta_m^n{\gamma}^p)\right] \tilde{\chi} \,, \nn \\
      &0=\left[-\frac{\ii}{2}g_{ij}\partial_m\phi^j {\gamma}^m - \frac{\ii}{2} \g \partial_i W_{(5)} + \frac{3}{8}\partial_i Y_I{\cF}_{mn}^I {\gamma}^{mn}\right] \tilde{\chi} \, .
\end{align}

We can define the following $D=5$ Killing spinor bilinears
\begin{equation}\label{dfiveksebiliniears}
	\cS \equiv \overline{\tilde{\chi}} \chi \, , \qquad \SUSYVec^m \equiv \overline{\tilde{\chi}} {\gamma}^m \chi \, , \qquad \cU_{mn} \equiv \ii \overline{\tilde{\chi}} {\gamma}_{mn} \chi \, .
\end{equation}
These satisfy a number of algebraic and differential relations, which can be found in appendix \ref{app:5d_Bilinears}.
In particular,  $\SUSYVec$ is a Killing vector that generates a symmetry of the full solution, i.e. we also have $\mathcal{L}_\cK\phi^i=\mathcal{L}_\cK \cF^I=0$. We can also construct equivariantly closed extensions of the field strengths, and of the $D=5$ on-shell action (in the gauge $\SUSYVec \hook \cA^I = - \ii Y^I \cS$).  
Generically, these bilinears are complex.
One way of obtaining real bilinears in $D=5$ is to consider a section of the Euclidean theory where it's consistent to take $\tilde{\chi} = \chi^c$ (see appendix \ref{app:5d_EuclideanSpinors} for details on the charge conjugation in $D=5$, and we note that $\chi^c$ is not the same as the Lorentzian charge conjugate spinor $\chi^c_L$). 
Consistency between the equations satisfied by $\chi^c$ and $\tilde{\chi}$ would then require
\begin{equation}
\label{eq:5d_Reality_Contour}
	\cA^I\in \R \, , \qquad Y^I \in \ii \R \, , \qquad \phi^i \in \R \, .
\end{equation}
As we shall see in section \ref{subsec:Reality}, though, these conditions are not consistent with the Kaluza--Klein (KK) ansatz that we shall use.

We also demand that the $D=5$ Killing spinors carry non-vanishing charge 
with respect to $\cK$, 
\begin{align}\label{lieckchi}
\cL_\cK \chi = \ii \cQ \chi\,,\qquad
\cL_\cK \tilde\chi = \ii \tilde \cQ \tilde\chi\,.
\end{align}
An expression for $\cQ$ is given in \eqref{cQueueexpapp} and one can show that
$\tilde\cQ =-\cQ $.
 
We are interested in Euclidean solutions that also admit non-trivial solutions to both \eqref{eq:E5dKSEepsilon}, \eqref{eq:E5dKSEepsilontilde} in order to have a non-trivial $\SUSYVec$, and will refer to these
as \emph{supersymmetric solutions} in the sequel.
A class of such solutions can be obtained by Wick rotating solutions of
the Lorentzian theory, which have been classified in \cite{Gauntlett:2003fk,Gutowski:2004yv}. 
It also seems likely that there are additional solutions that cannot be obtained this way; for example,
in a theory with both vector and hypermultiplets there are such Euclidean solutions with $\R^5$ topology, as discussed
in \cite{Bobev:2013cja, Bobev:2016nua}.

\subsection{Dimensional reduction}\label{sec:dimred}
Our goal is to compute the $D=5$ on-shell Euclidean action for supersymmetric solutions.
We will assume that in addition to the supersymmetric Killing vector, $\cK$, the $D=5$ solutions admit another Killing vector $\ell$.
We assume that $\ell$ is nowhere vanishing and generates a circle fibration of the $D=5$ spacetime with base $M_{(4)}$,
\begin{equation}
	\pi : M_{(5)} \to M_{(4)} \, .
\end{equation}
In general note that $M_{(4)}$ may have orbifold singularities.
We introduce local $D=5$ coordinates via $x^m=(x^\mu,x^5)$ with  
\begin{equation}\label{periodofthecircle}
	\KKVec = \partial_{x^5} \, , \qquad x^5 \sim x^5 + \Delta x^5 \, .
\end{equation}
We also assume that the pushforward of $\cK$ is not identically zero and write
\begin{equation}
\label{eq:Defn_xi}
	\FourdSUSYVec \equiv \pi_*(\SUSYVec) \, .
\end{equation}
In our setup below, we will see that $\xi$ is a supersymmetric Killing vector of a $D=4$ Euclidean supergravity theory. The fixed point set of $\xi$ in $M_{(4)}$ is precisely the 
subset over which $\SUSYVec$ and $\KKVec$ are aligned in $M_{(5)}$.

The $D=5$ metric on $M_{(5)}$ can be written in the form
\begin{align}\label{eq:Reduction_Ansatz1}
	\rd s^2_{(5)} &= \e^{-4 \lambda} \KKForm^2 + \e^{2\lambda} \rd s^2_{(4)} \,,
\end{align}
where $\rd s^2_{(4)}$ is a metric on $M_{(4)}$ and $\alpha\equiv \e^{4 \lambda} \KKVec^\flat$ is a globally-defined angular form on $M_{(5)}$, with $\KKVec^m\KKVec_m\equiv \e^{-4 \lambda}$. In local coordinates we can write
\begin{equation}
\label{eq:Reduction_Ansatz2}
	\KKForm = \rd x^5 - A^0 \, ,
\end{equation}
where $A^0$ is a $D=4$ gauge field.
We can also write the $D=5$ scalars and gauge fields in the form 
\begin{align}
\label{eq:Reduction_Ansatz3}
	 Y^I &= - \e^{2\lambda} z_2^I \, ,\nn \\ 
	 \cA^I &= A^I + z^I_1 \, \KKForm +\nvxi^I \rd x^5 \, ,\qquad  \Rightarrow \qquad
	 \cF^I = - \KKForm \wedge \rd z_1^I - z_1^I F^0 + F^I \, , 
\end{align}
where $A^I$ are $D=4$ gauge fields and $F^I=\rd A^I$.
The $\nvxi^I$ are constants that we discuss further below.
It is also convenient to define 
\begin{equation}
\label{eq:Reduction_Ansatz4}
	z^I \equiv z^I_1+\ii z^I_2  \, , \qquad \tilde z^I \equiv z^I_1-\ii z^I_2\, ,
\end{equation}
so $z^I_1 = \frac{1}{2}(z^I + \ztilde^I)$, $z^I_2 =\frac{1}{2\ii}(z^I - \ztilde^I)$
and we note that $z^I$ and $\tilde z^I$ are independent complex scalar fields in $D=4$.
Notice 
that we can express $\lambda$ in terms of $z^I$, $\ztilde^I$ via the cubic constraint \eqref{cubicconst}:
\begin{align}\label{lambdazrel}
\frac{1}{6} C_{IJK} z_2^I z_2^J z_2^K = - \e^{-6\lambda}\,.
\end{align}
At this stage we have the $D=4$ metric, $\rd s^2_{(4)}$, gauge fields $A^\Lambda=(A^0,A^I)$ and scalars
$z^I,\ztilde^I$, all of which are independent of $x^5$, and the constants $\nvxi^I$ parametrizing a $D=5$ flat connection which is not, in general, a global flat connection.

The KK ansatz \eqref{eq:Reduction_Ansatz1}-\eqref{eq:Reduction_Ansatz4}
is ambiguous, because we can perform various $D=5$ gauge transformations on $\cA^I$. 
First, the shift 
$\cA^I \to \cA^I + \diff \Gamma^I$, where $\Gamma^I$ is a local basic 
function for $\ell$ (so $\mathcal{L}_\ell\Gamma^I=0$), 
is associated
with a $D=4$ gauge transformation of $A^I$; that is,  $A^I \to A^I + \diff\Gamma^I$. Second, the coordinate redefinition $x^5 \to x^5 + \omega$ with $\omega$ a function on $M_{(4)}$, is associated with a gauge transformation for $A^0$, 
that is $A^0 \to A^0 + \rd\omega$. Finally, there is also the possibility of performing
transformations of the form $\cA^I \to \cA^I + c^I \, \rd x^5$, with $c^I$ constant,
which shifts $\nvxi^I\to \nvxi^I+c^I$.
Notice that in terms of the $D=4$ fields introduced in \eqref{eq:Reduction_Ansatz3},
the gauge transformation can be implemented
 via
$z_1^I \to z_1^I + c^I$, $A^I \to A^I + c^I A^0$. We find it convenient to
explicitly include $\nvxi^I$ in our ansatz \eqref{eq:Reduction_Ansatz3}, and 
then observe that the ansatz is invariant under the combined transformation
\begin{equation}
\label{eq:KK_Redundancy}
	 \nvxi^I \to \nvxi^I - c^I \, , \quad z^I \to z^I + c^I \, , \quad \ztilde^I \to \ztilde^I + c^I \, ,\quad  A^I \to A^I + c^I A^0 \, .
\end{equation}
Notice that the variables 
\begin{align}\label{checkvars}
\check{z}^I &\equiv z^I + \nvxi^I, \quad \check{\ztilde}^I \equiv \ztilde^I + \nvxi^I, \qquad
\check{z}_1^I \equiv z_1^I + \nvxi^I \,,
\nn\\
\check{A}^I &\equiv A^I + \nvxi^I A^0, \quad \check{F}^I \equiv F^I + \nvxi^I F^0,
\end{align}
are invariant under \eqref{eq:KK_Redundancy}, and so we anticipate that after dimensional reduction 
various $D=4$ quantities can be expressed in terms of these variables.
We highlight that in terms of these variables we can write $\cA^I$ in 
\eqref{eq:Reduction_Ansatz3} as 
\begin{align}
\label{eq:Reduction_Ansatz3nvs}
	 \cA^I &= \check A^I +  \check{z}_1^I \KKForm	 \, .
\end{align}
Observe that $\ell \hook \cA^I= \check{z}_1^I $ and so a $D=5$ gauge choice can impose a constraint on $\check{z}^I_1$.\footnote{For example, for solutions $M_{(4)}$ that have a non-trivial two-cycle, we may choose the $a^I$ to 
impose the constraint on the fluxes:
$\check{\mathfrak{p}}^I\equiv \mathfrak{p}^I+a^I\mathfrak{p}^0=0$, where the fluxes
$\mathfrak{p}^\Lambda$ are defined in  
\eqref{fluxesdefn}. We will make such a choice for the black hole example
in section \ref{examplesblackhole}. We won't 
make a choice of $a^I$ for the $AdS_5$ example in section \ref{examplesads5}.}

For later use we notice that if $\cA^I$ is a globally defined one-form, $\check{z}_1^I$ is globally defined (since $\ell \hook \cA^I= \check{z}_1^I $). Then given $\alpha$ is a globally defined one-form, we conclude that $\check A^I$ will be too.
We also note that
the $D=5$ spinor $\chi$ changes under gauge transformations. We assume that the $D=5$ spinors $\chi$ and $\tilde{\chi}$ satisfy the relevant spinor equations with the $D=5$ gauge field as in \eqref{eq:Reduction_Ansatz3}.

We next want to write the $D=5$ action in terms of a $D=4$ action. After substituting the ansatz
into \eqref{eq:Euclidean5dAction}, we find that we can rewrite the $D=5$ action as 
\begin{equation}
\label{eq:Reduction_5d_Lagrangian}
	I_{(5)} = I_{(4)} - \frac{{\Delta x^5}}{16\pi G_{(5)} } \int_{M_{(4)}} \Lambda_4 \, , 
\end{equation}
where $\Lambda_4$ is a basic four-form, which can be written,
locally, as $\Lambda_4= \rd \Lambda_3$ with
\begin{equation}
\label{eq:Lambda4_General}
	\Lambda_3
		= 2  *_{(4)} \rd\lambda +  \frac{\ii}{6}C_{IJK} \Big( 2 \check{z}^I_1 \check{A}^J \wedge \check{F}^K - \check{z}_1^I \check{z}_1^J \check{A}^K \wedge F^0  \Big)  \, ,
\end{equation} 
and $\frac{1}{G_{(4)}} = \frac{\Delta x^5}{G_{(5)}}$ relates the 
$D=4$ and $D=5$ Newton constants.

The $D=4$ action $I_{(4)}$ is given by the Euclidean $\mathcal{N}=2$, $D=4$ gauged supergravity theory, discussed in \cite{BenettiGenolini:2024lbj}. There are $n+1$ vector multiplets and $A^\Lambda\equiv (A^0,A^I)$ and we can write
\begin{align}\label{themainaction}
I_{(4)}	&= - \frac{1}{16\pi G_{(4)}} \int \Big[ \left( R - 2 \cG_{I\tilde{J}} \partial^\mu z^I \partial_\mu \ztilde^{\tilde{J}} - \g^2\cV_{(4)} (z, \ztilde)  \right)\vol_4 \nn\\
	& \qquad \qquad \qquad \qquad + \frac{1}{2} {\cI}_{\Lambda \Sigma} * F^\Lambda \wedge F^{\Sigma} - \frac{\ii}{2} {\cR}_{\Lambda \Sigma} F^\Lambda \wedge F^\Sigma \Big] \, .
\end{align}
The $D=4$ theory has $n+1$ vector multiplets and so $n+2$ gauge fields $A^\Lambda\equiv (A^0,A^I)$. There are $n+1$ scalars $z^I$ and their independent Euclidean counterparts $\ztilde^I$.
The metric on the scalar manifold and the potential are simply related to $D=5$ quantities via
\begin{align}
	\label{eq:4d_ReducedMetric_ScalarManifold}
	\cG_{I\tilde{J}} &= \frac{1}{2} \e^{4\lambda} G_{IJ}(z,\ztilde) \, , \nn\\
	 \cV_{(4)}(z,\ztilde) &= 
	 \e^{2\lambda} V_{(5)} (z,\ztilde) \, .
\end{align}
We also have
\begin{align}\label{thecalIs}
	\cI_{00} &=  - \e^{-6\lambda} \left(  1 + 4\cG_{IJ}z_1^I z_1^J \right) \, , \nn\\
	\cI_{0I} &= 4 \e^{-6 \lambda} \cG_{IJ} z_1^J \, ,\nn \\
	\cI_{IJ} &= - 4 \e^{-6\lambda} \cG_{IJ} \, ,
\end{align}
and
\begin{align}\label{thecalRs}
	\cR_{00} &= \frac{1}{3} C_{IJK} \Big( z^I_1 z^J_1 z^K_1 + \nvxi^I \nvxi^J \nvxi^K \Big) \, , \nn \\
	\cR_{0I} &= - \frac{1}{2} C_{IJK} \Big( z^J_1 z^K_1 - \nvxi^J \nvxi^K \Big) \, , \nn \\
	\cR_{IJ} &= C_{IJK} \left( z_1^K + \nvxi^K \right) \, .
\end{align}
Notice that ${\cR}_{\Lambda \Sigma}$ only depends on $z_1^I$, but 
$\cI_{\Lambda\Sigma}$ also depends on $z_2^I$, both via the 
warp factor $\e^{-6\lambda}$, as in \eqref{lambdazrel}, and $\cG_{IJ}$, as in \eqref{eq:4d_ReducedMetric_ScalarManifold}.
We can also inspect whether the action $I_{(4)}$ in \eqref{themainaction} is invariant under the shifts \eqref{eq:KK_Redundancy}, as it should be, since it combines with $\Lambda_4$, which in \eqref{eq:Lambda4_General} is locally written in terms of the invariant variables \eqref{checkvars}, to give $I_{(5)}$, which is invariant by definition. 
In fact one can check that the entire second line of \eqref{themainaction}, gauge fields included, can be written in terms of \eqref{checkvars}, and hence is also invariant under \eqref{eq:KK_Redundancy}. 

The $D=4$ action $I_{(4)}$ is governed by a special K\"ahler manifold with a prepotential that can be written as
(see \cite{Benini:2020gjh}):
	\begin{equation}
	\label{eq:4d_Prepotential}
		\cF(X) =  \frac{1}{6} \frac{C_{IJK} \check{X}^I \check{X}^J \check{X}^K}{X^0} \,, \qquad \check{X}^I \equiv X^I + \nvxi^I X^0 \, ,
	\end{equation}
	where the $n+2$ holomorphic sections $X^\Lambda=(X^0, X^I)$, with $\Lambda = 0, 1, \dots, n+1$ are homogeneous coordinates, and the physical complex scalars are given by $z^I \equiv X^I / X^0$. 
The K\"ahler potential is given by\footnote{We use the same letter for the K\"ahler potential and the $D=5$ Killing vector bilinear, but the meaning should be clear from the context.}
\begin{equation}
\label{eq:4d_Kahler_Potential}
	\cK = - \log ( 8  X^0 \tX^0 \e^{-6\lambda} )  \, ,
\end{equation}
and $\cG_{I\tilde{J}}\equiv \partial_I\partial_{\tilde J}\cK$, where here (and in \eqref{eq:EuclideanScalarPotential}
below) the derivative $\partial_I$ is with respect to $z^I=X^I / X^0$ (i.e. use \eqref{lambdazrel} and hold $X^0, \tilde X^0$ fixed).\footnote{Throughout the paper, and in particular  to recover the $D=4$ Killing spinor equations in the conventions of \cite{BenettiGenolini:2024lbj} from dimensional reduction, we choose $X^0 = \tX^0 = 1$  (see appendix \ref{appdimredkse}).\label{footnote:ChoiceSections}}
Furthermore the $D=4$ potential can be written
\begin{align}
\label{eq:EuclideanScalarPotential}
	\cV_{(4)} &= \e^{\cK} \left( \cG^{I \tilde{J}} \nabla_I W \nabla_{\tilde{J}} \tW - 3 W \tW \right) \, ,
\end{align}
where
\begin{equation}
\label{eq:Euclidean_Superpotentials}
	W \equiv \xinew_\Lambda^{(4)} X^\Lambda \, , \qquad \tW \equiv \xinew^{(4)}_\Lambda \tX^\Lambda \, .
\end{equation}
Here $\zeta^{(4)}_\Lambda$ are the $n+2$ $D=4$ FI parameters given by
	\begin{equation}
	\label{eq:4d_FI_Parameters}
		\g \zeta_0^{(4)} = - 4Q^{(\ell)} + 2\g  \zeta_I \nvxi^I \, , \qquad
		 \zeta_I^{(4)} = 2
		 \zeta_I \, .
\end{equation}
Here $Q^{(\ell)}$ is the charge of the Killing spinor with respect to the $D=5$ Killing vector $\ell$, i.e.
$\cL_\KKVec \chi = \ii Q^{(\ell)} \chi$, which we discuss below. 
The $D=4$ scalar potential $\cV_{(4)}$ is proportional to the $D=5$ scalar potential $V_{(5)}$ via \eqref{eq:4d_ReducedMetric_ScalarManifold}, and the latter only depends on the $D=5$ FI parameters, as in \eqref{W5d}, \eqref{P5d}. Therefore, the potential $\cV_{(4)}$ must only depend on $\zeta_I^{(4)}$ and be independent of $\zeta^{(4)}_0$: we now  explicitly check that this is indeed the case.
Using the definitions \eqref{eq:Cov_Derivates_Sections_app}, the terms in $\cV_{(4)}$ containing $\zeta_0^{(4)}$ are
\begin{align}
\label{eq:V4_supset_z0}
	\cV_{(4)} &\supset \e^{6\lambda} \Big[ \left( (\zeta_0^{(4)})^2 + 2\zeta_0^{(4)} \zeta_I^{(4)} z^I_1 \right) \left( 12 \cG^{K\tilde{L}} \partial_K\lambda \partial_{\tilde{L}}\lambda - 1 \right) \nn \\
	& \qquad \quad + 2\zeta_0^{(4)}\zeta_I^{(4)} \left( \cG^{I\tilde{J}} \partial_{\tilde{J}}\lambda + \cG^{J\tilde{I}}\partial_J\lambda \right) \Big] \, .
\end{align}
It follows from \eqref{lambdazrel} that
\begin{align}
\label{eq:app_ScalarPotKK_Useful_6}
    \partial_I \lambda &= - \frac{\ii}{24} C_{LMI} z^L_2 z^M_2 \e^{6\lambda} = - \frac{\ii}{4} \e^{2\lambda} Y_I \, , \nn\\ 
    \partial_{\tilde{I}} \lambda &= \frac{\ii}{24} C_{LMI} z^L_2 z^M_2 \e^{6\lambda} = \frac{\ii}{4} \e^{2\lambda} Y_I \, ,
\end{align}
where we have used
\begin{equation}
\label{eq:5d_Relations_Useful_for_Forms}
    Y_I \equiv \frac{2}{3} G_{IJ} Y^J \Big\rvert_{\cV_{(5)}=1} \, , \qquad Y_IY^I = 1 \, , \qquad C_{IJK} Y^JY^K = 6 Y_I \, .
\end{equation}
Then using $\cG^{K\tilde{L}} = 2 \e^{-4\lambda} G^{KL}$ the first line in \eqref{eq:V4_supset_z0} vanishes, and the second line vanishes because of the symmetry of $G^{IJ}$.
While the $D=4$ potential is independent
of $\zeta^{(4)}_0$, we will see that both $\zeta_I^{(4)}$ and $\zeta^{(4)}_0$ appear in the $D=4$ Killing spinor equations.

To summarize,
we are interested in
computing the on-shell $D=5$ action for a supersymmetric solution of the $D=5$ equations of motion, 
that lies within the ansatz \eqref{eq:Reduction_Ansatz1}-\eqref{eq:Reduction_Ansatz4}. The above result \eqref{eq:Reduction_5d_Lagrangian} gives the answer. We will later use the equivariant localization results of \cite{BenettiGenolini:2024lbj} which analysed supersymmetric solutions
to the $D=4$ theory with action $I_{(4)}$. We have checked that solutions of the $D=4$ equations of motion associated with $I_{(4)}$ will also give rise to solutions of the $D=5$ equations of motion.
Furthermore, we have checked that the solutions which also admit solutions to the
$D=4$ Killing spinor equations, to be discussed next, give rise to $D=5$ solutions
that admit solutions to the $D=5$ Killing spinor equations.

We now discuss the dimensional reduction of the Killing spinor equations. 
To dimensionally reduce the $D=5$ Killing spinor equations we introduce the obvious
$D=5$ orthonormal frame 
\begin{align}\label{Dequalsfiveframe}
{\rm E}^a = \e^{\lambda} \e^a\,,\qquad  {\rm E}^5 = \e^{-2\lambda} \KKForm\,,
\end{align} 
with
$\e^a$, $a=1, \dots, 4$,  an orthonormal frame for the $D=4$ metric $\rd s^2_{(4)}$.
For the $D=4$ gamma matrices we take $\gamma^a$, $a=1,\dots,4$ and since $\gamma_{1234}=\gamma^5=\gamma_5$, we see that $\gamma_5$ is also the $D=4$ chirality matrix. 
We assume that 
$\KKVec$ generates a symmetry of the entire $D=5$  solution, and therefore the $D=5$ Killing spinor equations are invariant under the action of $\ell$. Thus, the space of solutions will form a representation of the $U(1)$ symmetry generated by $\KKVec$. But since the irreducible representations are then one-dimensional, classified by their charge, it follows that we may take a solution $\chi$ with definite charge \cite{Ferrero:2021etw}, without loss of generality. One can check that in the frame we are using, acting on $D=5$ spinors we have
$\cL_{\KKVec} = \partial_{x_5}$.
Therefore, we can take
\begin{equation}\label{liederchiell}
	\cL_{\KKVec} {\chi} = \partial_{x_5}\chi= \ii Q^{(\ell)} \chi \,,\qquad
		\cL_{\KKVec} \tilde{\chi}= \partial_{x_5}\tilde \chi = \ii \tilde Q^{(\ell)} \tilde{\chi} \, .
\end{equation}

The reduction of the Killing spinor equations is described in appendix \ref{appdimredkse}. 
We introduce $D=4$ spinors $\epsilon$, $\tilde\epsilon$
via 
\begin{equation}\label{epschirels}
	\epsilon = {\ii} \e^{-\lambda/2} \e^{-\ii Q^{(\ell)}x^5} \e^{-\frac{\pi \ii}{4}\gamma_5} \chi \,,\qquad
		\tilde{\epsilon} = {\ii} \e^{-\lambda/2} \e^{-\ii \tilde{Q}^{(\ell)}x^5} \e^{\frac{\pi \ii}{4}\gamma_5} \tilde{\chi} \,,
\end{equation}
which only depend on the $D=4$ coordinates $x^\mu$ (i.e. $\partial_{x^5} \epsilon = \partial_{x^5} \tilde{\epsilon} = 0$). As anticipated in footnote \ref{footnote:ChoiceSections}, to match the conventions of \cite{BenettiGenolini:2024lbj}, we choose the sections such that
\begin{equation}
\label{eq:X0tX01}
    X^0 = \tilde{X}^0 = 1 \,.
\end{equation}
We then find that the $D=5$ Killing spinor equations precisely give those of the $D=4$ gauged supergravity theory
provided that we take
\begin{align}\label{qqtilderel}
Q^{(\ell)}=-\tilde Q^{(\ell)}\,,
\end{align}
with the FI parameters as given in \eqref{eq:4d_FI_Parameters}.
These $D=4$ reduced Killing spinor equations are given in appendix \ref{appdimredkse}.
Also, recalling the definition of the $D=5$ bilinears $\cS$ and $\cK$ in \eqref{dfiveksebiliniears},
we notice \eqref{qqtilderel} clearly implies $ \KKVec^m \partial_m\cS=\partial_{x^5}\cS=0$. Further using the
$D=5$ Killing spinor bilinear identity $\rd \cS = - \ii \SUSYVec \hook ( Y_I \cF^I )$
in appendix \ref{app:5d_Bilinears},
we deduce the $D=5$ condition
\begin{align}\label{QQtildeconstraint}
0 =  Y_I \KKVec^m \cF^I_{mn} \cK^n \,.
\end{align}
In fact, provided $\cS\ne 0$ this condition is actually equivalent to \eqref{qqtilderel}.\footnote{To prove the converse, one obtains expressions for $ \ii Q^{(\ell)}\chi$ using 
\eqref{eq:E5dKSEepsilon} and for $\ii \tilde Q^{(\ell)}\tilde \chi$ using 
\eqref{eq:E5dKSEepsilontilde} and then uses these to obtain expressions for
$Q^{(\ell)} \, \ii \overline{\tilde{\chi}}\chi$ and $\tilde{Q}^{(\ell)} \, \ii \overline{\tilde{\chi}} \chi$.} The condition \eqref{qqtilderel} is satisfied for all of the examples we study later. 
 We also highlight that \eqref{epschirels} allows us to relate the $D=5$ bilinears,  $\cS, \cK$
 and the $D=4$ bilinears $S, P, \xi$, as reviewed in appendix \ref{sec:usefulformulaed4} (see equation \eqref{eq:Bilinears_app}), finding 
\begin{align}
\label{eq:5d4dvectors}
	\cS = {-} \e^\lambda S \, , \qquad \SUSYVec = \xi^\mu \partial_\mu + (\xi \hook A^0 {-} \e^{3\lambda}P) \partial_{x^5} \, .
\end{align}

From now on, we will assume that we are in the gauge
\begin{align}\label{ggechoiceone}
    \cL_\xi A^0=0\,,
\end{align}
which is possible since $\cL_\xi F^0 = 0$, and, moreover,  
is consistent with \cite{BenettiGenolini:2024lbj}. In this gauge, we find that
\begin{equation}
\label{eq:Constant_Component_K}
    \rd (\xi \hook A^0 {-} \e^{3\lambda}P) = 0 \, , 
\end{equation}
so the component of $\SUSYVec$ along $\partial_{x^5}$ is a constant, which we label
\begin{equation}
\label{eq:Definition_Rup}
    \Rup \equiv \frac{2\pi}{\Delta x^5} \left(\xi \hook A^0 {-} \e^{3\lambda}P \right) \, .
\end{equation}
This also allows us to find the connection between $\cQ$, 
the charge of the spinor $\chi$ under $\mathcal{K}$, as in \eqref{lieckchi}, 
and the charge $Q^{(\xi)}$ of the spinor $\epsilon$ under $\xi$, namely
\begin{align}\label{qrels}
 \cQ = Q^{(\xi)} + (\xi \hook A^0 - \e^{3\lambda} P) Q^{(\ell)}\,.
\end{align}

At this stage we have shown that a supersymmetric solution of the $D=5$ gauged supergravity theory with 
an extra Killing vector $\ell$ and satisfying \eqref{qqtilderel}, is equivalent to a supersymmetric solution of 
a $D=4$ Euclidean gauged supergravity theory as described above.
For such solutions we can then import the equivariant localization results of \cite{BenettiGenolini:2024lbj}.
The $D=4$ gauge field strengths have an equivariantly closed completion given by
\begin{align}
\Phi^\Lambda= F^\Lambda+\Phi_0^\Lambda\,,\qquad \rd_\xi \Phi^\Lambda=0\,.
\end{align}
There is also an equivariantly closed four-form
\begin{align}
\Phi=\Phi_4+\Phi_2+\Phi_0\,, \qquad \rd_\xi \left( \Phi_4 + \Phi_2 + \Phi_0 \right) = 0\,.
\end{align}
On-shell, the $D=4$ Euclidean bulk action $I_{(4)}$ can be expressed as
an integral of $\Phi_4$ and hence evaluated using the BVAB theorem.
Explicit expressions for $\Phi_0^\Lambda$, as well as $\Phi_4$, $\Phi_2$ and $\Phi_0$, which
are constructed from the $D=4$ supergravity fields and the $D=4$ Killing spinor bilinears, can be found
in \cite{BenettiGenolini:2024lbj}, and in appendix \ref{sec:usefulformulaed4}.

\subsection{Evaluating the \texorpdfstring{$D=5$}{D=5} on-shell action}\label{ctoshelldimredact}

We are interested in computing the on-shell action 
for a supersymmetric solution of $D=5$ gauged supergravity. 
The total $D=5$ action computed using holographic renormalization consists of the bulk action $I_{(5)}$ supplemented with boundary contributions which we write as
\begin{align}\label{totalactos}
\Itot_{(5)}=
I_{(5)}[M_{(5)}]+I_{\mathrm{GHY}}^{\partial M_{(5)}}+I_{(5)}^{\partial M_{(5)}}\,.
\end{align}
Here $I_{\mathrm{GHY}}^{\partial M_{(5)}}$ is the standard Gibbons--Hawking--York term and
$I_{(5)}^{\partial M_{(5)}}$ are counterterms to remove divergences, as well as
finite counterterms, which are associated with choosing a renormalization scheme. Since we are interested in supersymmetric solutions, 
 we want to employ a supersymmetric scheme. In contrast to the $D=4$ setting, 
a general set of supersymmetry-preserving counterterms which are written in terms of the boundary metric and gauge field and preserving boundary diffeomorphisms, are not known in $D=5$. 
The issue has been studied in some detail for the special case when
$\partial M_{(5)} \cong S^1 \times M_3$, where $M_3$ is a Seifert manifold:  coupling a SCFT to this background using the R-symmetry multiplet 
and requiring a supersymmetric regularization, leads to an unambiguous definition of the Casimir energy of the SCFT obtained from the partition function on $S^1 \times M_3$ \cite{Assel:2015nca} (this has been especially studied in the case of $M_3$ being homeomorphic to $S^3$). However, in the presence of an 't Hooft anomaly for the $U(1)_R$ symmetry, coupling via the R-multiplet leads to an anomaly in the supercurrent, which has been argued to be related to the non-vanishing value of the supersymmetric Casimir energy and to the anomalous dependence of the partition function on the background geometry \cite{Papadimitriou:2017kzw, An:2017ihs, Papadimitriou:2019gel, Closset:2019ucb, Kuzenko:2019vvi, Bzowski:2020tue}. The latter issue had been originally flagged in \cite{BenettiGenolini:2016qwm, BenettiGenolini:2016tsn}, and a set of local finite counterterms was proposed that restored supersymmetry at the expense of invariance under boundary diffeomorphisms and gauge invariance (from the field theory side, a local functional on $M_3$ acting as a counterterm to preserve supersymmetry has also been suggested in \cite{Closset:2019ucb}). 
The supersymmetry-preserving counterterms suggested in \cite{BenettiGenolini:2016qwm, BenettiGenolini:2016tsn}, however, have been derived for supersymmetric solutions to minimal gauged supergravity and with a gauge field that is a global one-form. 

We will use the counterterms of \cite{BenettiGenolini:2016qwm, BenettiGenolini:2016tsn} to compute the on-shell action for Euclidean $AdS_5$ with $S^1\times S^3$ boundary in section \ref{examplesads5}, and recover the holographic expression for the 
supersymmetric Casimir energy using localization. However, we note that in section \ref{examplesblackhole}, where we compute the on-shell action for supersymmetric black holes, we will instead use background subtraction, which we discuss in section \ref{sec:bkgsub}.

The class of solutions we are interested in have 
an additional Killing vector $\ell=\partial_{x_5}$
and can be written in the form of the Kaluza--Klein (KK) ansatz \eqref{eq:Reduction_Ansatz1}-\eqref{eq:Reduction_Ansatz4}. The bulk on-shell action $I_{(5)}$ can then be written in the form 
\eqref{eq:Reduction_5d_Lagrangian} where $I_{(4)}$ is the bulk on-shell action for a $D=4$ gauged supergravity theory. We can now use the results of \cite{BenettiGenolini:2024lbj} to evaluate
$I_{(4)}$. As noted at the end of the last subsection, it was shown in \cite{BenettiGenolini:2024lbj} that the bulk $D=4$ action 
$I_{(4)}$ can be written as the integral of an equivariant closed form $\Phi=\Phi_4+\Phi_2+\Phi_0$
satisfying $\rd_\xi \left( \Phi_4 + \Phi_2 + \Phi_0 \right) = 0$,
which is canonically constructed from the $D=4$ supergravity fields and the $D=4$
Killing spinor bilinears. Using the BVAB theorem, the 
on-shell $D=4$ bulk action can then be written
\begin{align}
I_{(4)} = I_{(4)}^{\rm FP} + I_{(4)}^{\partial M_{(4)}}\,,
\end{align}
where $I_{(4)}^{\rm FP} $ is a contribution from the fixed point set of $\xi$ in $M_{(4)}$
and $I_{(4)}^{\partial M_{(4)}}$ is the $D=4$ boundary contribution given by
	\begin{equation}
		I_{(4)}^{\partial M_{(4)}} = - \frac{1}{8 \pi G_{(4)}} \int_{\partial M_{(4)}} \FourdSUSYForm \wedge \left( \Phi_2 + \Phi_0 \, \rd \FourdSUSYForm \right) \, , 
	\end{equation} 
	where the one-form $\eta$ on $M_{(4)}$ is given by $\FourdSUSYForm \equiv |\xi|^{-2} \FourdSUSYVec^\flat$. It was shown in \cite{BenettiGenolini:2024lbj} that for supersymmetric 
	solutions of a $D=4$ supergravity theory with $M_{(4)}$ asymptotically approaching
	Euclidean $AdS_4$ this boundary term exactly cancels the associated 
	$D=4$ counterterms (in a $D=4$ supersymmetric scheme). Here, however, $M_{(4)}$ does not asymptotically approach Euclidean $AdS_4$
	and moreover the boundary terms that are relevant for our purposes are the $D=5$ terms
$I_{(5)}^{\partial M_{(5)}}$ in \eqref{totalactos}.

Putting these ingredients together we can therefore write the $D=5$ total on-shell action \eqref{totalactos}
as
\begin{align}
\label{eq:totac_fp_bdy}
	\Itot_{(5)} 	= I_{(4)}^{\rm FP} &- \frac{1}{8\pi G_{(5)}} \int_{\partial M_{(5)}} \KKForm \wedge \FourdSUSYForm \wedge \left( \Phi_2 + \Phi_0 \, \rd \FourdSUSYForm \right) - \frac{\Delta x_5}{16\pi G_{(5)}} \int_{M_{(4)}} \Lambda_4 \nn\\ &+I_{\mathrm{GHY}}^{\partial M_{(5)}}+ I_{(5)}^{\partial M\fixme{_{(5)}}}\,,
\end{align}
where we recall\footnote{On-shell we can use the $D=4$ equation of motion for $\lambda$ to rewrite
$\rd \Lambda_3$, but the resulting expression is not very illuminating; see \eqref{eq:Lambda4_Minimal} for the case of minimal gauged supergravity.}
that $\Lambda_4= \rd \Lambda_3$ with $\Lambda_3$ given in \eqref{eq:Lambda4_General}.
Remarkably we find that for certain classes of solutions, and for a certain choice of Killing vector 
$\ell$, the last four terms on the right hand side of this expression exactly cancel and we have
$I_{(5)} 	= I_{(4)}^{\rm FP}$. However, for different choices of $\ell$ this is no longer the
case and one finds that all four terms combine with the fixed point contribution to give the same total result for $I_{(5)}$.  

The expression for $I_{(4)}^{\rm FP}$ can be obtained from the results of 
\cite{BenettiGenolini:2024lbj} for the specific $D=4$ gauged supergravity theory obtained from the dimensional reduction. As explained in \cite{BenettiGenolini:2024lbj} the fixed point set of $\xi$ on $M_{(4)}$ consists of isolated fixed points (nuts) or fixed surfaces (bolts) and on them the Killing spinor
has definite $\pm$ chirality with respect to the $D=4$ chirality operator $\gamma_5$.
Specifically, we can write 
\begin{align}
\label{eq:5d_Conjectural_OSAction}
	 I^{\rm FP}_{(4)}&= \frac{\Delta x^5\pi}{\g^2 G_{(5)}} \Bigg[ \sum_{\rm nuts_{\pm}} \frac{1}{d_{F_0}} \frac{(b_1 \mp b_2)^2}{b_1b_2}\ii {\cF}({u}_\pm) 
	 \nn \\
	& \qquad \quad + \sum_{\rm bolts_{\pm}} \frac{1}{d_{F_2}} \left( \pm\kappa \ii {\cF}_\Lambda(u_\pm) \mathfrak{p}^\Lambda - \ii {\cF}(u_\pm) \int_{\rm bolt_\pm} c_1(L) \right) \Bigg] \,,
\end{align}
where $\cF$ is given in \eqref{eq:4d_Prepotential},
 $b_1, b_2$ are the weights of the $D=4$ R-symmetry Killing vector $\xi$ and
${u}_\pm$ are given by 
\begin{equation}
\label{eq:defn_u_pm}
	u_+^\Lambda = \left. \frac{\tilde{X}^\Lambda}{\zeta_\Sigma^{(4)} \tilde{X}^\Sigma} \right\vert_+ \,,\qquad
	u_-^\Lambda = \left. \frac{X^\Lambda}{\zeta_\Sigma^{(4)}X^\Sigma} \right\vert_- \, .
\end{equation}
For the contribution from bolts $\Sigma_\pm$, $c_1(L)$ is the first Chern class of the normal bundle to $\Sigma_\pm$
and $\mathfrak{p}^\Lambda$ are the fluxes through the bolt
\begin{equation}\label{fluxesdefn}
	\mathfrak{p}^\Lambda \equiv \frac{1}{4\pi} \int_{\Sigma} \g \, F^\Lambda \, , 
	\end{equation}
subject to the constraint
\begin{equation}
\label{eq:Constraint_Fluxes}
	\zeta_\Lambda^{(4)} \mathfrak{p}^\Lambda_\pm = \kappa \int_{\rm \Sigma_\pm} \left[ \pm c_1(L) - c_1(T\Sigma_\pm) \right]\,,
\end{equation}
where $c_1(T\Sigma_\pm)$ is the first Chern class of the tangent bundle to $\Sigma_\pm$. 
We similarly define
\begin{equation}
\label{eq:p0}
	\mathfrak{p}^0 \equiv - \frac{1}{4\pi} \int_{\Sigma} \g \, \rd \KKForm \,.
\end{equation}
From  \eqref{eq:Reduction_Ansatz2} we see that this times $4\pi \g^{-1}/\Delta x^5$ 
is the 
Chern number of the $U(1)$ KK bundle over $\Sigma$ associated to reduction along   
$\ell$.
The constant $\kappa = \pm 1$ is a sign that can be fixed for each connected component of the fixed point set, as discussed in various examples in \cite{BenettiGenolini:2024hyd, BenettiGenolini:2024lbj}.
Moreover, we have allowed for the possibility that the normal space to the nuts and bolts can have 
orbifold singularities with the order of the orbifold being $d_{F_{0,2}}$, respectively.

In fact, we can rewrite \eqref{eq:5d_Conjectural_OSAction} in terms of the gauge field equivariant forms, which will be useful in the examples later. As reviewed in appendix \ref{sec:usefulformulaed4} (see equation \eqref{eq:u_Phi0_nut_app}), at an isolated fixed point with chirality $\chi$ and sign $\kappa$, we have
\begin{equation}
\label{eq:4d_RelationEquivariant_v1}
	\g \Phi^\Lambda_0 = - 2\kappa (b_1 - \chi b_2) u^\Lambda_\chi \, ,
\end{equation}
from which we have
\begin{equation}
\label{eq:4d_RelationEquivariant_v2}
\begin{split}
	\g \Phi^0_0 = - 2\kappa (b_1 - \chi b_2) u^0_\chi \, , \qquad \g \check{\Phi}^I_0 = - 2\kappa (b_1 - \chi b_2) \check{u}^I_\chi \, , 
\end{split}
\end{equation}
where we have introduced $\check{u}^I_\chi \equiv u^I_\chi + \nvxi^I u^0_\chi$ and $\check{\Phi}^I_0 \equiv \Phi^I_0 + \nvxi^I \Phi^0_0$, in analogy with $\check{X}^I$ in \eqref{eq:4d_Prepotential}.
By the definition of $u^\Lambda$ in \eqref{eq:defn_u_pm}, these are constrained by 
\begin{equation}
\label{eq:Rsymmconst}
 	\g \zeta^{(4)}_\Lambda \Phi^\Lambda_0 = - 2\kappa (b_1 - \chi b_2) \, ,
 \end{equation}
or equivalently
\begin{equation}
\label{eq:4d_Constraint_v1}
	- 4 Q^{(\ell)} \Phi^0_0 + \g \zeta_I^{(4)} \check{\Phi}^I_0 = - 2 \kappa (b_1 - \chi b_2) \, ,
\end{equation}
where we have used the definition of $\zeta_0^{(4)}$ from \eqref{eq:4d_FI_Parameters}. Therefore, the contribution from isolated fixed points to \eqref{eq:5d_Conjectural_OSAction} can be rewritten as 
\begin{equation}
\label{eq:I_FP_Nuts}
\begin{split}
I^{\rm FP}_{(4)} 
	&= \frac{\Delta x^5\pi}{G_{(5)}} \frac{\ii}{6} \sum_{\rm nuts} \frac{1}{d_{F_0}} \frac{1}{4b_1b_2} \frac{C_{IJK} \check{\Phi}^I_0 \check{\Phi}^J_0 \check{\Phi}^K_0 }{\Phi^0_0}\,,
\end{split}
\end{equation}
subject to the constraint \eqref{eq:4d_Constraint_v1}.

Something similar can be done for fixed surfaces. From \eqref{eq:u_Phi0_bolt_app} we find that, at a bolt with chirality $\chi$ such that the action of $\FourdSUSYVec$ on the normal plane has weight $b$
\begin{equation}
	u^\Lambda_\chi = \frac{\kappa}{2b} \chi \g \Phi^\Lambda_0 \, ,
\end{equation}
constrained by
\begin{equation}
\label{eq:4d_Constraint_bolt}
	- 4 Q^{(\ell)} \Phi^0_0 + \g \zeta_I^{(4)} \check{\Phi}^I_0 = 2 \kappa \chi b \, .
\end{equation}
Therefore, the contribution from bolts to \eqref{eq:5d_Conjectural_OSAction} is
\begin{align}
\label{eq:I_FP_Bolts}
	I^{\rm FP}_{(4)} &= \frac{\Delta x^5 \pi}{\g G_{(5)}} \frac{\ii}{12} \sum_{\rm bolts} \frac{1}{d_{F_2}} \frac{1}{b} \frac{ C_{IJK} \check{\Phi}^J_0 \check{\Phi}^K_0 }{ \Phi^0_0} \Bigg[ - \frac{\check{\Phi}^I_0 }{\Phi^0_0} \mf{p}^0 + 3 \check{\mf{p}}^I - \frac{\g}{2b} \check{\Phi}^I_0  \int_{\rm bolt} c_1(L) \Bigg] \, ,
\end{align}
where 
$\check{\mf{p}}\equiv \mf{p}^I+a^I{\mf{p}}^0$.
In both cases, note that, since the ``checked'' variables are invariant under the transformation \eqref{eq:KK_Redundancy}, the final answer is invariant under it too.

Many of the examples considered in sections \ref{examplesads5} and \ref{examplesblackhole} are supersymmetric solutions to $D=5$ minimal gauged supergravity. In this case, we have only one $D=5$ gauge field, so the constraints \eqref{eq:Constraint_Fluxes}, \eqref{eq:4d_Constraint_v1}, \eqref{eq:4d_Constraint_bolt} can be solved, and the answer written only in terms of the data from the isometry and $\Phi^0_0$. In the conventions of appendix \ref{app:minsugra}, we find that the contribution from fixed sets to the on-shell action of a supersymmetric solution of $D=5$ minimal gauged supergravity is
\begin{align}
\label{eq:I_FP_Minimal}
	I^{\rm FP}_{(4)} &= \frac{2\pi \Delta x^5}{27 \g^3 G_{(5)}} \ii \Bigg\{ \sum_{\rm nuts} \frac{1}{d_{F_0}} \frac{1}{8b_1b_2} \frac{1}{\Phi^0_0} \left( - \kappa (b_1 - \chi b_2) + 2 Q^{(\ell)} \Phi^0_0 \right)^3 \nn \\
	& \qquad + \frac{1}{4} \sum_{{\rm bolts} \ \Sigma } \frac{1}{d_{F_2}} \frac{1}{b} \frac{ 1 }{ \Phi^0_0} \left( \kappa \chi b + 2 Q^{(\ell)} \Phi^0_0 \right)^2 \Bigg[ - \frac{1 }{\Phi^0_0} \frac{1}{\g} \left( \kappa \chi b + 2 Q^{(\ell)} \Phi^0_0 \right) \mf{p}^0  \nn \\
	& \qquad \qquad \quad \quad +  6 \frac{Q^{(\ell)}}{\g} \mf{p}^0 + \left( \kappa \chi - \frac{Q^{(\ell)}}{b} \Phi^0_0 \right) \int_{\Sigma} c_1(L) - \frac{3}{2} \kappa \int_{\Sigma} c_1(T\Sigma)  \Bigg] \Bigg\} \, .
\end{align}
Here, we have isolated as an overall factor the action of Euclidean $AdS_5$ with $S^1\times S^3$ boundary, computed in \cite{BenettiGenolini:2016tsn}. 
On-shell, 
the expression for $\Lambda_4=\rd \Lambda_3$, with
$\Lambda_3$ as in \eqref{eq:Lambda4_General},
can be written for minimal gauged supergravity as
\begin{align}
	\label{eq:Lambda4_Minimal}
	\Lambda_4 \rvert_{\rm OS} &= \frac{\e^{-2\lambda}}{2} \left( ( \check{z}_1^2 + \e^{-4\lambda}) * F^0 \wedge F^{0} -2 \check{z}_1 * F^0 \wedge \check{F}^{1} + * \check{F}^1 \wedge \check{F}^{1} \right) \nn\\
	& \qquad \ - \e^{4\lambda} * \rd \check{z}_1 \wedge \rd \check{z}_1 + 4 \g^2 \e^{2\lambda} \vol_4 \nn\\
	&\ + \ii \Big( 2 \rd \check{z}_1 \wedge \check{A}^1 \wedge \check{F}^1 + 2 \check{z}_1 \check{F}^1 \wedge \check{F}^1 - 2 \check{z}_1 \, \rd \check{z}_1 \wedge \check{A}^1 \wedge F^0 
- \check{z}_1^2 \check{F}^1 \wedge F^0 \Big)\,,
\end{align} 
where we have eliminated the $\rd *_{(4)} \rd\lambda$ term using its equation of motion.

Finally, we remark that $\Phi^0_0$ at a fixed set is determined by the geometry of the KK reduction, as can be argued analogously to \cite{BenettiGenolini:2024kyy}. Recall that the assumptions on the KK fibration of the $D=5$ solution means that
\begin{equation}
	S^1 \hookrightarrow M_{(5)} \rightarrow M_{(4)}\,,
\end{equation}
and $S^1$ is fibred as an associated bundle for the $U(1)$ action on $S^1 \subset \R^2$, with connection one-form $-A^0$ (as can be gleaned from \eqref{eq:Reduction_Ansatz1}-\eqref{eq:Reduction_Ansatz2}). The condition for having a good fibration corresponds to  $4\pi \mf{p}^0\g^{-1}/\Delta x^5$ being an integer (or, in presence of an orbifold, an integer divided by the order of the orbifold group). Because of the existence of the Killing spinors, $M_{(5)}$ has an additional Killing vector $\SUSYVec$, which upon reduction becomes $\FourdSUSYVec$ (as is clear from \eqref{eq:5d4dvectors}).  Using \eqref{eq:Definition_Rup}, over each fixed point $p$ of $\FourdSUSYVec$, we can write
\begin{equation}
\label{eq:SUSYVec_FixedPoint}
	\SUSYVec {\rvert_p} = \frac{\Delta x^5}{2\pi} \Rup \partial_{x^5} {\Big\rvert_p} \, .
\end{equation}
On the base $M_{(4)}$, we know that for an equivariant Chern class evaluated at a fixed point
\begin{equation}
	c_1^{\FourdSUSYVec} (L) \Big|_p = \frac{\epsilon_p(L)}{2\pi} \, , 
\end{equation}
where $\epsilon_p(L)$ is the weight of the action of $\FourdSUSYVec$ on the complex line bundle $L$ over the fixed point $p$. Since $\Phi^\Lambda/2\pi$ are by construction representatives of the cohomology class of $c_1^{\FourdSUSYVec} (L^\Lambda)$, we find that by construction $\g\Phi^\Lambda_0 \rvert_p = \epsilon_p(L^\Lambda)$. 
In particular, $\g \Phi^0_0\rvert_p$, which recall is equal to the evaluation of the equivariant completion of $\g F^0 = - \g \smallspace \rd\KKForm$ at the fixed point, corresponds to the weight of the \textit{lift} of $\FourdSUSYVec$ in $M_{(5)}$, that is, $\SUSYVec$. 
Therefore, with the appropriate rescaling due to the definition \eqref{eq:Reduction_Ansatz2} of $A^0$ 
\begin{equation}
\label{eq:Uplift_Phi00}
	\g \Phi^0_0 \rvert_p = \frac{\Delta x^5 \g}{2\pi} \Rup \, .
\end{equation}
Note that this identification requires the gauge field $A^0$ to be in the regular gauge at the fixed point
\begin{equation}\label{engelbert}
   \xi \hook A^0 \rvert_p = 0 \, ,
\end{equation}
which is a more stringent condition than the gauge choice $\cL_\xi A^0=0$ in \eqref{ggechoiceone}, used to derive \eqref{eq:Constant_Component_K}.
Indeed, substitution in \eqref{eq:5d4dvectors} gives us
\begin{equation}
   \Phi^0_0 \rvert_p = (\xi \hook A^0 - \e^{3\lambda}P) \rvert_p \, , 
\end{equation}
and in the regular gauge, the result $ \Phi^0_0\rvert_p = - \e^{3\lambda}P\rvert_p$ is consistent with our choice of holomorphic sections (see \eqref{eq:Equivariant_PhiF_app}).

\subsection{Reality conditions}
\label{subsec:Reality}

We briefly pause to make various comments regarding reality conditions that are associated with the derivation of the key result \eqref{eq:5d_Conjectural_OSAction}.

First, we recall that the derivation in \cite{BenettiGenolini:2024lbj} assumed a real section of the $D=4$ Euclidean theory with
\begin{align}
\label{eq:4d_Reality_Original}
	g_{(4)} &\in \mathbb{R}\,,  &\qquad A^\Lambda &\in \mathbb{R}\,, &\qquad z^I, \tz^{\tilde{I}} &\in \mathbb{R}\, , \nn\\
\cA &\in \ii \mathbb{R} \, , &\qquad X^\Lambda, \tilde{X}^\Lambda &\in  \mathbb{R} \, , &\qquad \cF(X) &\in \ii \R \, , \quad \tilde{\cF}(\tilde{X}) = - \cF(\tilde{X}) \, .
\end{align}
These assumptions guarantee that the $D=4$ Euclidean Killing spinor $\tilde{\epsilon}$ can be 
taken to be $\epsilon^c$, and thus the $D=4$ spinor bilinears are real and define a (real) identity structure on the frame of $M_{(4)}$. We want to compare these conditions with the $D=5$ reality conditions \eqref{eq:5d_Reality_Contour}, mentioned earlier, and the dimensional reduction ansatz \eqref{eq:Reduction_Ansatz3}. We first notice that \eqref{eq:4d_Reality_Original} means $A^I$ and $z_1^I$ (as defined in \eqref{eq:Reduction_Ansatz4}) are real, and so are $\cA^I$ in \eqref{eq:Reduction_Ansatz3}, provided $\nvxi^I\in\R$. So, \eqref{eq:4d_Reality_Original} is consistent with the reality of the $D=5$ gauge fields required in \eqref{eq:5d_Reality_Contour}. Reality of $A^0$, instead, is guaranteed if $\KKForm$ in \eqref{eq:Reduction_Ansatz2} and $x^5$ are both real. We then notice that $z_2^I$ is instead pure imaginary, and then so is the power of the warp factor $\e^{-6\lambda}$, because of the cubic constraint \eqref{lambdazrel} (with $C_{IJK}\in\R$). However, this means that its third root $\e^{2\lambda}$ cannot be real, and so the condition $Y^I = - \e^{2\lambda} z^I_2$ is not consistent with the $D=5$ reality conditions, since \eqref{eq:5d_Reality_Contour} requires $Y^I\in \ii\R$. Moreover, if $\e^{-6\lambda}$ is pure imaginary, then the K\"ahler potential \eqref{eq:4d_Kahler_Potential} is real if and only if $X^0 \tilde{X}^0\in \ii\R$, which is not consistent with the assumptions in \eqref{eq:4d_Reality_Original}
(and indeed also not with our choice \eqref{eq:X0tX01}). Thus, the 
$D=4$ reality conditions \eqref{eq:4d_Reality_Original} are not consistent with the $D=5$ reality conditions \eqref{eq:5d_Reality_Contour}. 

In fact, the form of the prepotential \eqref{eq:4d_Prepotential} already means that we are outside the class of gauged supergravities to which the derivation of \cite{BenettiGenolini:2024lbj} strictly applies: indeed we have $\tilde{\cF}(\tilde{X}) = \cF(\tilde{X})$, which is not consistent with \eqref{eq:4d_Reality_Original} and the symplectic constraint between the prepotential and the K\"ahler potential \eqref{eq:4d_Kahler_Potential}. We do not believe this is a 
fundamental issue \cite{BenettiGenolini:2024lbj}: there are many  
supersymmetric solutions which violate the reality conditions \eqref{eq:4d_Reality_Original}, and yet the equivariant localization expression for the $D=4$ action remains valid. This suggests that the localization result applies far beyond the restrictive assumptions \eqref{eq:4d_Reality_Original}, which were made to simplify the computations. 
In \cite{BenettiGenolini:2024lbj} it was highlighted how the final results are
expected to change if the $D=4$ reality assumptions are relaxed (e.g. see
eq. (3.34) of \cite{BenettiGenolini:2024lbj}) and this has been used in obtaining 
\eqref{eq:5d_Conjectural_OSAction}.

We end this subsection with an observation regarding reality conditions for
Wick-rotations of {real} stationary, Lorentzian, supersymmetric $D=5$ solutions;
this is relevant for the $AdS_5$ example in section \ref{examplesads5} (but not the black hole example in section~\ref{examplesblackhole}).
If we assume that we can take $\KKVec$ to be the Wick-rotated timelike Killing vector $\partial_t$, i.e. we set $x^5 = \ii t$, then, from the reality of the original Lorentzian fields we infer the following reality properties for the KK ansatz \eqref{eq:Reduction_Ansatz1}-\eqref{eq:Reduction_Ansatz3}:
\begin{align}
\label{eq:4d_Reality_New}
	g_{(4)} &\in \mathbb{R}\,,  \qquad A^0 \in \ii\mathbb{R}\,, \qquad A^I \in \mathbb{R} \, , \nn\\
	z_1^I &\in \ii \R \, , \qquad  z_2^I \in \R \, , \qquad \nvxi^I \in \ii \R \, .
\end{align}
Moreover, $Y^I \in \R$. The $D=5$ metric \eqref{eq:Reduction_Ansatz1} is then complex unless the original spacetime is static (as in the $AdS_5$ example treated in section \ref{examplesads5}). As it stands, this ansatz is outside both the $D=5$ reality condition \eqref{eq:5d_Reality_Contour} and the $D=4$ condition \eqref{eq:4d_Reality_Original}. However, we observe it is consistent with studying $D=4$  solutions with Euclidean spinors related by $\tilde{\epsilon}=\gamma_5 \epsilon^c$ 
(cf. \eqref{appekse1}-\eqref{appekse3}),
provided we use the following choice of parametrization of the holomorphic sections
\begin{equation}
\label{eq:ParametrizationHolomorphicSections}
	z^I = \frac{X^I}{X^0} \, , \qquad \tilde{z}^I = \frac{\tilde{X}^I}{\tilde{X}^0} \, ,
\end{equation}
and we take $X^I , \tilde{X}^I \in \ii\R$, $X^0, \tilde{X}^0 \in \R$, so that $\cI_{00}, \cI_{IJ} \in \R$, $\cI_{0I} \in \ii\R$ (note the consistency with \eqref{thecalIs} and with \eqref{eq:X0tX01}). Moreover, we require $Q^{(\KKVec)} \in \ii \R$, to have $\zeta_0^{(4)} \in \ii\R$ and $\zeta_I^{(4)}\in \R$ (compare with \eqref{eq:4d_FI_Parameters}). These assumptions provide a well-defined $D=4$ ``real'' contour in the Euclidean theory. Thus, they correspond to having complex $D=5$ spinor bilinears 
and imposing a $D=4$ reality condition different from the one studied in \cite{BenettiGenolini:2024lbj}. We also remark that \eqref{eq:4d_Reality_New}, and setting $X^0$ real with $X^I$ pure imaginary with our parametrization of the sections corresponds to the reality condition imposed in \cite{Cacciatori:2009iz}.

\subsection{Background subtraction}
\label{sec:bkgsub}

In section \ref{ctoshelldimredact} we described how to compute the $D=5$ on-shell action using 
localization for classes of solutions where the supersymmetric boundary counterterms are known; we will illustrate this formalism for Euclidean $AdS_5$ with $S^1\times S^3$ boundary and recover the supersymmetric Casimir energy in section \ref{examplesads5}. 

In general, such supersymmetric
boundary counterterms are not known. In such cases, a practical procedure to obtain a finite
result for the $D=5$ on-shell action is to regulate using background subtraction, with the hope of obtaining a result that
is scheme independent. This effectively computes the regularized 
free energy of the solution relative to that of a suitable reference background. 
Instead of \eqref{totalactos}, one introduces a suitably chosen
supersymmetric $D=5$ solution $N_{(5)}$ and defines
\begin{align}
\label{totalactos_backgroundsubtraction}
\Itot_{(5), {\rm bs}} = I_{(5)}[M_{(5)}]+I_{\mathrm{GHY}}^{\partial M_{(5)}} - I_{(5)}[N_{(5)}]- I_{\mathrm{GHY}}^{\partial N_{(5)}}  \, .
\end{align}
There is certainly some arbitrariness in the choice of $N_{(5)}$,
but one definite requirement that is needed to cancel the divergences is that $\partial N_{(5)}$ has the same geometry as $\partial M_{(5)}$. In the most well-studied examples $N_{(5)}$ is simply taken to be the Euclidean $AdS_5$ vacuum.  

It should be noted that in the case of Euclidean $AdS_5$ with $S^1\times S^3$ boundary, this background subtraction procedure will lead to a vanishing result for the $D=5$ on-shell action,  
and hence
will not give the known non-vanishing result for the supersymmetric Casimir energy.
On the other hand, using background subtraction for complex supersymmetric black hole
solutions in minimal gauged supergravity leads to a result for the action which agrees with the supersymmetric index for $\mathcal{N}=4$ super-Yang-Mills theory \cite{Cabo-Bizet:2018ehj, Choi:2018hmj,Benini:2018ywd, Aharony:2021zkr}.

Now suppose that the GHY terms in \eqref{totalactos_backgroundsubtraction} cancel: in fact
this is known to be the case\footnote{
For minimal gauged supergravity,
by carrying out a Fefferman--Graham expansion, one finds that the evaluation of the GHY terms only depends on the intrinsic geometry and hence will also cancel. This will not be the case
with vector multiplets: if the GHY terms in \eqref{totalactos_backgroundsubtraction} do not cancel, it is possible that at least for some classes of solutions,
they will cancel with the boundary terms in 
\eqref{eq:totac_fp_bdy}, again leading to the result \eqref{bsfptformula}, assuming the vanishing of the $\Lambda_4$ contributions as discussed below.} 
for the black hole solution 
\cite{Chen:2005zj}, that we consider in section \ref{examplesblackhole}.  
Then, the evaluation of
\eqref{totalactos_backgroundsubtraction} boils down to evaluating the $D=5$ action on the 
\emph{closed} manifold $M_{(5)}\cup (-N_{(5)})$. In this case we can evaluate 
\eqref{totalactos_backgroundsubtraction} by choosing a Killing vector $\ell$ and using the BVAB theorem in $D=4$. Now recalling \eqref{eq:totac_fp_bdy} we will get fixed point contributions
in $M_{(4)}$ and $N_{(4)}$ and an integral of $\Lambda_4$ over $M_{(4)}\cup (-N_{(4)})$. 
Now $\Lambda_4$ is closed but not necessarily exact, and moreover it is
gauge dependent, in general (recall $\Lambda_4= \rd \Lambda_3$ with
$\Lambda_3$ given in \eqref{eq:Lambda4_General}). 
However, there are classes of solutions and choice of $\ell$, where the integrals of $\Lambda_4$
would give no contribution: for example, if $\Lambda_4$ was globally defined 
and exact on both\footnote{Taking $N_{(5)}$ to be the $AdS_5$ vacuum, as we will for the black hole example later, then $\Lambda_4$ would be exact and globally defined on $N_{(4)}$.} $M_{(4)}$ and $N_{(4)}$, it would be exact on $M_{(4)}\cup (-N_{(4)})$. 
For such classes the final answer for the $D=5$ on-shell action will be 
\begin{align}\label{bsfptformula}
\Itot_{(5), {\rm bs}} = I_{(4)}^{\rm FP}[M_{(4)}] - I_{(4)}^{\rm FP}[N_{(4)}]\,.
\end{align}
where $I_{(4)}^{\rm FP}$ is given by \eqref{eq:5d_Conjectural_OSAction}.

Bearing all these comments regarding background subtraction in mind, we
will take \eqref{bsfptformula} as a concrete prescription for evaluating the $D=5$ on-shell action.  
In section \ref{examplesblackhole}, we will see that for $D=5$ black hole solutions 
in minimal gauged supergravity, taking $N_{(5)}$ to be the $AdS_5$ vacuum,
 it gives a result (without using the explicit solution), 
for various choices of $\ell$, that precisely gives the supersymmetric index in the dual field theory. Furthermore, it can be used in theories with arbitrary vector multiplets to give a result which
has the natural form to be the supersymmetric index of the dual field theories.
It would certainly be of interest to precisely characterize the validity of \eqref{bsfptformula}, including an understanding of how to deal with the gauge fields when there is not a global choice of gauge, but we leave this to future work.

\section{Euclidean \texorpdfstring{$AdS_5$}{AdS5} with \texorpdfstring{$S^1\times S^3$}{S1xS3} boundary}\label{examplesads5}

The first example we consider is $D=5$ hyperbolic space i.e. Euclidean $AdS_5$. As in \cite{BenettiGenolini:2016tsn} the solution has vanishing scalars and flat gauge fields, with metric given by
\begin{equation}
\label{eq:EAdS5_Metric}
	\rd s^2 = \frac{1}{\g^2} \left[ \frac{\rd\rho^2}{\rho^2} + \left(\frac{1}{\rho} + \frac{\rho}{4r_3^2}\right)^2
	\frac{\beta^2}{4\pi^2} \rd \tau^2 + \left(\frac{1}{\rho} - \frac{\rho}{4r_3^2}\right)^2 r_3^2 \, \rd s^2(S^3) \right] \, ,
\end{equation}
where $\rd s^2(S^3)$ has unit radius and recall that the $AdS_5$ radius is given by $L=\g^{-1}$.
Here $\tau$ parametrizes an $S^1$ with
$\Delta \tau=2\pi$ and $\rho\in (0,2r_3]$, with $\rho\to 0$ giving
the conformal boundary $S^1_\beta\times S^3_{r_3}$.
The topology of the $D=5$ solution is $S^1\times \mathbb{R}^4$. 

The essential features of utilizing our localization techniques for this example are
covered by considering $D=5$ \emph{minimal gauged supergravity}, with a single $D=5$ gauge field $\cA$ (see appendix \ref{app:minsugra}). In order to have  spinors that are independent of $\tau$, we take the $D=5$ gauge field to be
\begin{align}
\label{eq:EAdS5_GaugeField}
	\cA = \frac{\ii }{4 \pi \sqrt{3} \g }\frac{\beta}{r_3}\rd\tau \, ,
\end{align}
(in models with additional vector multiplets this would be the form of the R-symmetry gauge field $\frac{1}{2}\zeta_I \cA^I$).
The on-shell action for this solution was explicitly computed in \cite{BenettiGenolini:2016tsn} using novel supersymmetry-preserving counterterms, and matched to the supersymmetric Casimir energy defined in \cite{Assel:2015nca}.
The result is 
\begin{align}\label{osactsoimp}
\Itot_{(5)} = \frac{2\pi}{27 \g^3 G_{(5)}} \frac{\beta}{r_3} \,,
\end{align}
which only depends on $\beta, r_3$ through their ratio, which is conformally invariant.

This $D=5$ solution has been generalized to one with $S^1\times \mathbb{R}^4$ topology, but allowing for the 
boundary to be the product of $S^1_\beta$ and a squashed three sphere.
In \cite{Cassani:2014zwa} metrics were constructed numerically that approach 
$S^1_\beta$ and a biaxially squashed three-sphere 
$S^3_{r_3, v}$ with metric given by
$\rd s^2(S^3_{r_3,v})=r_3^2\left[\sigma_1^2+\sigma_2^2+v^2\sigma_3^2\right]$, for some constant $v$. 
More generally, one can consider a conformal boundary that is a Hopf surface with topology $S^1\times S^3$. These solutions have a boundary supersymmetric Killing vector on $S^1_\beta\times S^3_{r_3,v}$
of the form $\partial_\tau+\ii \frac{\beta}{2\pi} ( b_1\partial_{\phi_1}+b_2\partial_{\phi_2})$ 
which for $b_1= b_2=1/r_3$ 
reduces to the case of \eqref{eq:EAdS5_Metric}.
Although no solutions have been analytically constructed with this boundary and $S^1\times \mathbb{R}^4$ topology, 
it was also shown \cite{BenettiGenolini:2016tsn} how to compute the on-shell action with result\footnote{Reference \cite{BenettiGenolini:2016tsn} implicitly made some assumptions about sign conventions in the weights $b_i$. We shall be more precise below.}
\begin{align}
\label{eq:SquashedSolution_Old_Result}
\Itot_{(5)}= \frac{2\pi}{27 \g^3 G_{(5)}} \beta \frac{(b_1+b_2)^3}{ 8 b_1 b_2 }  \,,
\end{align}
which agrees with \eqref{osactsoimp} after setting $b_1=b_2=1/r_3$, 
and also agrees with the field theory prediction of \cite{Assel:2014paa}.

The goal is to now see how this result can be obtained using the localization and dimensional reduction technology developed in section \ref{ctoshelldimredact}.
We first compute the result only using the data from topology and geometry: 
for a particular choice of reduction vector $\ell$, we precisely recover \eqref{eq:SquashedSolution_Old_Result} from contributions
from the $D=4$ fixed point set. However, for another choice of $\ell$ we find that we do not recover 
\eqref{eq:SquashedSolution_Old_Result} from contributions
from the $D=4$ fixed point set, so necessarily there are additional contributions arising from the last four terms in \eqref{eq:totac_fp_bdy}.
We explicitly check that this is what is going on by analysing the explicit form of the metric and Killing spinors.
Before starting, though, we briefly pause to record some facts about $S^3$, which will also be useful
for the black hole solutions considered in the next section.

\subsection{Some \texorpdfstring{$S^3$}{S3} results}
\label{essthrefacts}

The round metric on a unit radius $S^3$ can be written as
\begin{equation}\label{s3coords}
	\rd s^2(S^3) = \rd\vartheta^2 + \cos^2\vartheta \, \rd\varphi_1^2 + \sin^2\vartheta \, \rd\varphi_2^2 \, ,
\end{equation}
with $\vartheta\in [0,\pi/2]$ and $\Delta\varphi_i=2\pi$.
It is also useful to use the coordinates 
\begin{equation}
\label{eq:S3Angles_EulerHopf}
	\psi = \varphi_1 + \varphi_2 \,, \qquad \phi \equiv - \varphi_1 + \varphi_2 \, , \qquad \theta = 2 \vartheta \, ,
\end{equation}	
with $\Delta\psi=4\pi$, $\Delta\phi=2\pi$ and $\theta\in [0,\pi]$
which is associated with the Hopf fibration:
\begin{align}
	\rd s^2(S^3) &= \frac{1}{4} \left[ \rd\theta^2 + \sin^2\theta \, \rd\phi^2 + ( \rd\psi - \cos\theta \, \rd\phi )^2 \right] \,,\nn\\
	&=\frac{1}{4}\left[\sigma_1^2+\sigma_2^2+\sigma_3^2\right]\,,
\end{align}
where $\sigma_i$ are left-invariant one forms on $S^3$ with $\sigma_3=\rd\psi - \cos\theta \, \rd\phi$.
Half of the 4 Killing spinors on $S^3$ are constant in the left-invariant frame, and are solutions to the Killing spinor equation $\nabla_a \upsilon = \frac{\ii}{2} \gamma_a^{(3)} \upsilon$, where $\gamma^{(3)}$ are elements of $\rm{Cliff}(3)$, with $a=1,2,3$. These 
two spinors satisfy, respectively,
\begin{equation}
\label{eq:Lie_KS_S3}
	\cL_{\partial_{\varphi_1}} \upsilon = \rsigma \frac{\ii}{2} \upsilon\,,\qquad  \cL_{\partial_{\varphi_2}} \upsilon = \rsigma \frac{\ii}{2} \upsilon \,,\qquad 
    \rsigma=\pm1\,,
\end{equation}
implying that they are charged under $\partial_\psi$ with charge $ \frac{1}{2}\rsigma$, and uncharged under $\partial_\phi$.

\subsection{Simple reduction}

Ignoring for a moment the explicit form of the metric and gauge field in \eqref{eq:EAdS5_Metric}-\eqref{eq:EAdS5_GaugeField}, we assume the existence of a supersymmetric solution on $M_{(5)}$ with topology and bilinear Killing vector given by
\begin{equation}
\label{eq:EAdS5_Data}
	M_{(5)} = S^1 \times \R^4 \, , \qquad \cK = \partial_{\tau} + \varepsilon_1 \partial_{\varphi_1} + \varepsilon_2 \partial_{\varphi_2} \, ,
\end{equation}
where $\varphi_1$, $\varphi_2$ are the polar angles in the two orthogonal planes in $\R^4$, and $\tau$ parametrizes $S^1$, and they all have period $2\pi$. 
As noted below \eqref{dfiveksebiliniears}, $\cK$ is generically complex, so $\varepsilon_1, \varepsilon_2 \in \mathbb{C}$.

We first consider carrying out the reduction with respect to the Killing vector $\ell$ along the $S^1$
given by
\begin{align}\label{simplekkell}
	\KKVec &= \g\partial_{\tau} \, .
\end{align}
Here, we have fixed $\Delta x^5 = 2\pi\g^{-1}$ using the possibility of rescaling $x^5$ via a redefinition of $\e^{4\lambda}$ in the ansatz \eqref{eq:Reduction_Ansatz1}. 
We now assume there exists a gauge field such that the spinors are independent of $\tau$ (and in particular periodic) so the charge, $Q^{(\ell)}$, of the $D=5$ Killing spinor $\chi$ with respect to the KK reduction vector $\ell$ 
vanishes:
\begin{align}
Q^{(\ell)}=0\,.
\end{align}

It is clear that $M_{(4)} \cong \R^4$, and from \eqref{eq:5d4dvectors}, we have 
\begin{equation}
	\FourdSUSYVec = \varepsilon_1 \partial_{\varphi_1} + \varepsilon_2 \partial_{\varphi_2} \, , 
\end{equation}
which has an isolated fixed point at the origin with weights $\varepsilon_1$ and $\varepsilon_2$, and chirality $\chi$. We identify $\Rup= 1$ in \eqref{eq:SUSYVec_FixedPoint}, and so $\g\Phi^0_0 \rvert_{\rm nut} = 1 $ from \eqref{eq:Uplift_Phi00}. 
So, \eqref{eq:I_FP_Minimal} immediately gives
\begin{equation}
	I^{\rm FP}_{(4)} = \frac{2\pi }{27 \g^3 G_{(5)}} (-\kappa \ii 2\pi) \frac{( \varepsilon_1 - \chi \varepsilon_2)^3}{8 \varepsilon_1 \varepsilon_2} \, .
\end{equation}
Note that this matches the value for the on-shell action \eqref{eq:SquashedSolution_Old_Result}, which had been constructed for solutions with Hopf surface boundary, with the following identification
\begin{equation}
\label{eq:Identifications_AdS}
	\varepsilon_1 = \ii \frac{\beta}{2\pi} b_1 \, , \qquad \varepsilon_2 = - \ii \frac{\beta}{2\pi} \chi b_2 \, , \qquad \chi = - \kappa \, .
\end{equation}
Note that this identification requires a choice of chirality for the $D=4$ spinor, which in analogous analysis for specific backgrounds of $D=4$ minimal gauged supergravity was shown to be due to regularity \cite{Farquet:2014kma} (see the discussion in \cite{BenettiGenolini:2024lbj}). Moreover, we observe that with this identification the Killing vector $\SUSYVec$ is generically complex and matches the one computed from the actual solution and mentioned above \eqref{eq:SquashedSolution_Old_Result}.

Thus, with the choice of KK Killing vector in \eqref{simplekkell}, we have $\Itot_{(5)} = I^{\rm FP}_{(4)}$.
We will later confirm 
that the last four terms in \eqref{eq:totac_fp_bdy} indeed vanish and so the $D=5$ on-shell
action for this choice of KK Killing vector is given precisely by the $D=4$ fixed point set contribution.
While this is a pleasing result, interestingly, it does not generalize to generic choices of the KK Killing vector, as we shall now see.

\subsection{Reduction with a \texorpdfstring{$q$}{q}-twist using the Hopf fibre}

We now carry out the KK reduction of the solution associated with \eqref{eq:EAdS5_Data} using the Killing vector $\ell$ given by
\begin{equation}\label{thetwisteellads5}
	\KKVec = \g \left( \partial_\tau + 2q \partial_\psi \right) \, ,
\end{equation}
with $q$ an integer.
We refer to this as a ``$q$-twist'' of the previous reduction, that uses also the Hopf fibre of the $S^3$ (we are using the Euler angles defined in \eqref{eq:S3Angles_EulerHopf}); setting $q=0$ returns us to the simple reduction just considered.
As before, we assume that the gauge field is flat and that the spinor is periodic around $S^1_\tau$. Without loss of generality, we focus on
\begin{equation}
	\cL_{\partial_\psi} \chi = - \frac{\ii}{2} \chi \, ,
\end{equation}
so that
\begin{equation}
	Q^{(\ell)} = - \g q \, .
\end{equation}
To proceed it is useful to introduce periodic coordinates $\mu,\nu$ via
\begin{equation}
\label{eq:EAdS5_Twisted_Coords}
	\tau = \nu \, , \qquad \psi = 2(q\nu + \mu) \, ,
\end{equation}
with $\Delta\mu=\Delta\nu=2\pi$, in terms of which
\begin{equation}
\label{eq:EAdS5_Twisted_Vecs}
	\KKVec = \g \partial_\nu \, , \qquad \SUSYVec = \partial_\nu + \left( -q + \frac{\varepsilon_1+\varepsilon_2}{2} \right) \partial_\mu - ( \varepsilon_1 - \varepsilon_2) \partial_\phi \, .
\end{equation}

For simplicity, we now just consider $\varepsilon_1 = \varepsilon_2 \equiv \ii\frac{\beta}{2\pi r_3}$, which will allow us to see 
the key feature that using this KK reduction, the contribution from the ``boundary'' terms (the last four terms in \eqref{eq:totac_fp_bdy}) to the $D=5$ on-shell action will not vanish.
Reducing along $\ell$ we again find that $M_{(4)} \cong \R^4$, and from \eqref{eq:5d4dvectors} we have
\begin{equation}
	\FourdSUSYVec = \left( -q + \ii\frac{\beta}{2\pi r_3} \right) \partial_\mu  \, .
\end{equation}
Modulo the caveats about the connection between weights, chirality, and regularity, discussed in the previous subsection, this $D=4$ supersymmetric Killing vector again has an isolated fixed point at the origin of $M_{(4)}$,
with weights equal to $(-q + \ii\frac{\beta}{2\pi r_3})$, negative chirality, and $\kappa = 1$. From \eqref{eq:I_FP_Minimal} we now obtain the fixed point contribution
\begin{align}
\label{eq:EAdS5_qtwisted}
	I^{\rm FP}_{(4)} &= \frac{2\pi}{27 \g^3 G_{(5)}} \frac{\beta}{r_3} - q \frac{8\pi^2 \ii \beta }{27 \g^3 G_5} \frac{ \beta + \pi \ii q r_3 }{ (\beta + 2 \pi \ii  q r_3 )^2} \, .
\end{align}

Thus, for the $q$-twist KK reduction vector, the $D=5$ on-shell action is not entirely given by the fixed point contribution in the reduced $D=4$ action. Indeed, we shall now show from a concrete computation for the explicit metric \eqref{eq:EAdS5_Metric}
that the last four terms in \eqref{eq:totac_fp_bdy} do not vanish and contribute precisely the term proportional to $q$ above.

\subsection{Reduction with a \texorpdfstring{$q$}{q}-twist: analytic computation}

We now consider the explicit metric \eqref{eq:EAdS5_Metric}, verifying that when one does a $q$-twist reduction, the boundary terms give a non-trivial contribution to the $D=5$ on-shell action that precisely cancels the term in \eqref{eq:EAdS5_qtwisted} proportional to $q$.

Writing \eqref{eq:EAdS5_Metric} according to the reduction ansatz \eqref{eq:Reduction_Ansatz1} and using the 
coordinates $\nu,\mu$ in \eqref{eq:EAdS5_Twisted_Coords}, we find
\begin{align}
	\e^{-2\lambda} &= \sqrt{ \frac{\beta^2}{4\pi^2} \cW_+^2 + q^2 r_3^2 \cW_-^2} \, , \nn \\ 
	\KKForm &= \frac{1}{\g} \left( \rd \nu + q\frac{ r_3^2 \cW_-^2}{\frac{\beta^2}{4\pi^2 } \cW_+^2 + q^2 r_3^2 \cW_-^2 } \left( \rd\mu - \frac{1}{2} \cos\theta \, \rd\phi \right) \right) \, , \nn \\
	\rd s^2_{(4)} &= \frac{1}{\g^2} \sqrt{ \frac{\beta^2}{4\pi^2} \cW_+^2 + q^2 r_3^2 \cW_-^2} \Bigg[ \frac{\rd\rho^2}{\rho^2} \nn \\
	\label{eq:EAdS5_Twisted_4dMetric}
	& \qquad + \cW_-^2 \frac{r_3^2}{4} \left( \frac{\frac{\beta^2}{4\pi^2} \cW_+^2 }{ \frac{\beta^2}{4\pi^2} \cW_+^2 + q^2 r_3^2 \cW_-^2} 4 \left( \rd \mu - \frac{1}{2} \cos\theta \, \rd\phi \right)^2 + \rd s^2(S^2) \right) \Bigg] \, ,
\end{align}
where
\begin{equation}
\label{eq:EAdS5_Metric_cW}
	\cW_\pm \equiv \frac{1}{\rho} \pm \frac{\rho}{4r_3^2} \, .
\end{equation}
From \eqref{eq:Reduction_Ansatz3}, we have
\begin{equation}
	z_1 + \nvxi = \frac{2}{\sqrt{3}} \ell \hook \cA \, , \qquad z_2 = - \e^{-2\lambda} \, ,
\end{equation}
so, using the gauge field in \eqref{eq:EAdS5_GaugeField}, we find
\begin{equation}
	\label{eq:EAdS5_Twisted_Reduced_ScalarFields}
	z_1 + \nvxi = \frac{\ii\beta}{6 \pi r_3} \, , \qquad z_2 = - \sqrt{ \frac{\beta^2}{4\pi^2} \cW_+^2 + q^2 r_3^2 \cW_-^2} \, .
\end{equation}
Note that, despite the fact that the $D=5$ gauge field is flat, $\cF=0$ (so the consistency condition \eqref{QQtildeconstraint} is trivially satisfied). Despite this the $D=4$ gauge fields are not flat:
\begin{align}
	A^0 &= - \frac{q}{\g} \frac{ r_3^2 \cW_-^2}{\frac{\beta^2}{4\pi^2 } \cW_+^2 + q^2 r_3^2 \cW_-^2} \left( \rd\mu - \frac{1}{2} \cos\theta \, \rd\phi \right) \, , \nn \\
	\label{eq:EAdS5_Twisted_Reduced_GaugeFields}
	A^1 &= z_1 A^0\,.
\end{align}

It is worth studying the geometry of the $D=4$ space $M_{(4)}$. First, note that the metric and fields are clearly smooth everywhere outside $\rho = 2r_3$. As $\rho \to 2r_3$, we have $\cW_+ = \frac{1}{r_3^2} + o(\rho - 2r_3)$ and $\cW_- = \frac{(\rho - r_3)^2}{4r_3^2} + o((\rho-2r_3)^2)$, so the metric and gauge fields are regular, so the topology is that of $\R^4$. On the other hand, near the conformal boundary $\rho\to 0$, we can introduce $r=1/\rho$ to find
\begin{align}
	\rd s^2_{(4)} &\sim r\frac{1}{\g^2} \sqrt{ \frac{\beta^2}{4\pi^2} + q^2 r_3^2 } \Bigg[ \frac{\rd r^2}{r^2} \nn \\
	& \qquad \qquad \qquad \quad + r^2 \frac{r_3^2}{4} \left[ \frac{\frac{\beta^2}{4\pi^2} }{ \frac{\beta^2}{4\pi^2} + q^2 r_3^2 } 4 \left( \rd \mu - \frac{1}{2} \cos\theta \, \rd\phi \right)^2 + \rd s^2(S^2) \right] \Bigg] \, .
\end{align}
This is conformal to $AlAdS$, with boundary being a squashed three-sphere preserving $SU(2)\times U(1)$ symmetry. This is consistent with the fact that the $D=4$ scalar potential obtained reducing minimal gauged supergravity is $\cV_{(4)} = 12/z_2$, which does not have a minimum, so there is no $AdS_4$ vacuum.

In order to study the supersymmetric structure, we introduce the $D=5$ frame
\begin{align}
\label{eq:EAdS5_Twisted_Frame}
	{\rm E}^1 &= \frac{1}{\g}\cW_- \frac{r_3}{2} \, \rd\theta \, , \nn \\
	{\rm E}^2 &= \frac{1}{\g} \cW_- \frac{r_3}{2} \, \sin\theta \, \rd\phi \, , \nn \\
	{\rm E}^3 &= \frac{1}{\g} \cW_- \frac{r_3}{2} \frac{\frac{\beta}{2\pi} \cW_+ }{ \sqrt{ \frac{\beta^2}{4\pi^2} \cW_+^2 + q^2 r_3^2 \cW_-^2 } } 2 \left( \rd \mu - \frac{1}{2} \cos\theta \, \rd\phi \right) \, , \nn \\
	{\rm E}^4 &= \frac{1}{\g} \frac{\rd\rho}{\rho} \, , \nn \\
	{\rm E}^5 &= \frac{1}{\g} \sqrt{ \frac{\beta^2}{4\pi^2} \cW_+^2 + q^2 r_3^2 \cW_-^2} \left( \rd \nu + \frac{ q r_3^2 \cW_-^2}{ \frac{\beta^2}{4\pi^2 } \cW_+^2 + q^2 r_3^2 \cW_-^2 } \left( \rd\mu - \frac{1}{2} \cos\theta \,  \rd \phi \right) \right) \, ,
\end{align}
which is consistent with \eqref{Dequalsfiveframe}, and the generators of the Clifford algebra
\begin{equation}
\label{eq:EAdS_GammaMatrices}
	\gamma_{1,2,3} = - \sigma_2 \otimes \sigma_{1,2,3} \, , \quad \gamma_4 = \sigma_3 \otimes \identity_2 \, , \quad \gamma_5 = \sigma_1 \otimes \identity_2 \, , \quad \gamma_{12345} = \identity \, .
\end{equation}
The $D=5$ supersymmetry equations for Euclidean minimal gauged supergravity are (cf. appendix \ref{app:minsugra})
\begin{align}
\label{eq:E5dKSEepsilon_Minimal_text}
      &0=\left[\nabla_m - \sqrt{3} \ii \g  {\cA}_m+\frac{1}{2} \g {\gamma}_m +\frac{\ii}{4\sqrt{3}}  {\cF}_{np} ({\gamma}_{m}{}^{np}-4\delta_m^n{\gamma}^p)\right]\chi\,,\nn\\
            &0=\left[\nabla_m+ \sqrt{3} \ii \g  {\cA}_m-\frac{1}{2} \g {\gamma}_m +\frac{\ii}{4\sqrt{3}}  {\cF}_{np} ({\gamma}_{m}{}^{np}-4\delta_m^n{\gamma}^p)\right]\tilde{\chi} \, .
\end{align}
With the choices made above, these equations are solved by
\begin{align}
\label{eq:EAdS5_Twisted_5dSpinors}
	\chi &= M \begin{pmatrix}
		 \mathtt{p} \frac{2 r_3}{\sqrt{\rho }} \e^{ \ii (\mu + q \nu )} \\ \mathtt{q} \frac{2 r_3}{\sqrt{\rho }} \e^{ - \ii (\mu + q\nu )} \\ - \mathtt{p} \sqrt{\rho } \e^{ \ii (\mu + q\nu )} \\ - \mathtt{q} \sqrt{\rho }  \e^{ - \ii (\mu + q \nu )}
	\end{pmatrix} \, , \qquad
	\tilde{\chi} = M \begin{pmatrix}
		- \tilde{\mathtt{p}} \sqrt{\rho } \e^{ \ii (\mu + q\nu )} \\ - \tilde{\mathtt{q}} \sqrt{\rho }  \e^{ - \ii (\mu + q \nu )} \\
		 \tilde{\mathtt{p}} \frac{2 r_3}{\sqrt{\rho }} \e^{ \ii (\mu + q \nu )} \\ \tilde{\mathtt{q}} \frac{2 r_3}{\sqrt{\rho }} \e^{ - \ii (\mu + q\nu )}
	\end{pmatrix} \, ,
\end{align}
where $\mathtt{p},\mathtt{q},\tilde{\mathtt{p}},\tilde{\mathtt{q}}$ are constants and $M$ is the matrix given by
\begin{equation}
    M = \begin{pmatrix}
		f^{-1} & & & \\ & f & & \\ & & f & \\ & & & f^{-1}
	\end{pmatrix} \, , \qquad f \equiv 
    \left( \frac{\frac{\beta}{2 \pi } \cW_+ + \ii q r_3 \cW_- }{\frac{\beta}{2 \pi } \cW_+ - \ii q r_3 \cW_- } \right)^{1/4}\,.
\end{equation}

Out of these spinors we can compute the $D=5$ spinor bilinears \eqref{dfiveksebiliniears}. In particular, in order to match \eqref{eq:EAdS5_Twisted_Vecs} with $\varepsilon_1 = \varepsilon_2 = \ii\frac{\beta}{2\pi r_3}$, we restrict to the case  
\begin{equation}
	\mathtt{q} \tilde{\mathtt{p}} = \frac{\beta}{8\pi \g r_3^2} \, , \qquad \mathtt{p} = \tilde{\mathtt{q}} = 0 \, ,
\end{equation}
so
\begin{equation}
	\cS = - \frac{\beta}{2\pi \g r_3} \, , \qquad \SUSYVec = \partial_\nu + \left( - q + \ii \frac{\beta}{2\pi r_3} \right) \partial_\mu \, .
\end{equation}
Moreover, we find that the Killing spinors $\chi$ and $\tilde{\chi}$ are charged under $\SUSYVec$ and $\KKVec$, as required in \eqref{lieckchi}, with 
\begin{equation}
	\cQ = - \ii \frac{\beta}{2\pi r_3} = - \tilde{\cQ} \, , \qquad Q^{(\KKVec)} = - \g q = - Q^{(\KKVec)}\,.
\end{equation}
Note that this solution does not satisfy the reality condition \eqref{eq:5d_Reality_Contour}, and indeed $\SUSYVec$ is manifestly complex.

We now consider the supersymmetry structure induced on the $D=4$ solution with metric \eqref{eq:EAdS5_Twisted_4dMetric} and fields \eqref{eq:EAdS5_Twisted_Reduced_ScalarFields}, \eqref{eq:EAdS5_Twisted_Reduced_GaugeFields}. The $D=4$ frame is constructed from the $D=5$ frame \eqref{Dequalsfiveframe} via
$\{\e^a = \e^{-\lambda} {\rm E}^a \}$,
and the $\gamma$ matrices are the same as the above. It is straightforward to check that the spinors $\epsilon$ and $\tepsilon$ constructed from \eqref{eq:EAdS5_Twisted_5dSpinors} as in \eqref{epschirels} solve the Killing spinor equations for the $D=4$ model obtained reducing minimal gauged supergravity, namely (using $z^I = X^I$, $\tilde{z}^I = \tilde{X}^I$, which is consistent with \eqref{eq:X0tX01} and \eqref{eq:ParametrizationHolomorphicSections}):
\begin{align}
	0 &= \Bigg[ \nabla_\mu + \frac{3\ii }{4z_2} \partial_\mu z_1  \gamma_5 -  \frac{\ii}{4} \g \left( \left( - 4 \frac{Q^{(\KKVec)}}{\g} + 6 a \right) A^0_\mu + 6 A^1_\mu \right) \nn \\
	& \ \ \ - \g \frac{\ii}{4}  \frac{1}{z_2 \sqrt{z_2}}\gamma_\mu \left( \left( - 2 \frac{Q^{(\KKVec)}}{\g} + 3 \check{z}_1 \right)  \identity - 3 \ii z_2 \gamma_5 \right)\nn\\
\label{eq:Euclidean_KSE_epsilon_Model}
& \ \ \  - \frac{\sqrt{z_2}}{16} \left[ \left( z_2 \identity - 3 \ii z_1 \gamma_5 \right) F^{0}_{\nu\rho} + 3 \ii \gamma_5 F^{1}_{\nu\rho} \right] \gamma^{\nu\rho}\gamma_\mu \Bigg] \epsilon \, , 
\end{align}
\begin{align}
	0 &= \Bigg[ - \frac{z_2 \sqrt{z_2}}{4} \gamma^{\nu\rho} \left[  F^{0}_{\nu\rho} \left( - z_1 \identity + \ii z_2 \gamma_5 \right) + F^1_{\nu\rho} \, \identity \right] + \gamma^\mu \left( \partial_\mu z_1  \identity - \ii \partial_\mu z_2  \gamma_5 \right) \nn \\
& \ \ \  - \g \frac{\ii}{\sqrt{z_2}} \left( z_2 \identity + \ii \left( 3 \check{z}_1 - 2 \frac{Q^{(\KKVec)}}{\g} \right) \gamma_5 \right) \Bigg] \epsilon \, ,
\end{align}
\begin{align}
	0 &= \Bigg[ \nabla_\mu + \frac{3\ii }{4z_2} \partial_\mu z_1  \gamma_5 +  \frac{\ii}{4} \g \left( \left( - 4 \frac{Q^{(\KKVec)}}{\g} + 6 a \right) A^0_\mu + 6 A^1_\mu \right) \nn \\
	& \ \ \ - \g \frac{\ii}{4}  \frac{1}{z_2 \sqrt{z_2}}\gamma_\mu \left( \left( - 2 \frac{Q^{(\KKVec)}}{\g} + 3 \check{z}_1 \right)  \identity - 3 \ii z_2 \gamma_5 \right) \nn \\
\label{eq:Euclidean_KSE_tepsilon_Model}
	& \ \ \  + \frac{\sqrt{z_2}}{16} \left[ \left( z_2 \identity - 3 \ii z_1 \gamma_5 \right) F^{0}_{\nu\rho} + 3 \ii \gamma_5 F^{1}_{\nu\rho} \right] \gamma^{\nu\rho}\gamma_\mu \Bigg] \tepsilon \, , 
	\end{align}
    and
\begin{align}
	0 &= \Bigg[ \frac{z_2 \sqrt{z_2}}{4} \gamma^{\nu\rho} \left[  F^{0}_{\nu\rho} \left( - z_1 \identity + \ii z_2 \gamma_5 \right) + F^1_{\nu\rho} \, \identity \right] + \gamma^\mu \left( \partial_\mu z_1  \identity - \ii \partial_\mu z_2  \gamma_5 \right) \nn \\
\label{eq:Euclidean_Algebraic_tepsilon_Model}
	& \ \ \  - \g \frac{\ii}{\sqrt{z_2}} \left( z_2 \identity + \ii \left( 3 \check{z}_1 - 2 \frac{Q^{(\KKVec)}}{\g} \right) \gamma_5 \right) \Bigg] \tepsilon \, .
\end{align}
From these, we can construct the $D=4$ bilinears \eqref{eq:Bilinears_app}, finding
\begin{align}
	S &= \frac{\beta}{ 2 \pi \g r_3\e^\lambda} \, , \qquad P = - \frac{\beta}{2\pi \g} \e^\lambda \left(\frac{\beta}{2 \pi } \cW_+^2 + \ii q r_3 \cW_-^2\right) \, , \nn \\
	\FourdSUSYVec &= \left( - q + \ii \frac{\beta}{2\pi r_3} \right) \partial_\mu \, .
\end{align}
We then check that indeed the only fixed point is at $\rho = 2r_3$, where $S = - P$, thus confirming the negative chirality necessary to obtain a non-zero on-shell action, and the weights are $b_1 = b_2 = -q + \ii \frac{\beta}{2\pi r_3}$. We also immediately confirm the relations between the $D=5$ and $D=4$ spinor bilinears given in \eqref{eq:5d4dvectors}, and compute the charge of $\epsilon$ under $\FourdSUSYVec$, obtaining
\begin{equation}
		Q^{(\FourdSUSYVec)} = q - \ii \frac{ \beta }{2 \pi r_3 } = \cQ - (\FourdSUSYVec \hook A^0 - \e^{3\lambda} P ) Q^{(\KKVec)}  = \cQ - \frac{1}{\g} Q^{(\KKVec)} \, ,
\end{equation}
as in \eqref{qrels}.
Moreover, we can compute the equivariant Chern classes $\Phi^I$, confirming that $\Phi^I_0$ are not constant, but at $\rho = 2r_3$
\begin{equation}
		\g \Phi^0_0 = 1 \, , \qquad \g \check{\Phi}^1_0 = - \ii \frac{\beta}{3\pi r_3} \, ,
\end{equation}
which is consistent both with \eqref{eq:4d_Constraint_v1} with $\kappa = 1$ , and with \eqref{eq:Uplift_Phi00} with $R^{\rm up} = 1$.

Finally, we move to checking the rewriting of the $D=5$ action given in \eqref{eq:totac_fp_bdy}. First, we compute the contribution of the single fixed point at the origin of $\R^4$, which is given by \eqref{eq:I_FP_Minimal}, and indeed, precisely matches \eqref{eq:EAdS5_qtwisted}. We are then left with the evaluation of the ``boundary" terms, the last four terms in \eqref{eq:totac_fp_bdy}, all of which are non-zero and separately divergent. We evaluate them considering a cutoff spacetime $\{\rho \geq \cutoff\}$ with boundary $\partial M_{\cutoff}$ and induced metric $h$. For the first two terms arising from the BVAB theorem we get
\begin{align}
	& - \frac{1}{8\pi G_{(5)}} \int_{\partial M_{(5)}} \alpha \wedge \eta \wedge ( \Phi_2 + \Phi_0 \rd\eta ) = - \frac{3 \pi  \beta  r_3^3 }{8 \g^3 G_{(5)} \cutoff^4} + \frac{ \pi \beta r_3 ( 3 \beta^2 + 8 \pi^2 q^2 r_3^2)}{ 8 \g^3 G_{(5)} \left( \beta^2 + 4 \pi ^2 q^2 r_3^2\right) \cutoff^2 } \nn \\
	& \qquad \qquad \ - \frac{\pi  \beta  \left(115 \beta ^4+2160 \pi ^4 q^4 r_3^4+1224 \pi ^2 \beta ^2 q^2 r_3^2+512 \pi \ii \beta ^3 q r_3\right)}{1728 r_3 \g^3 G_{(5)} \left(\beta ^2+4 \pi ^2 q^2 r_3^2\right)^2 } + o(1) \, , 
    \end{align}
and
\begin{align}
	&- \frac{\Delta x^5}{16 \pi G_{(5)}} \int_{M_{(4)}} \Lambda_4 = \frac{\pi  \beta  r_3^3}{8 \g^3 G_{(5)} \cutoff^4} -\frac{\pi  \beta ^3 r_3}{ 8 \g^3 G_{(5)} \left( \beta^2 + 4 \pi^2 q^2 r_3^2\right) \cutoff^2} \nn \\
	& \qquad \qquad \qquad \qquad \qquad +\frac{ \pi \beta \left( 81 \beta^4 - 432 \pi^4 q^4 r_3^4 - 584 \pi ^2 \beta ^2 q^2 r_3^2 \right)}{1728 r_3 \g^3 G_{(5)} \left(\beta ^2+4 \pi ^2 q^2 r_3^2\right)^2} + o(1) \, , 
 \end{align}
while the Gibbons--Hawking--York and counterterm actions give
\begin{align}
    I_{\mathrm{GHY}}^{\partial M_{(5)}}
    &= \frac{1}{8 \pi G_{(5)}}  \int_{\partial M_\cutoff}  K \, \vol_h
    = \frac{\pi  \beta  r_3^3 }{\g^3 G_{(5)} \cutoff^4} - \frac{\pi  \beta  r_3 }{4 \g^3 G_{(5)} \cutoff^2} + o(1) \, ,
    \end{align}  
and
\begin{align}
	I^{\partial M}_{(5)} &= -\frac{1}{8 \pi G_{(5)}} \Bigg[ 
	 \int_{\partial M_\cutoff} \left( 3 \g + \frac{1}{4\g }R + \frac{1}{\g^3}\left( E - C_{ijkl}C^{ijkl} + 8\cF_{ij}\cF^{ ij} \right) \log \cutoff \right) \vol_h + \Delta S_{\rm new} \Bigg] \nn \\
	&  = - \frac{3 \pi  \beta  r_3^3 }{4 \g^3 G_{(5)} \cutoff^4} + \frac{17 \pi  \beta }{864 r_3 \g^3 G_{(5)} } + o(1) \, .
\end{align}
Here the counterterm action contains the logarithmically divergent Weyl anomaly expressed in terms of
the Euler scalar $E$, the Weyl tensor $C_{ijkl}$ and the gauge field, which vanishes for this background. It also
contains the finite counterterms $\Delta S_{\rm new}$, which can be found in \cite{BenettiGenolini:2016tsn}, to obtain 
a scheme that preserves supersymmetry.
The total sum of \eqref{eq:EAdS5_qtwisted} and these four expressions does indeed match \eqref{osactsoimp}, confirming that indeed the rewriting \eqref{eq:totac_fp_bdy} is consistent.

We also note that the boundary terms sum to zero when $q=0$, so they don't contribute to the untwisted reduction leading to the purely fixed point contributions in $D=4$. It is interesting to highlight, though, that each one separately is non-zero. In particular, the integral of $\Lambda_4$ is non-zero, essentially because of the term coming from the $\rd *_{(4)}\rd \lambda$ from the reduction of the $D=5$ Ricci scalar. More specifically, for $AdS_5$ the $D=5$ gauge field is flat, so the $D=4$ gauge are related by 
$\check{A}^1 = \check{z}_1 A^0$, and the on-shell expression of $\Lambda_4$ given in \eqref{eq:Lambda4_Minimal} simplifies further to
\begin{equation}
\begin{split}
 \Lambda_4 \rvert_{\rm OS} &= 2 \Big[ \frac{\e^{-6\lambda}}{4} *_{(4)} F^0 \wedge F^{0} + 2 \g^2 \e^{2\lambda} \vol_4 + \frac{\ii}{2} \check{z}_1^3 {F}^0 \wedge F^0 \Big] \, .
\end{split}
\end{equation} 
Thus, we see that it is non-zero even for the untwisted reduction along $\partial_\tau$, where $F^0=0$.

Finally, we recall that the supersymmetric solution \eqref{eq:EAdS5_Metric}, \eqref{eq:EAdS5_GaugeField} is the Wick-rotation of Lorentzian $AdS_5$, and its reduction satisfies the $D=4$ reality conditions \eqref{eq:4d_Reality_New}, though not the original ones \eqref{eq:4d_Reality_Original} used in \cite{BenettiGenolini:2024lbj}.

\section{The black hole}\label{examplesblackhole}

We now look at the complex solutions described in \cite{Cabo-Bizet:2018ehj}, which represent supersymmetric complex non-extremal deformations of the supersymmetric extremal black holes of \cite{Gutowski:2004ez, Chong:2005hr}. The deformation away from extremality allows one to avoid the bulk divergence associated with the infinite near-horizon throat.
The topology of the $D=5$ solution is $\mathbb{R}^2\times S^3$, and supersymmetry is preserved via the Killing spinor being anti-periodic around the circle shrinking at the origin of the  $\R^2$ factor, when considered in a regular gauge for the $U(1)$ gauge field.

We start by writing the $D=5$ bilinear Killing vector as
\begin{equation}
\label{eq:CCLPsusyKV}
	\cK = \partial_\tau + \varepsilon_1 \partial_{\varphi_1} + \varepsilon_2 \partial_{\varphi_2} \, ,
\end{equation}
where $\varphi_1,\varphi_2$ are coordinates on $S^3$, as in \eqref{s3coords}, and $\tau$ is a coordinate on $S^1$, the ``thermal circle", 
with $\Delta\tau=\Delta\varphi_1=\Delta\varphi_2=2\pi$. As before, $\cK$ is generically complex, 
with $\varepsilon_1, \varepsilon_2 \in \mathbb{C}$. The horizon of the black hole is generated by
\begin{equation}
	V_H = \partial_\tau \, ,
\end{equation}
and note, in general, $\cK \neq V_H$. Using a regular gauge for the $U(1)$ gauge field, the anti-periodic spin structure of the $\R^2$ disc factor implies that the Killing spinor $\chi$ satisfies 
\begin{equation}
\label{eq:Lie_CCLP_tau}
	\cL_{\partial_\tau} \chi = \JS\frac{\ii}{2} \chi \, ,\qquad
    \JS=\pm 1\,.
\end{equation} 
Next, recalling \eqref{eq:Lie_KS_S3}, 
we also have
\begin{equation}
\label{eq:Lie_CCLP_S3}
    \cL_{\partial_{\varphi_1}}\chi = \rsigma \frac{\ii}{2}\chi \,, \qquad \cL_{\partial_{\varphi_2}}\chi = \rsigma \frac{\ii}{2}\chi \,,
    \qquad \rsigma=\pm 1\,.
\end{equation}

In contrast to section \ref{examplesads5}, we will compute the on-shell action for the black holes by utilizing background subtraction, as discussed in section \ref{sec:bkgsub}. We will take the background solution $N_{(5)}$ to be the vacuum $AdS_5$ solution and then compute the on-shell action for different
choices of KK vectors $\ell$ using the specific formula \eqref{bsfptformula}.
We first consider black holes in minimal $D=5$ gauged supergravity, before considering black holes with multiple charges in section \ref{multisec}. 

\subsection{Reduction with a \texorpdfstring{$p$}{p}-twist using the thermal circle; minimal gauged supergravity}
\label{sec:BH_ptwist}

We first carry out a KK reduction with respect to the Killing vector
\begin{equation}
\label{eq:CCLP_ell}
	\ell = \g \left( 2 \tinyspace \partial_{\psi} + p \tinyspace \partial_{\tau} \right) \, ,
\end{equation}
where $p$ is an arbitrary integer. Doing so leads to a $D=4$ base with topology
$M_{(4)} \cong \mathcal{O}(-p) \rightarrow S^2$. 
We shall see that the $D=5$ on-shell action, regularized by background subtraction, is independent of $p$. Note for $p=0$ the reduction is simply along the Hopf fibre in the $S^3$, as done in \cite{Hosseini:2017mds,Benini:2020gjh}.
Interestingly, 
we find that in this case the contribution from the subtraction manifold vanishes, and the black hole on-shell action \eqref{bsfptformula} is recovered solely from the fixed point contributions of $M_{(4)}$. As discussed around equation \eqref{engelbert}, we proceed assuming we are in a gauge with $\mathcal{L}_\xi A^0=0$, and such that the gauge field is regular at the fixed points, with $\xi=\pi_*(\SUSYVec)$.

Given \eqref{eq:Lie_CCLP_tau}-\eqref{eq:Lie_CCLP_S3}, the charge of the $D=5$ Killing spinor $\chi$ with respect to $\ell$ is given by 
\begin{equation}
\label{eq:CCLP_Qell}
	Q^{(\ell)} =  \frac{\g}{2} \left( p \tinyspace \JS + 2 \rsigma \right) \,,
\end{equation}
where recall that the charges in \eqref{eq:Lie_CCLP_tau}-\eqref{eq:Lie_CCLP_S3} are fixed by assuming a regular gauge for the $D=5$ gauge field.
To proceed, we introduce periodic coordinates $\mu,\nu$ via
\begin{equation}
\label{eq:CCLP_coordtransform}
    \tau = p\nu+\mu\,,\qquad \psi = 2\nu \,,
\end{equation}
with $\Delta\mu=\Delta\nu=2\pi$. In the new coordinates we have
\begin{equation}
\label{eq:CCLP_ell_nu}
\ell = \g\partial_\nu \, ,
\end{equation}
and so from \eqref{periodofthecircle} we have $\Delta x^5 = 2\pi \g^{-1}$.
Also, 
the $D=5$ supersymmetric Killing vector is given by
\begin{equation}
    \cK  = \left( 1 - \frac{\varepsilon_1 + \varepsilon_2}{2} p \right) \partial_\mu - ( \varepsilon_1-\varepsilon_2 ) \partial_{\phi} + \frac{\varepsilon_1 + \varepsilon_2}{2} \partial_{\nu} \,.
\end{equation}

The pushforward of $\cK$, $\xi = \pi_*(\cK)$, can then be read off from \eqref{eq:5d4dvectors}. 
In order to compute the weights of $\xi$ at the fixed points of $M_{(4)}$, we first diagonalize the $U(1)^2$ action at (say) the south pole of the zero-section $S^2\subset M_{(4)}\cong \mathcal{O}(-p)\rightarrow S^2$. This leads to the coordinate transformation
\begin{equation}
	\partial_{\gamma_1} = \partial_\phi + \frac{p}{2} \partial_\mu \,, \qquad \partial_{\gamma_2} = \partial_\mu \,,
\end{equation}
where $\Delta\gamma_1=\Delta\gamma_2=2\pi$, so that we have
\begin{equation}
\label{eq:CCLP_xi}
	\xi 
	= - (\varepsilon_1 - \varepsilon_2) \partial_{\gamma_1} + ( 1 - \varepsilon_2 \, p ) \partial_{\gamma_2} \,.
\end{equation}
In this basis we then have the standard toric 
data for this manifold \cite{BenettiGenolini:2024hyd}:
\begin{equation}
\label{eq:CCLP_toric}
	\vec{v}_0 = (-1 , 0 ) \,, \qquad \vec{v}_1 = ( 0 , -1 ) \,, \qquad \vec{v}_2 = ( 1 , -p ) \,.
\end{equation}
Here $\vec{v}_1=-\partial_{\gamma_2}$ rotates the fibre of $\mathcal{O}(-p)\rightarrow S^2$, fixing the zero-section, while 
$\vec{v}_0=-\partial_{\gamma_1}$, $\vec{v}_2 = \partial_{\gamma_1} - p \partial_{\gamma_2}$ rotate the tangent spaces in $S^2$ at the south ($\theta=\pi$) and north ($\theta=0$) poles, respectively.
As explained in \cite{BenettiGenolini:2024hyd}, the weights $b_i^{N,S}$ of $\xi$ at the fixed north and south poles can be computed from the toric data. Given \eqref{eq:CCLP_xi}, \eqref{eq:CCLP_toric},
these are
\begin{align}
\label{eq:BH_weights}
	(b_1^N , b_2^N) & = \left( -\det (\vec{v}_2,\xi) , \, \det (\vec{v}_1, \xi) \right)\nn \\
	& = ( -1 + \varepsilon_1 p , - \varepsilon_1 + \varepsilon_2 ) \, , \nn\\
	(b_1^S, b_2^S) & = \left( - \det (\vec{v}_1 , \xi) , \, \det (\vec{v}_0 , \xi) \right)\nn \\
	& = ( \varepsilon_1 - \varepsilon_2 , -1 + \varepsilon_2 p) \, .
\end{align}
Notice that, when $\varepsilon_1 = \varepsilon_2$, the $S^2$ becomes an isolated fixed surface, i.e. a ``bolt" for the $D=4$ supersymmetric Killing vector $\xi$. The fixed point contribution to the action is then evaluated using the bolt formula in \eqref{eq:I_FP_Minimal}. Noting that the $\varepsilon_1 \rightarrow \varepsilon_2$ limit is smooth, in the sequel we assume $\varepsilon_1 \neq \varepsilon_2$, noting that the ``bolt limit" can be easily recovered, as explained in \cite{BenettiGenolini:2024hyd}. 

Next, from \eqref{eq:Uplift_Phi00}, we have that $\g\Phi_0^0$ at the fixed points give the weights of the lift of $\xi$ in $M_{(5)}$, i.e. $\cK$. The circles generated by $\varphi_2$ and $\varphi_1$ degenerate at the north and south poles, respectively, so from \eqref{eq:CCLPsusyKV} we have
\begin{equation}
\label{eq:BH_Phi00}
	\left. \g \Phi_0^0 \right\vert_N = \varepsilon_1 \,, \qquad \left. \g \Phi_0^0 \right\vert_S = \varepsilon_2 \,.
\end{equation}
Since we are working with Euclidean $D=5$ minimal gauged supergravity, the constraint \eqref{eq:4d_Constraint_v1} uniquely determines $\check{\Phi}_0^1$, and the fixed point contribution to the action is simply given by \eqref{eq:I_FP_Minimal}. Specifically, from the constraint we have
\begin{align}
\label{eq:BH_Phi01}
	\left. \g \check{\Phi}_0^1 \right\vert_N & = \frac{1}{3} \Big[ - \kappa_N (-1 + \varepsilon_1 p + (\varepsilon_1 - \varepsilon_2) \chi_N ) ) + (p \tinyspace  \JS+2\rsigma) \varepsilon_1 \Big] \, , \nn\\
	\left. \g \check{\Phi}_0^1 \right\vert_S & = \frac{1}{3} \Big[ - \kappa_S ( \varepsilon_1 - \varepsilon_2 + (1 - \varepsilon_2 p ) \chi_S ) + (p \tinyspace  \JS + 2\rsigma) \varepsilon_2 \Big] \, .
\end{align}
Note that these are invariant under the following symmetry
\begin{equation}
    \kappa_{S} \leftrightarrow -\kappa_{S} \, , \quad
	\chi_{N,S} \leftrightarrow - \chi_{N,S} \,, \quad \varepsilon_{1,2} \leftrightarrow - \varepsilon_{1,2} \,, \quad p \leftrightarrow -p \,, \quad \rsigma \leftrightarrow - \rsigma \, .
\end{equation}
The KK reduction introduces the signs $\kappa_{N,S}$, 
$\chi_{N,S}$, in addition to the signs $\JS$, $\rsigma$ associated to the charge of the original $D=5$ Killing spinor under the $U(1)^3$ isometry, \eqref{eq:Lie_CCLP_tau}, \eqref{eq:Lie_CCLP_S3}. These signs are not all independent: 
some are related, 
as explained in \cite{BenettiGenolini:2024hyd}, 
and some may be absorbed via coordinate redefinitions and choices of convention. 

In order to analyze this carefully, we begin by recalling from \cite{BenettiGenolini:2024hyd} that to each edge $\vec{v}_a$ of the toric diagram, $a=0,1,2$, 
we have an associated sign $\sigma_a\in \{\pm 1\}$. The latter specifies (twice) the charge of the  
Killing spinor under the $U(1)$ subgroup 
defined by $\vec{v}_a$, which must be $\pm 1$ in order that the spinor is non-singular along the subspace corresponding to that edge. This data is related to $\kappa_{N,S}$, $\chi_{N,S}$ via
\begin{equation}\label{terrytheturnip}
\sigma_0=-\kappa_S\, , \quad \sigma_1=-\kappa_N\, , \quad 
\sigma_2 = -\chi_N\sigma_1\, ,
\end{equation}
where $\chi_S = -\sigma_0\sigma_1$, $\chi_N=-\sigma_1\sigma_2$. In particular, it's immediate from these relations that $\kappa_N=-\chi_S\kappa_S$. 

Further constraints on the signs are 
 most easily fixed by looking at the gauge field fluxes.
These are given by \eqref{eq:BH_weights}, and we apply the BVAB theorem on the equivariantly closed form $\Phi_{(F)}^\Lambda \equiv F^\Lambda + \Phi_0^\Lambda$ to obtain
\begin{equation}
\label{eq:BH_BVAB_F}
	\mf{p}^\Lambda = - \frac{1}{2(\varepsilon_1 - \varepsilon_2)} \left( \g \Phi_0^\Lambda \rvert_N - \g \Phi_0^\Lambda \rvert_S \right) \,.
\end{equation}
From \eqref{eq:Rsymmconst}, it follows that the $D=4$ R-symmetry gauge field flux is constrained by
\begin{equation}
\label{eq:BH_Rsymm_v1}
	\zeta_\Lambda^{(4)} \mf{p}^\Lambda = \frac{1}{(\varepsilon_1 - \varepsilon_2)} \left[ \kappa_N (b_1^N - \chi_N b_2^N) - \kappa_S (b_1^S - \chi_S b_2^S) \right] \, .
\end{equation}
Substituting the weights \eqref{eq:BH_weights}, together with the relation $\kappa_N = - \chi_S \kappa_S$, \eqref{eq:BH_Rsymm_v1} reads
\begin{equation}
    \zeta_\Lambda^{(4)} \mf{p}^\Lambda = -\kappa_S (1+\chi_N\chi_S +\chi_S p ) \, .
\end{equation}
Now, substituting the FI parameters \eqref{eq:4d_FI_Parameters}, this can be re-written as:
\begin{equation}
\label{eq:BH_Rsymm_v2}
    - \frac{4}{\g} Q^{(\KKVec)} \mf{p}^0 + 6 \check{\mf{p}}^1 = -\kappa_S (1+\chi_N\chi_S +\chi_S p ) \, .
\end{equation}
We can then choose $\nvxi^1 = -\mf{p}^1/\mf{p}^0$, such that $\check{\mf{p}}^1=0$ (recalling that these combinations are invariant under the redundancy in the reduction ansatz \eqref{eq:KK_Redundancy}).
On the other hand, $\mf{p}^0$ is proportional to the Chern number of the KK bundle over $S^2$, as noted below \eqref{eq:p0}. Since $\Delta x^5 = 2\pi\g^{-1}$, it must be that $\mf{p}^0$ is half the Chern number. To compute it, we substitute \eqref{eq:BH_Phi00} in \eqref{eq:BH_BVAB_F}, obtaining
\begin{equation}
	\mf{p}^0 = - \frac{1}{2} \, .
\end{equation}
Note that this relation gives the same Chern number as the Hopf fibration over $S^2$, which is precisely as expected for the ``$p$-twist'' we have done. 
Substituting this value for $\mf{p}^0$, 
along with $Q^{(\KKVec)}$ from \eqref{eq:CCLP_Qell}, into \eqref{eq:BH_Rsymm_v2}, we find
\begin{equation}
\label{eq:BH_z0p0}
    p \tinyspace \JS + 2 \rsigma = -\kappa_S (1+\chi_N\chi_S +\chi_S p )\,.
\end{equation}
We thus conclude that $\chi_N\chi_S=1$.
This is consistent with the fact that when $\varepsilon_1 = \varepsilon_2$, which is associated with $D=5$ black holes with 
the two angular momenta equal to each other, the reduction along the ($p=0$) Hopf-fibre direction is then known to lead to supersymmetric $D=4$ static solutions with $\R^2 \times S^2$ topology, 
that preserve supersymmetry 
via a topological twist (see also \cite{Hristov:2018spe}). 
We also observe that 
$\kappa_N = - \chi_S \kappa_S = \JS$ and $\kappa_S = - \rsigma$. Thus, we have
\begin{equation}
\label{eq:BH_kappaNS}
	\chi_N=\chi_S=\JS\tinyspace \rsigma\,,\qquad \kappa_S=-c_J,\qquad \kappa_N=c_R\,.
\end{equation}
In terms of the sign variables $\sigma_a$ in \eqref{terrytheturnip}, these are $\sigma_0=\sigma_2=\rsigma, \sigma_1=-\JS$.

Taking these signs in \eqref{eq:BH_Phi01}, we may now evaluate \eqref{eq:I_FP_Minimal}. A short computation shows that 
 the contribution of the $D=5$ on-shell action in \eqref{bsfptformula}  
from
$M_{(4)}$ is given by
\begin{align}
 I_{(4)}^{\rm FP}[M_{(4)}] =I^{\rm FP}_{(4) N} +I^{\rm FP}_{(4) S} \,,
\end{align}
where
\begin{align}
\label{eq:BH_FP_NS}
	I^{\rm FP}_{(4) N} & = \frac{\ii \pi^2}{54 \g^3 G_{(5)}} \JS\frac{[1 + \JS \rsigma(\varepsilon_1 + \varepsilon_2)]^3}{\varepsilon_1 (\varepsilon_1 - \varepsilon_2) (1 - p \tinyspace\varepsilon_1)} \,, \nn\\
	I^{\rm FP}_{(4) S} & =  - \frac{\ii \pi^2}{54 \g^3 G_{(5)}} \JS\frac{[1 + \JS \rsigma(\varepsilon_1 + \varepsilon_2)]^3}{\varepsilon_2 (\varepsilon_1 - \varepsilon_2) (1 - p \tinyspace\varepsilon_2)} \,.
\end{align}

In order to carry out background subtraction,
we next need to compute $I_{(4)}^{\rm FP}[N_{(4)}]$. Here $N_{(4)}$ arises by starting with 
$N_{(5)}$ being the $AdS_5$ vacuum,
with the same $D=5$ supersymmetric Killing vector $\cK$ in \eqref{eq:CCLPsusyKV}. 
We perform the same KK reduction with respect to $\ell$, which we remind the reader is
\begin{equation}
\label{eq:bkgsub_ell}
    \ell = \g (2 \tinyspace \partial_\psi + p \tinyspace \partial_\tau ) \,.
\end{equation}
Note the difference with the KK reduction of the $AdS_5$ example considered in section \ref{examplesads5} in
equation \eqref{thetwisteellads5}.
Equations \eqref{eq:Lie_CCLP_tau} and \eqref{eq:Lie_CCLP_S3} for spinor charges hold as before, 
and hence so does 
the expression 
for $Q^{(\ell)}$ 
given in \eqref{eq:CCLP_Qell}.

We then compute $I_{(4)}^{\rm FP}[N_{(4)}]$
using $D=4$ localization with respect to the $D=4$ supersymmetric Killing vector $\xi$ \eqref{eq:CCLP_xi}. Dimensional reduction of $AdS_5$ gives $N_{(4)} \cong \R^4/\Z_p$, for which the toric data is specified by (in the conventions of \cite{BenettiGenolini:2024hyd})
\begin{equation}
	\vec{v}_0 = (-1 , 0 ) \,, \qquad \vec{v}_1 = (1 , -p) \,.
\end{equation}
Notice this is the toric data \eqref{eq:CCLP_toric}, 
where the old vector $\vec{v}_1$ has 
been discarded (corresponding to collapsing the 
$S^2$ horizon of the $D=4$ black hole/black bolt), 
and relabeling $\vec{v}_2\mapsto \vec{v}_1$. 
We can then correspondingly immediately identify
the signs for this KK reduction of $AdS_5$: $\kappa=-\sigma_0=\kappa_S=-\rsigma$, $\chi=-\sigma_0\sigma_2=-1$. In particular, notice that 
the chirality of the spinor is fixed to be negative,\footnote{That it is negative rather than positive can be traced back to the fact that we took both signs in \eqref{eq:Lie_CCLP_S3} to be $c_J$, rather than $c_J$ and $-c_J$.}
independently of other unfixed signs that may be specified freely. 
The geometry has a single fixed point at the origin of $\R^4/\Z_p$. It follows that the weights of $\xi$ \eqref{eq:CCLP_xi} at the fixed point are given by
\begin{align}
	b_1 & = - \frac{1}{p} \det(\vec{v}_1,\xi) = - \frac{1}{p} + \varepsilon_1 \,, \nn\\
	b_2 & = \frac{1}{p} \det(\vec{v}_0,\xi) = - \frac{1}{p} + \varepsilon_2 \,.
\end{align}
Again, $\Phi_0^0$ at the fixed point is given by \eqref{eq:Uplift_Phi00}. Since the entire $S^3$ degenerates at the centre of $AdS_5$, from \eqref{eq:CCLPsusyKV} we have
\begin{equation}
	\left. \g\Phi_0^0\right\vert_{\rm{centre}} = \frac{1}{p} \, ,
\end{equation}
where the order of the orbifold group appears, as the complex line bundle $L$ is now over the orbifold singularity at the centre of $\R^4/\Z_p$.
Finally, the constraint \eqref{eq:4d_Constraint_v1} uniquely determines $\check{\Phi}_0^I$ at the fixed point.
Collecting everything, we find that $I_{(4)}^{\rm FP}[N_{(4)}]$ using \eqref{eq:I_FP_Minimal} evaluates to 
\begin{equation}
	I_{(4)}^{\rm FP}[N_{(4)}] = \JS \frac{\ii \pi^2}{54 \g^3 G_{(5)}} \frac{[1 + \JS \rsigma( \varepsilon_1 + \varepsilon_2)]^3}{(1 - p \varepsilon_1) (1 - p \varepsilon_2) } p^2 \, .
\end{equation}

Thus, we find that the total $D=5$ action, using background subtraction as in \eqref{bsfptformula},
leads to
\begin{align}
\label{eq:CCLP_ans_v1}
\Itot_{(5), {\rm bs}} &= \JS \frac{\ii \pi^2}{54 \g^3 G_{(5)}} \left[ \frac{(1 + \varepsilon_1 + \varepsilon_2)^3}{\varepsilon_1 (\varepsilon_1 - \varepsilon_2)} - \frac{(1 + \varepsilon_1 + \varepsilon_2)^3}{\varepsilon_2 (\varepsilon_1 - \varepsilon_2)} \right ] \, ,\nn\\
	&= - \JS \frac{\ii \pi^2}{2 \g^3 G_{(5)}} \left[ \frac{[1 + \JS\tinyspace\rsigma(\varepsilon_1 + \varepsilon_2)]^3}{27 \varepsilon_1 \varepsilon_2} \right ] \, .
\end{align}

Before discussing this expression further, we note that to use \eqref{bsfptformula} we require 
the vanishing of the $\Lambda_4$ contributions for both $M_{(4)}$ and $N_{(4)}$. As discussed in section~\ref{sec:bkgsub}, this requires $\Lambda_4$ to be globally defined (and hence exact) on both $M_{(4)}$ and $N_{(4)}$. First, from the reduction of $AdS_5$, we have $N_{(4)} \cong \R^4/\Z_p$, on which any closed two-form is 
exact. 
Next, for the black hole, in the regular gauge $V_H \hook \cA$ vanishes at the horizon, where the $\R^2$ disc factor smoothly caps off. It follows that $\cA$ is a globally defined one-form, and so is $\check{A}^1$, as noted below \eqref{eq:Reduction_Ansatz3nvs}. This shows \eqref{eq:Lambda4_General} is globally defined, which implies that $\Lambda_4$ is exact on $M_{(4)}$. We thus conclude that \eqref{bsfptformula} is valid for this example, for any choice of $\ell$.

Notice that all $p$ dependence has dropped out of the final result \eqref{eq:CCLP_ans_v1}. We also notice that
the first expression is in the form of gravitational blocks \cite{Hosseini:2019iad}; for $p=0$, when
the KK reduction is just with respect to the Hopf direction, the contributions are coming
from the north and south pole fixed point contributions to
$I_{(4)}^{\rm FP}[M_{(4)}]$ (with $I_{(4)}^{\rm FP}[N_{(4)}]=0$). When $p\ne 0$ the gravitational origin of the block formula is more obscure, 
although formally 
the additional contribution of $I_{(4)}^{\rm FP}[N_{(4)}]$ appears 
as another ``gravitational block,'' as part of the way the on-shell action has been regularized.

We have recovered the correct expression for the on-shell action, suitably regularized, to 
be dual to the supersymmetric index of the dual SCFT. 
For example, 
we can compare with e.g. \cite{Aharony:2021zkr}, where one should identify
\begin{equation}
\label{eq:BH_angvel_id}
	 \sigma_g \leftrightarrow - \rsigma \varepsilon_1 \,, \qquad \tau_g \leftrightarrow - \rsigma \varepsilon_2 \, ,
\end{equation}
and introduce $\Delta_g$, which satisfies\footnote{To compare with \cite{Cabo-Bizet:2018ehj}, note that $\omega_1^{\rm{there}} = 2\pi \ii \sigma_g$, $\omega_2^{\rm{there}} = 2\pi \ii \tau_g$, and $\varphi^{\rm{there}} = 3 \pi \ii \Delta_g$ . We then have $\Itot_{(5),{\rm bs}} = \frac{2\pi}{27\g^3 G_{(5)}} \frac{\varphi^3}{\omega_1\omega_2}$, with constraint $\omega_1 + \omega_2 - 2\varphi = 2\pi\ii \JS$.}
\begin{equation}
\label{eq:BH_constraint}
	\sigma_g + \tau_g - 3 \Delta_g = \JS \, .
\end{equation}
With this constraint imposed, \eqref{eq:CCLP_ans_v1} is then equivalent to
\begin{equation}
\label{eq:CCLP_ans_fin}
	\Itot_{(5), {\rm bs}} = \frac{\ii \pi^2}{2 \g^3 G_{(5)}} \frac{\Delta_g^3}{\sigma_g \tau_g} \,.
\end{equation}
We remark that the sign $\JS = \pm 1$ in \eqref{eq:BH_constraint}, associated to the two ``branches" of solutions \cite{Aharony:2021zkr}, is precisely the discrete data specified by the charge of the $D=5$ Killing spinor with respect to $V_H$ \eqref{eq:Lie_CCLP_tau}.

At this point it is interesting to compare with the recent results of \cite{Colombo:2025ihp}, where the odd-dimensional generalization of the BVAB theorem given in  \cite{Goertsches:2015vga} was used to localize the same black hole action in $D=5$ minimal gauged supergravity. The method consists of constructing a polyform using the $D=5$ supersymmetric Killing vector, $\xi^{\rm{there}}$, which is equivariantly closed with respect to another arbitrary Killing vector field $X^{\rm{there}}$. Subsequently, their localization is performed with respect to $X^{\rm{there}}$. Our approach is inherently different, in that we perform a dimensional reduction with respect to an arbitrary Killing vector $\ell$, and localize with respect to the $D=4$ supersymmetric Killing vector $\xi$. Interestingly, the two localization computations can be matched, term by term, if the arbitrary Killing vectors in the two prescriptions are identified. Note our $(\tau, \varphi_1, \varphi_2)$ basis is equivalent to their ``$\phi$-basis", in which
\begin{align}
	\xi_0^{\rm{there}} & = \frac{1}{\beta} (-2\pi \ii, -\omega_1,-\omega_2)\,,\quad \xi_1^{\rm{there}} = \frac{1}{\beta} ( -\omega_1,-\omega_2,2\pi \ii)\,, \nonumber\\
 \xi_2^{\rm{there}}&  = \frac{1}{\beta} (-\omega_2,2\pi \ii,-\omega_1)\,.
\end{align}
Next, we identify the arbitrary vector $X^{\rm{there}}$ in their prescription with our KK reduction vector $\ell = (p,1,1)$. Specifically, we have
\begin{equation}
    X_0^{\rm{there}} = (-p,1,1)\,,\quad X_1^{\rm{there}} = (1,1,p)\,,\quad X_2^{\rm{there}} = (1,p,1)\,.
\end{equation}
It follows that (see (4.25) there)
\begin{align}
\label{eq:colombo_ans}
	\frac{1}{16\pi\ii G_{(5)}} \cI^{\rm{there}}_1[X] & = \frac{\pi^2 L^3}{54 G_{(5)}} \frac{(2\pi \ii - \omega_1 - \omega_2)^3}{\omega_1(2\pi - \ii p\mskip2mu  \omega_1)(\omega_1-\omega_2)} \, , \nn\\
	\frac{1}{16\pi\ii G_{(5)}} \cI^{\rm{there}}_2[X] & = - \frac{\pi^2 L^3}{54 G_{(5)}} \frac{(2\pi \ii - \omega_1 - \omega_2)^3}{\omega_2(2\pi - \ii p\mskip2mu  \omega_2)(\omega_1-\omega_2)} \, , \nn\\
	\frac{1}{16\pi\ii G_{(5)}} \cI^{\rm{there}}_0[X] & = - \frac{\pi L^3}{108 G_{(5)}} \frac{(2\pi \ii - \omega_1 - \omega_2)^3}{ (2\pi - \ii p\mskip2mu \omega_1) (2\pi - \ii p\mskip2mu  \omega_2)} p^2 \,.
\end{align}
Each line in \eqref{eq:colombo_ans} agrees with our $I^{\rm{FP}}_{(4)N},I^{\rm{FP}}_{(4)S}$, and $I_{(4)}^{\rm FP}[N_{(4)}]$, respectively, where one should identify
\begin{equation}
	\omega_1^{\rm there} \leftrightarrow - 2 \pi \ii \mskip2mu  \varepsilon_1\, \qquad \omega_2^{\rm there} \leftrightarrow - 2 \pi \ii\mskip2mu  \varepsilon_2 \, ,
\end{equation}
 and set $\JS = \rsigma = 1$.

\subsection{Spindle reduction; multi-charge black holes}\label{multisec}

Families of supersymmetric extremal Lorentzian black hole solutions are also known in $D=5$ gauged supergravity coupled to an arbitrary number of vector multiplets \cite{Gutowski:2004yv,Kunduri:2006ek}. 
To deal with the infrared divergence associated with the $AdS_2$ factor in the near-horizon, one can
compute their on-shell action using the limiting procedure presented in \cite{Cabo-Bizet:2018ehj}, which
requires \fixme{ constructing a non-extremal, supersymmetric (and complex) deformation.
For the $U(1)^3$ model, this could be obtained using the non-supersymmetric, Lorentzian (and real) 
black hole solutions with two angular momenta and three electric charges found in \cite{Wu:2011gq}, but
this has not yet been done. However, 
when two angular momenta are equal, such deformations have been constructed and analyzed in \cite{Cassani:2019mms} using the Lorentzian solutions of \cite{Cvetic:2004ny, Cvetic:2005zi}.}

Here, we use dimensional reduction and equivariant localization to find the on-shell action of supersymmetric solutions with topology $\mathbb{R}^2 \times S^3$ and $U(1)^3$ isometry in $D=5$ gauged supergravity coupled to an arbitrary number of vector multiplets, finding 
\begin{equation}
\label{eq:BH_os_multisec}
    \Itot_{(5), {\rm bs}} = - \frac{\ii \pi^2}{2G_{(5)}} \frac{C_{IJK}}{6} \frac{\left. \check{\Phi}_0^I \check{\Phi}_0^J \check{\Phi}_0^K \right\vert_{\rm IR}}{\varepsilon_1 \varepsilon_2}  \, ,
\end{equation}
generalizing \eqref{eq:CCLP_ans_v1} for minimal gauged supergravity.
As we will remark at the end of the section, after using some generalization of the ``UV-IR'' relations of \cite{BenettiGenolini:2024lbj}, this result is consistent with field theory conjectures in the literature.

In the previous subsection, we have shown that for minimal gauged supergravity we obtain
the $D=5$ on-shell action solely from fixed point contributions, given that one performs a reduction along the Hopf direction in the $S^3$, by setting $p=0$. We therefore anticipate that
we can obtain the on-shell action for the multi-charge black holes in the same way. While this is indeed the case, 
instead of considering $\ell$ to be a 
Hopf reduction, where the KK reduction vector is simply $\ell = 2 \g \partial_\psi$, we perform a reduction along
\begin{equation}
	\ell = \g ( n_N \partial_{\varphi_1} + n_S \partial_{\varphi_2} ) \, ,
\end{equation}
with $n_N, n_S \in \mathbb{N}$ and coprime; setting $n_N=n_S=1$ returns to the Hopf reduction. Reducing along this Killing vector
we obtain a spindle black hole in $D=4$ with topology $M_{(4)} \cong \R^2 \times \mathbb{WCP}^1_{[n_N,n_S]}$, and accordingly we will refer to this as the ``spindle reduction" in the sequel. 
Interestingly, as in the Hopf reduction, we find that there is no contribution from the background  
manifold $N_{(4)}$ obtained by doing a spindle reduction of $N_{(5)}=AdS_5$; it would be interesting to know if this is the most general Killing vector $\ell$ with this property.

We start by considering the following coordinate transformation \cite{Arav:2025jee}: 
\begin{equation}
    \nu = m_S \varphi_1 - m_N \varphi_2 \, , \qquad \mu = n_N \varphi_2 - n_S \varphi_1 \, ,
\end{equation}
where $m_N, m_S \in \Z$ are any solutions to $n_N m_S - n_Sm_N = 1$, which exist by B\'ezout's lemma.
As before, the coordinate transformation is such that in the new coordinates we have
\begin{equation}
	\ell = \g \partial_\nu \, .
\end{equation} 
Given \eqref{eq:Lie_CCLP_S3}, the charge of the $D=5$ Killing spinor $\chi$ with respect to $\ell$ is then given by
\begin{equation}
\label{eq:spindle_BH_Qell}
    Q^{(\ell)} =  \g \rsigma \frac{n_N+n_S}{2} \,.
\end{equation}
The $D=5$ supersymmetric Killing vector \eqref{eq:CCLPsusyKV} reads
\begin{equation}
    \SUSYVec = \partial_\tau - ( \varepsilon_1 n_S - \varepsilon_2 n_N ) \partial_\mu + ( \varepsilon_1 m_S - \varepsilon_2 m_N) \partial_\nu \, , 
\end{equation}
and from \eqref{eq:5d4dvectors} it follows that
\begin{equation}
\label{eq:spindle_xi}
	\xi = \partial_\tau - (\varepsilon_1 n_S - \varepsilon_2 n_N) \partial_\mu \, ,
\end{equation}
where $\partial_\tau$ rotates the $\R^2$ direction, and $\partial_\mu$ generates azimuthal rotations of $\mathbb{WCP}^1_{[n_N,n_S]}$. In the conventions of \cite{BenettiGenolini:2024xeo}, the toric data specifying $\R^2 \times \mathbb{WCP}^1_{[n_N,n_S]}$ is given by
\begin{equation}
	\vec{v}_0 = (-n_S , 0 ) \,, \qquad \vec{v}_1 = ( 0 , -1 ) \,, \qquad \vec{v}_2 = ( n_N , 0 ) \,.
\end{equation}
Using the above data, one can now compute the weights of $\xi$ at the fixed points, located at the north and south poles of $\mathbb{WCP}^1_{[n_N,n_S]}$:
\begin{align}
\label{eq:spindle_BH_weights}
	(b_1^N , b_2^N) & = \left( - \det(\vec{v}_2,\xi) / n_N , \det(\vec{v}_1, \xi)/n_N \right) \nn\\
	& =  \left( -1 , - \frac{\varepsilon_1 n_S - \varepsilon_2 n_N}{n_N} \right) \, , \nn\\
	(b_1^S, b_2^S) & = \left( - \det(\vec{v}_1,\xi) / n_S , \det(\vec{v}_0, \xi)/n_S \right) \nn\\
	& =  \left( \frac{\varepsilon_1 n_S - \varepsilon_2 n_N}{n_S} , -1 \right) \, ,
\end{align}
where now the orders of the orbifold singularities appear, given by $n_N = \det(\vec{v}_1,\vec{v}_2)$ and $n_S = \det(\vec{v}_0,\vec{v}_1)$, respectively. Finally, by \eqref{eq:Uplift_Phi00}, $\Phi_0^0$ at the fixed points are the weights of the lift of $\xi$ in $M_{(5)}$, that is, $\cK$.
The circles generated by $\varphi_2$ and $\varphi_1$ degenerate at the north and south poles, respectively, so from \eqref{eq:CCLPsusyKV} we have
\begin{equation}
\label{eq:BH_vec_Phi00}
	\left. \g \Phi_0^0 \right\vert_N = \frac{\varepsilon_1}{n_N} \,, \qquad \left. \g \Phi_0^0 \right\vert_S = \frac{\varepsilon_2}{n_S} \,,
\end{equation}
where, again, the complex line bundle $L$ is over fixed points with orbifold singularities, thus dressing the weights of $\cK$ by $n_N,n_S$.

Collecting everything, we can now evaluate the fixed point contribution to the on-shell action, using \eqref{eq:I_FP_Nuts}, to find
\begin{equation}
\label{eq:BH_vec_os_v1}
 I_{(4)}^{\rm FP}[M_{(4)}] = \ii \frac{\Delta x^5 \pi}{4G_{(5)}} \frac{C_{IJK}}{6} \frac{1}{\varepsilon_1 n_S - \varepsilon_2 n_N} \left[ \left.\frac{\check{\Phi}_0^I\check{\Phi}_0^J\check{\Phi}_0^K}{\Phi_0^0}\right\vert_N - \left.\frac{\check{\Phi}_0^I\check{\Phi}_0^J\check{\Phi}_0^K}{\Phi_0^0}\right\vert_S \right] \,,
\end{equation}
where we have yet to substitute in \eqref{eq:BH_vec_Phi00}. Before we do that, we pause to consider the flux of the gauge fields through $\mathbb{WCP}^1_{[n_N,n_S]}$. Applying the BVAB theorem on the equivariantly closed form $\Phi^\Lambda_{(F)} \equiv F^\Lambda + \Phi_0^\Lambda$, we have
\begin{equation}
\label{eq:spindle_flux_BVAB}
	\mf{p}^\Lambda = - \frac{1}{2(\varepsilon_1 n_S - \varepsilon_2 n_N)} \left( \left. \g \Phi_0^\Lambda \right\vert_N - \left. \g \Phi_0^\Lambda \right\vert_S \right) \, .
\end{equation}
Moreover, using \eqref{eq:Rsymmconst} we find that the R-symmetry flux gauge field is constrained by
\begin{equation}
\label{eq:BH_Rsymm_const_v0}
	\zeta_\Lambda^{(4)} \mf{p}^\Lambda = \frac{1}{\varepsilon_1 n_S - \varepsilon_2 n_N} \Big( \kappa_N (b_1^N - \chi_N b_2^N) - \kappa_S (b_1^S - \chi_S b_2^S) \Big) \, .
\end{equation}
To ensure that the spinor is globally defined on $\mathbb{WCP}^1_{[n_N,n_S]}$, we require $\kappa_N = - \chi_S\kappa_S$ (cf. \eqref{terrytheturnip}). Substituting the weights, \eqref{eq:BH_Rsymm_const_v0} then reads 
\begin{equation}
\label{eq:BH_multisec_Rsymm_const}
	\zeta_\Lambda^{(4)} \mf{p}^\Lambda = - \kappa_S \frac{n_N + \chi_N\chi_S n_S}{n_N n_S} \, .
\end{equation}
Now, substituting the FI parameters \eqref{eq:4d_FI_Parameters}, this can be re-written as:
\begin{equation}
\label{eq:BH_multisec_Rsymm_const_v2}
    -\frac{4}{\g} Q^{(\KKVec)} \mf{p}^0 + \zeta_I^{(4)} \check{\mf{p}}^I = - \kappa_S \frac{n_N + \chi_N\chi_S n_S}{n_N n_S} \,,
\end{equation}
where notice that both sides of this equation are invariant under the redundancy \eqref{eq:KK_Redundancy}. Utilising the freedom to choose the flat connection piece in the reduction ansatz \eqref{eq:Reduction_Ansatz3}, we now set $a^I = -\mf{p}^I/\mf{p}^0$, such that $\check{\mf{p}}^I = 0$. 
Next, substituting \eqref{eq:BH_vec_Phi00} into \eqref{eq:spindle_flux_BVAB}, we find
\begin{equation}
	\mf{p}^0 = - \frac{1}{2n_N n_S} \,,
\end{equation}
which is half the Chern number of the reduction fibration, as expected. Substituting in \eqref{eq:BH_multisec_Rsymm_const_v2} the flux $\mf{p}^0$, and $Q^{(\ell)}$ from \eqref{eq:spindle_BH_Qell}, we deduce (cf. \eqref{eq:BH_kappaNS})
\begin{align}
  \chi_N\chi_S = 1\,,\qquad
 \kappa_S = - \rsigma\,.
\end{align}
The first condition implies that supersymmetry is preserved via a topological twist.

Turning back to \eqref{eq:spindle_flux_BVAB}, from $\check{\mf{p}}^I = 0$ we have
\begin{equation}
	\left. \check{\Phi}_0^I \right\vert_N = \left. \check{\Phi}_0^I \right\vert_S \equiv \left. \check{\Phi}_0^I \right\vert_{\rm IR} \, .
\end{equation}
Therefore, the constraints on the fluxes allow one to rewrite the fixed point contribution to the on-shell action \eqref{eq:BH_vec_os_v1}. 
 For our choice of $\ell$ we get no contribution from
background subtraction so the full on-shell action
is then 
\begin{equation}
\label{eq:BH_vec_os_v2}
	\Itot_{(5), {\rm bs}} =I_{(4)}^{\rm{FP}}[M_{(4)}] = - \frac{\ii \pi^2}{2G_{(5)}} \frac{C_{IJK}}{6} \frac{\left. \check{\Phi}_0^I \check{\Phi}_0^J \check{\Phi}_0^K \right\vert_{\rm IR}}{\varepsilon_1 \varepsilon_2} \, .
\end{equation}

A field theory-based conjecture for the on-shell action of the complex deformations of multi-charge black holes was made in \cite{Hosseini:2018dob}, finding an entropy function $I$ given by
\begin{equation}
\label{eq:MultiCharge_QFT}
    I = \frac{\ii \pi^2}{2 \g ^3G_{(5)}} \frac{C_{IJK}}{6} \frac{\Delta_g^I \Delta_g^J \Delta_g^K}{\sigma_g \tau_g}\, ,
\end{equation}
with the variables explained below.
It was proved in \cite{Cassani:2019mms} that the Legendre transform of \eqref{eq:MultiCharge_QFT} leads to the entropy of the supersymmetric extremal black holes of \cite{Gutowski:2004yv,Kunduri:2006ek}, provided the variables are constrained by
\begin{equation}
\label{eq:BH_const_multisec}
	\sigma_g + \tau_g - \zeta_I \Delta^I_g = \JS \, ,
\end{equation}
where, recall $c_R=\pm 1$.
For a number of SCFTs with holographic duals, this expression for the entropy function \eqref{eq:MultiCharge_QFT} has been recovered by computing the Cardy-like limit of the large $N$ limit of the superconformal index (see \cite{Cassani:2025sim} for a recent review of the results).

We now compare our result \eqref{eq:BH_vec_os_v2} for the on-shell action of the complex deformations of multi-charge black holes with the field theory conjecture in \eqref{eq:MultiCharge_QFT}. 
Given that $\sigma_g,\tau_g$ are chemical potentials conjugate to angular momenta, we may identify
\begin{equation}
	\sigma_g \leftrightarrow - \rsigma \varepsilon_1 \,, \qquad \tau_g \leftrightarrow - \rsigma \varepsilon_2 \,,
\end{equation}
as we did in \eqref{eq:BH_angvel_id}. While the identification is relating bulk and boundary quantities, such an identification is natural: while $\varepsilon_{1,2}$ are associated with angular velocities of the black hole, they are also off-diagonal components of the boundary metric on $\partial M_{(5)} \cong S^1 \times S^3$.
It is then suggestive that we should also identify
\begin{equation}
\label{eq:UV-IR}
	\left. \g \check{\Phi}_0^I \right\vert_{\rm IR} = - \Delta^I_g \,,
\end{equation}
such that \eqref{eq:BH_vec_os_v2} agrees with \eqref{eq:MultiCharge_QFT}.  However, 
we should note that $\check\Phi_0^I \rvert_{\rm IR}$ are IR fixed point quantities, whereas $\Delta_g^I$ are associated with UV background holonomies of the boundary gauge fields. 
In the context of $AdS_4/CFT_3$, novel ``UV-IR relations" were constructed by considering submanifolds of topology $\R^2$, whose boundary ends on $\partial M_4 = M_3$ \cite{BenettiGenolini:2024xeo,BenettiGenolini:2024lbj}. This can be generalized to the present setting, and for constant holonomies this will lead to a precise agreement with the field theory conjecture. Black hole solutions may also exist when the holonomies are not constant and this leads to more intricate UV-IR relations, as discussed in \cite{BenettiGenolini:2024xeo,BenettiGenolini:2024lbj}.

At this point, it is interesting to highlight the role of the constraint \eqref{eq:4d_Constraint_v1}, which is a constraint on the IR fixed point quantities. Using the values of $Q^{(\ell)}$, $\Phi_0^0 \rvert_{N,S}$, and the weights ($b_1^{N,S},b_2^{N,S}$) computed above, the constraint \eqref{eq:4d_Constraint_v1} (at both poles) reads
\begin{equation}
	\left. \g \zeta_I \check{\Phi}_0^I \right\vert_{\rm IR} = \JS + \rsigma(\varepsilon_1 + \varepsilon_2) \,,
\end{equation}
with $\chi_N = \chi_S = \JS \rsigma$.
Under the assumption that the suggested identification \eqref{eq:UV-IR} is true, this is precisely equivalent to the field theory constraint \eqref{eq:BH_const_multisec}. 
It is remarkable that this constraint can be \emph{derived} from \eqref{eq:4d_Constraint_v1}, just assuming that the solutions exist.

\section{Final comments}\label{sec:fincomms}

We have presented a formalism which allows one to
compute the on-shell action for supersymmetric solutions of $D=5$ gauged supergravity coupled to $n$ vector multiplets using equivariant localization. In addition to the $D=5$ R-symmetry Killing vector $\cK$, the approach requires the existence of an extra Killing vector $\ell$, which we assume to be nowhere vanishing.
We use $\ell$ to carry out a dimensional reduction to a $D=4$ gauged supergravity coupled to
$n+1$ vector multiplets.
We then deploy the results of \cite{BenettiGenolini:2024xeo,BenettiGenolini:2024lbj} 
for computing the $D=4$ on-shell action using equivariant localization.
We showed how the formalism can be used to recover the on-shell action for the Euclidean 
$AdS_5$ vacuum solution with $S^1\times S^3$ boundary, recovering the known result for the
supersymmetric Casimir energy, and also for 
the complex locus of supersymmetric black hole solutions of \cite{Cabo-Bizet:2018ehj}, 
obtaining expressions for the supersymmetric index of the dual SCFT. 

For $D=4$ gauged supergravity, the on-shell action can be expressed entirely in terms of the fixed point data of the R-symmetry Killing vector \cite{BenettiGenolini:2023kxp,BenettiGenolini:2024xeo,BenettiGenolini:2024lbj}, which is a remarkable 
result. Specifically, this allows one to compute the on-shell action after just inputting some topological information regarding the R-symmetry Killing vector and the fixed point set.
By contrast, things are considerably more involved for $D=5$ gauged supergravity.
A key issue is that using the BVAB theorem in the $D=4$ KK reduced space $M_{(4)}$
gives rise to boundary terms on $\partial M_{(4)}$ (as in \eqref{eq:totac_fp_bdy}), as well as a 
contribution from integrating $\Lambda_4=\dd\Lambda_3$, with $\Lambda_3$ (as in \eqref{eq:Reduction_5d_Lagrangian})
not globally defined or gauge invariant, in general. 
Determining exactly when these terms contribute to the $D=5$ on-shell action for particular classes solutions, and how to carefully deal with $\Lambda_3$ globally, are important open issues that we hope to return to in future work. The origin of the gauge dependence of $\Lambda_3$ is associated
with the Chern--Simons term in the $D=5$ action, and so resolving this issue will certainly involve anomalies of the boundary theory. 
Similar issues will also arise in the approaches of \cite{Cassani:2024kjn,Colombo:2025ihp}.

For the example of the $AdS_5$ vacuum solution, using the supersymmetric boundary counterterms
of \cite{BenettiGenolini:2016qwm, BenettiGenolini:2016tsn} we have shown 
there exists a choice of KK reduction vector $\ell$ whereby
the computation of the $D=5$ action is entirely given by
contributions from the
fixed point set of the $D=4$ R-symmetry Killing vector on $M_{(4)}$.
However, for other choices of $\ell$ we have seen that there are additional
boundary contributions, and the formalism appears to be
less useful. We do not have a good understanding of what characterizes the special
choice of $\ell$ that leads to just fixed point contributions in this example.
More generally, given a class of solutions, it would be most desirable to know \emph{a priori}
whether or not such an $\ell$ exists.

For the black hole example, we used a background subtraction technique to regulate the $D=5$ action; although less satisfactory than holographic renormalization, this can be a helpful
pragmatic approach. After discussing some additional assumptions, we presented a concrete prescription for computing the $D=5$ on-shell action 
using localization in \eqref{bsfptformula}. For the black hole solution, \eqref{bsfptformula}
gave the correct result for the $D=5$ on-shell action to be identified with the supersymmetric index of the dual
 SCFT. Moreover, for this example, we showed that the result for the action
 was obtained from fixed point data (both in $M_{(4)}$ and $N_{(4)}$ in \eqref{bsfptformula}), for various choices of $\ell$, including reductions
on the Hopf fibre, reductions on the Hopf fibre twisted by the thermal circle, as well
as a spindle generalization of the Hopf fibre reduction for which $M_{(4)}$ has a different topology than the other two reductions. It would also be interesting to have a better understanding why in this example all of these different reductions just involved fixed point contributions.
More generally, given a class of solutions, it would be most desirable to know \emph{a priori}
when \eqref{bsfptformula} will compute the correct physical result.

With the above comments in mind, our formalism
can nonetheless be used to construct the on-shell action
of other classes of solutions with different topologies, either using holographic renormalization or background subtraction,
and we aim to report on this in the future.

The results we have obtained are relevant for arbitrary $D=5$ gauged supergravity theories coupled to vector multiplets. In a few cases, these theories arise from a consistent KK truncation of $D=10,11$ supergravity. For example minimal gauged supergravity can be uplifted on an arbitrary $SE_5$ manifold \cite{Buchel:2006gb}
 or an $M_6$ solution of \cite{Gauntlett:2004zh}, as shown in \cite{Gauntlett:2007ma}. In addition the STU model can be uplifted on $S^5$ to $D=10$ \cite{Cvetic:1999xp}. In such cases
our results for the $D=5$ on-shell action-immediately lead to
an associated exact result for the corresponding $D=10,11$ solutions. However, in the $D=4$ context it was argued that one can obtain exact top-down holographic results even if there is not a consistent KK truncation and some concrete examples were discussed. 
Without a consistent KK truncation one should be studying
the lower-dimensional gauged supergravity coupled to an infinite tower of KK multiplets and yet it seems that  
the latter are not relevant for computing
the on-shell action via equivariant localization, at least for some examples.
While we hope that a general proof of this can be obtained, it is natural to conjecture that the same is also true for computing the $D=5$ on-shell action. In particular we anticipate that our result
for the on-shell action for multiply charged black holes in \eqref{eq:BH_vec_os_v2} 
will be an exact result for infinite classes of SCFTs with holographic duals.

Finally, it would also be interesting to 
incorporate hypermultiplets into the formalism as well as higher derivatives. In particular, the latter would help to clarify 
the puzzle highlighted in\cite{Hristov:2025ygr}, where a $D=4$ fixed point result obtained via dimensional reduction was shown to disagree with a direct $D=5$ computation.

\section*{Acknowledgements}

\noindent 
We thank Davide Cassani, Seyed Morteza Hosseini and Sameer Murthy for helpful discussions.  
This work was supported in part by STFC grants ST/X000575/1
and ST/X000761/1, and SNSF Ambizione grant PZ00P2\_208666.
PBG was supported by a short term scientific mission grant from the COST action CA22113 THEORY-CHALLENGES.
JPG is supported as a Visiting Fellow at the Perimeter Institute. 
JP is supported by a Dean's PhD studentship at Imperial College.
JPG and JFS would like to thank the Centro de Ciencias de Benasque Pedro Pascual for hospitality while this work was being completed.

\appendix

\section{Conventions}\label{appconvs}

\subsection{\texorpdfstring{$D=5$}{D=5} Lorentzian}
\label{app:5d_LorentzianSpinors}

In $D=5$ with Lorentzian signature a basis of Cliff$(1,4)$ is given by ${\Gamma}_m$ satisfying 
\begin{align}
\label{eq:CliffordAlgebra_Lorentzian}
	{\Gamma}_m^\dagger &= {\Gamma}_0 {\Gamma}_m {\Gamma}_0 \, , \nn \\ 
	{\Gamma}_m^T &= {\cC} {\Gamma}_m {\cC}^{-1} \, , \, \, \quad \qquad {\cC}^T = - {\cC} \, , \nn \\
	 {\Gamma}_m^* &= - {\cB}_L {\Gamma}_m {\cB}_L^{-1} \, , \qquad {\cB}_L = \ii {\cC} {\Gamma}^0 \, ,
\end{align}
with ${\cB}_L^T = - {\cB}_L$.
The charge conjugate of a $D=5$ spinor $\lambda$ is defined by
\begin{equation}
\label{eq:cc_Lorentzian}
	\lambda^c_L \equiv {\cB}_L^{-1} \lambda^*= \ii {\Gamma}^0 {\cC}^{-1} \lambda^* \, ,
\end{equation}
and we have
\begin{equation}
		\overline{\lambda^c_L} \equiv (\lambda^c_L)^T\cC = - \ii \lambda^\dagger \Gamma^0  \, .
\end{equation}
On Cliff$(1,4)$ we take
\begin{equation}\label{prodgammas}
	{\Gamma}_{01234} = \ii\, ,
\end{equation}
so ${\Gamma}_{m_1\dots m_5}=\ii\epsilon_ {m_1\dots m_5}$ with $\epsilon_{01234}=+1$.

\subsection{\texorpdfstring{$D=5$}{D=5} Euclidean }
\label{app:5d_EuclideanSpinors}

In $D=5$ with Euclidean signature a basis of Cliff$(5)$ is given by ${\gamma}_m$ satisfying 
\begin{align}
	{\gamma}_m^\dagger &= {\gamma}_m \, ,\nn \\ 
	{\gamma}_m^T &= {\cC} {\gamma}_m {\cC}^{-1} \, , &\qquad {\cC}^T &= - {\cC} \, , \nn\\
	 {\gamma}_m^* &= {\cB}_E {\gamma}_m {\cB}_E^{-1} \, , &\qquad {\cB}_E &= - {\cC}  \, .
\end{align}
The charge conjugate spinor is defined by
\begin{equation}
	\lambda^c \equiv {\cB}_E^{-1} \lambda^* = - {\cC}^{-1} \lambda^* \, , \qquad \overline{\lambda^c}\equiv (\lambda^c)^T\cC = \lambda^\dagger \, .
\end{equation}

We obtain the $\gamma_m$ from the Lorentzian $\Gamma_m$ via
$\gamma_{1,2,3,4}=\Gamma_{1,2,3,4}$ and $\gamma_5=-\ii \Gamma_0$ and so
on Cliff$(5)$ we take
\begin{equation}\label{defgamsd5e}
	{\gamma}_{12345} = 1\,,
\end{equation}
and $\epsilon_ {m_1\dots m_5}=+1$.
Notice also that we can use the same $\cC$ in both the Lorentzian and Euclidean theory,
but $\cB_E\ne \cB_L$ and so one cannot identify $\lambda^c_L$ with $\lambda^c$.

\subsection{\texorpdfstring{$D=4$}{D=4} Euclidean}
\label{app:4d_EuclideanSpinors}

In $D=4$ with Euclidean signature a basis of Cliff$(4)$ is given by ${\gamma}_\mu$ satisfying 
\begin{align}
	{\gamma}_\mu^\dagger &= {\gamma}_\mu \, , \nn\\ 
	{\gamma}_\mu^T &= {\cC} {\gamma}_\mu {\cC}^{-1} \, , \qquad\qquad {\cC}^T = - {\cC} \, , \nn\\
	 {\gamma}_\mu^* &= {\cB}_E {\gamma}_\mu {\cB}_E^{-1} \, , \qquad\qquad {\cB}_E = - {\cC}  \, .
\end{align}
The charge conjugate spinor is defined by
\begin{equation}
	\lambda^c \equiv {\cB}_E^{-1} \lambda^* = - {\cC}^{-1} \lambda^* \, , \qquad \overline{\lambda^c} = \lambda^\dagger \, .
\end{equation}

We obtain the $D=4$ Euclidean $\gamma_\mu$ from the $D=5$ Euclidean 
$\gamma_m$ via
$\gamma_{1,2,3,4}=\gamma_{1,2,3,4}$ and notice from \eqref{defgamsd5e} we have
$\gamma_5= \gamma_1\gamma_2\gamma_3\gamma_4$ and so $\gamma_5$ is also the $D=4$ chirality matrix.
Notice also that we can use the same $\cC$ in both the $D=5$ and the $D=4$ 
Euclidean theories.

\section{Minimal gauged supergravity}
\label{app:minsugra}

Here we briefly discuss the special case of $D=5$ minimal gauged supergravity, connecting
with the conventions of \cite{Gauntlett:2003fk}. We focus on the $D=5$ Lorentzian theory.

The minimal case is when there are no vector multiplets, so $n=0$. The only
bosonic fields are the metric and a real gauge field $\cA^1$. The real scalar $Y^1$ is a constant.
Without loss of generality we can take
\begin{align}
Y^1=1\,,\qquad \zeta_1 = 3\,,\qquad \cA^1 \equiv  \frac{2}{\sqrt{3}} \cA\,.
\end{align}
The $D=5$ Lorentzian action \eqref{eq:Lorentzian5dAction} then takes the form
\begin{equation}
\label{eq:Lorentzian5dAction_Minimal_v2}
\begin{split}
	S_5 &= \frac{1}{16\pi G_{(5)}} \int_{M_{(5)}} \Big[ \left( R + 12 \g^2 -  \cF^2 \right) \vol_{(5)}  - \frac{8 }{3\sqrt{3}} \cF \wedge \cF \wedge \cA \Big] \, ,
\end{split}
\end{equation}
with $ \cF^2= \cF_{mn}\cF^{mn}$.
This theory admits an $AdS_5$ vacuum with radius $L=\g^{-1}$.
We have $W_5 = 3$ and
the Lorentzian Killing spinor equations \eqref{L5dKSE}, \eqref{L5dKSEc} become
\begin{align}
\label{eq:E5dKSEepsilon_Minimal}
      &0=\left[\nabla_m- \sqrt{3} \ii \g  {\cA}_m+\frac{1}{2} \g {\Gamma}_m +\frac{\ii}{4\sqrt{3}}  {\cF}_{np} ({\Gamma}_{m}{}^{np}-4\delta_m^n{\Gamma}^p)\right]\chi\,,\nn\\
            &0=\left[\nabla_m+ \sqrt{3} \ii \g  {\cA}_m-\frac{1}{2} \g {\Gamma}_m +\frac{\ii}{4\sqrt{3}}  {\cF}_{np} ({\Gamma}_{m}{}^{np}-4\delta_m^n{\Gamma}^p)\right]\chi^c\,.
\end{align}
To match \cite{Gauntlett:2003fk}, which had the opposite signature, choose
\begin{equation}
	\chi^{\rm there} = 2\sqrt{3}\g \, , \qquad (\gamma^m)^{\rm there} = (\ii {\Gamma}^m)^{\rm here} \,  , \qquad g_{mn}^{\rm there} = - g_{mn}^{\rm here} \, .
\end{equation}
Note that in \cite{Gauntlett:2003fk} they used $\varepsilon_{m_1\dots m_5} = (\gamma_{m_1\dots m_5})^{\rm there}$ with $\varepsilon_{01234}=+1$
and with $(\gamma_m)^{\rm there} = (-\ii {\Gamma}_m)^{\rm here}$ this is
in agreement with \eqref{prodgammas}.

\section{Bilinear relations in \texorpdfstring{$D=5$}{D=5}}
\label{app:5d_Bilinears}

In this appendix, we consider a supersymmetric solution of the $D=5$ Euclidean supergravity theory with bosonic action \eqref{eq:Euclidean5dAction}. Such a solution supports two non-vanishing Dirac spinors $\chi$ and $\tilde{\chi}$ satisfying \eqref{eq:E5dKSEepsilon} and \eqref{eq:E5dKSEepsilontilde}. We also define bilinears as in \eqref{dfiveksebiliniears}:
\begin{equation}
	\cS \equiv \overline{\tilde{\chi}} \chi \, , \qquad \SUSYVec^m \equiv \overline{\tilde{\chi}} {\gamma}^m \chi \, , \qquad \cU_{mn} \equiv \ii \overline{\tilde{\chi}} {\gamma}_{mn} \chi \, .
\end{equation}
Note that in absence of any choice of ``reality contour'' for the spinors, each of these bilinears is complex.
Using Fierz identities, one checks that
\begin{equation}
\label{eq:5d_Fierz}
	\cS^2 = \SUSYVec_m \SUSYVec^m  \, , \qquad 4\cS^2 = \cU_{mn}\cU^{mn} \, , \qquad \cS \cU = - \SUSYVec \hook *_{(5)} \cU \, .
\end{equation}

From the equations derived from the variations of the gravitino, we find that $\SUSYVec$ is a Killing vector, and that
\begin{align}
	\rd \cS &= - \ii \SUSYVec \hook ( Y_I \cF^I ) \, , \nn \\
\label{eq:5d_dK}
	\rd \SUSYVec^\flat &= - \frac{2\ii}{3}\g W_{(5)} \, \cU + \ii \SUSYVec \hook (*_{(5)} Y_I\cF^I) + 2\ii \cS \, Y_I \cF^I \, , \nn \\
\rd \cU&=0\,.
\end{align}
From the variations of the gaugino, we immediately see that $\cL_\SUSYVec \phi^i = 0$, and that
\begin{equation}
\label{eq:5d_dphi}
	\cS \, \rd \phi^i = \frac{3}{2} \ii g^{ij} \partial_j Y_I \, (\SUSYVec \hook \cF^I ) \, ,
\end{equation}
as well as
\begin{equation}
    \ii g_{ij}\mathrm{d}\phi^j\wedge\mathcal{K}^\flat=-\g\partial_iW_{(5)}\cU+\frac{3}{2}\partial_iY_I\Bigl(\cS\cF^I-*_{(5)}(\mathcal{K}^\flat\wedge \cF^I)\Bigr)\,.
\end{equation}

Combining these equations, we find that
\begin{align}
	\rd (Y^I \cS) &= Y^I \, \rd\cS + \partial_j Y^I \, \cS \, \rd\phi^j \nn \\
	&= - \ii \left( Y^I Y_J - \frac{3}{2} g^{ij} \partial_j Y^I \partial_j Y_J \right) (\SUSYVec \hook \cF^J) \, .
\end{align}
However, it follows from the definitions of the sections and metric reviewed in section~\ref{subsec:5d_Lorentzian} that
\begin{equation}
\label{eq:Useful_5d_Relations}
	\partial_i Y_I = - \frac{2}{3}G_{IJ} \partial_i Y^J \,, \qquad g^{ij} \partial_i Y^I \partial_j Y^J = G^{IJ} - \frac{2}{3} Y^I Y^J \, ,
\end{equation}
which allows us to conclude that
\begin{equation}
\label{eq:5d_KinF}
	\rd ( Y^I \cS) = - \ii \SUSYVec \hook \cF^I \, .
\end{equation}
From this, it follows that $\cL_{\SUSYVec}\cF^I=0$, so the vector $\SUSYVec$ generates a symmetry of the entire solution. Moreover, the following polyform is equivariantly closed
\begin{equation}
\label{eq:5d_PhiI}
	\Phi^I = \cF^I + \ii Y^I \cS \, ,
\end{equation}
satisfying $(\rd-\SUSYVec \hook\, )\Phi^I =0$.

Although not used in this paper, one can also construct an equivariantly closed extension of the 
$D=5$ on-shell action,
\begin{equation}
    \Phi=\Phi_5+\Phi_3+\Phi_1\,,
\end{equation}
with
\begin{align}
\label{eq:app_5d_Polyforms}
    \Phi_5 &= -\frac{2}{3}\g^2V_{(5)}\vol_{(5)} +\frac{2}{3}G_{IJ}*_{(5)}\cF^I\wedge\cF^J+\frac{\ii}{6}C_{IJK}\cF^I\wedge\cF^J\wedge \cA^K \, , \nn \\
    \Phi_3 &= \frac{2}{3}\ii\g W_{(5)}*_{(5)}\cU-\ii Y_I\cF^I\wedge \cK^\flat + \frac{2}{3} G_{IJ} \Phi^I_0 *_{(5)} \cF^J + \frac{\ii}{3}C_{IJK} \Phi_0^I \, \cF^J\wedge \cA^K \nn \\
    \Phi_1 &= \cS \left( \SUSYVec^\flat - \frac{2}{3} G_{IJ} \Phi^I_0 \cA^J \right) \, ,
\end{align}
{where we have introduced $\Phi^I_0 \equiv \ii Y^I \cS$ from \eqref{eq:5d_PhiI}, and to show equivariant closure with these representatives, we needed to choose the gauge
\begin{equation}
\label{eq:Gauge_Choice_Closure}
    \SUSYVec \hook \cA^I = - \Phi^I_0 \, ,
\end{equation}
which we remark is the same gauge that was chosen in the analysis of the minimal theory in \cite{Gauntlett:2003fk}.
Notice that, in general, the expressions \eqref{eq:app_5d_Polyforms} depend on the choice of gauge for $\cA^I$, but are otherwise globally defined.
We can also write an alternative expression for
$\Phi_3$ to emphasize the relation with Wald's formalism \cite{Wald:1993nt}, already highlighted in \cite{BenettiGenolini:2024lbj}. In fact, recalling that the 
$D=5$ gauge equations of motion represents the conservation of the electric current (associated to a Page charge)
\begin{equation}
    0 = \rd *_{(5)} \left[ G_{IJ} \cF^J + \frac{\ii}{4} *_{(5)} \left( C_{IJK} \cF^J \wedge \cA^K \right) \right] \equiv \rd *_{(5)} \cJ_I \, ,
\end{equation}
we can write 
\begin{equation}
\label{eq:Wald_Phi3_5d}
    \Phi_3 = - *_{(5)} \rd \SUSYVec^\flat + 2 \Phi^I_0 *_{(5)} \cJ_I - \frac{\ii}{6} C_{IJK}\Phi^I_0 \, \cF^J \wedge \cA^K \, .
\end{equation}}

Finally, we also comment on the relation with the equivariantly closed extension of the $D=5$ on-shell action appearing in \cite{Cassani:2024kjn} for ungauged supergravity (i.e. $\g=0$) and for non-supersymmetric solutions. First, note that we can rewrite the forms in \eqref{eq:app_5d_Polyforms} as
\begin{align}
    \Phi_5 &= \frac{2}{3} \left( - \g^2 V_{(5)} \, \vol_{(5)} + \cF^I \wedge *_{(5)} \cJ_I \right) \, , \nn \\
    \Phi_3 &= \frac{2}{3} \left( \ii \g W_{(5)} *_{(5)} \cU + \Phi^I_0 *_{(5)} \cJ_I + \nu_I \wedge \cF^I  \right) \, , \nn \\
    \Phi_1 &= \frac{2}{3} \Phi^I_0 \nu_I \, ,
\end{align}
with
\begin{align}
\label{eq:app_nuI}
    \nu_I &= - \frac{3\ii}{2} \left( Y_I \SUSYVec^\flat - \frac{1}{6} C_{IJK} \Phi^J_0 \cA^K \right) \nn \\
    &= \frac{\ii}{4} C_{IJK} \Phi_0^J \left( \frac{1}{\cS^2}  \Phi_0^K \, \cK^\flat + \cA^K \right)\, ,
\end{align}
where we have used \eqref{eq:5d_Relations_Useful_for_Forms}.
These correspond to (4.55) of \cite{Cassani:2024kjn}, provided we identify
\begin{equation}
\label{eq:app_Identification}
    G_I^{\rm there} = *_{(5)} \cJ_I \, , \qquad (- \xi \hook \cA^I + c^I)^{\rm there} = \Phi^I_0 \, ,
\end{equation}
and set $\g=0$.
The Killing vector $\xi^{\rm there}$
of \cite{Cassani:2024kjn}, should be identified with the $D=5$ R-symmetry Killing vector $\cK$. 
Here, for supersymmetric solutions, we have found \emph{canonical} expressions for the forms $\Phi_5$, $\Phi_3$, $\Phi_1$ appearing in \cite{Cassani:2024kjn} in terms of Killing spinor bilinears and supergravity fields. The authors prove that their expression for $\Phi$ is equivariantly closed provided the following conditions hold
\begin{equation}
    \rd \nu_I = (\xi \hook G_I)^{\rm there} \, , \qquad (\xi \hook \nu_I)^{\rm there} = 0 \, .
\end{equation}
We note that with the expression \eqref{eq:app_nuI} for $\nu_I$ and using $\xi^{\rm there} = \SUSYVec$, the first condition holds,
and the second one is implied by
 \eqref{eq:Gauge_Choice_Closure}. Conversely, from \eqref{eq:app_nuI} we have
 \begin{equation}
     Y^I ( \SUSYVec \hook \nu_I) = - G_{IJ} Y^I \cS \left( \Phi_0^J + \SUSYVec \hook \cA^J \right) \, ,
 \end{equation}
 using \eqref{eq:5d_Relations_Useful_for_Forms},
 which implies \eqref{eq:Gauge_Choice_Closure} provided we can take $Y^I \neq 0$.
Moreover, when combined with \eqref{eq:app_Identification}, this suggests that $(c^I)^{\rm there} = 0$ if we choose $(\xi)^{\rm there} = \SUSYVec$. We remark that this condition is different from the choice of constants $(c^I)^{\rm there}$ used in \cite{Cassani:2024kjn}, which was justified by the choice of setting to zero the boundary contribution to the action.

A solution of the Killing spinor equations transforms in a one-dimensional irreducible representation of the $U(1)$ symmetry generated by $\SUSYVec$, so we can take $\chi$ to transform with definite charge $\cQ$:
\begin{align}
\cL_{\SUSYVec} \chi &\equiv \SUSYVec^m \nabla_m \chi + \frac{1}{8} (\rd \SUSYVec^\flat)_{mn} \gamma^{mn} \chi \, ,\nn\\
 & =\ii \cQ \chi \,.
\end{align}
To compute the charge $\cQ$, we take the product with $\overline{\tilde{\chi}}$, use the Killing spinor equation \eqref{eq:E5dKSEepsilon}, and substitute \eqref{eq:5d_dK} and the relations \eqref{eq:5d_Fierz}, finding 
\begin{equation}\label{cQueueexpapp}
	\cQ = \frac{1}{2} \g \zeta_I \left( \SUSYVec \hook \cA^I +\Phi^I_0 \right) \, .
\end{equation}
It is straightforward to check that the charge of $\tilde{\chi}$ is $\tilde{\cQ} = - \cQ$. Notice that in the gauge chosen in \eqref{eq:Gauge_Choice_Closure}, we have $\cQ=0$. 
In the $D=4$ context, the analogous gauge condition that gives rise to the vanishing of the charge of the Killing spinor was called a ``supersymmetric gauge'' \cite{BenettiGenolini:2024lbj},
which is generically a weaker requirement than \eqref{eq:Gauge_Choice_Closure}.

\section{\texorpdfstring{$D=4$}{D=4} Euclidean supergravity}
\label{sec:usefulformulaed4}

In this appendix, we review some useful formulae for $D=4$ Euclidean supergravity coupled to an arbitrary number of vector multiplets, as constructed in \cite{BenettiGenolini:2024lbj}.

We consider Euclidean  $D=4$, $\cN=2$ gauged supergravity coupled to $n+1$ Abelian vector multiplets, with $n+2$ Abelian gauge fields $A^\Lambda$, $\Lambda = 0, \dots, n+1$, and $2(n+1)$ complex scalar fields $z^I$ and $\tz^I$, $I=1, \dots, n+1$.\footnote{Note that we use the labelling of the main text, which is slightly different from that used in \cite{BenettiGenolini:2024lbj}.} On the manifold parametrized by the scalars we define a (Euclidean) K\"ahler potential $\cK(z, \tz)$, $2(n+2)$ sections $X^\Lambda (z)$, $\tilde{X}^\Lambda(\tz)$, and we assume the existence of a prepotential $\cF=\cF(X)$ with Euclidean counterpart $\tilde{\cF}(\tilde{X})$. They are related by the constraint
\begin{equation}
\label{eq:Symplectic_Constraint_app}
	\e^{\cK} \left( X^\Lambda \frac{\partial \widetilde{\cF}}{\partial \tilde{X}^\Lambda} - \frac{\partial \cF}{\partial X^\Lambda} \tilde{X}^\Lambda \right) = \ii \, .
\end{equation}
From the prepotential, we can construct the matrices representing the kinetic and $\theta$ terms for the gauge fields as follows
\begin{align}
	N_{\Lambda\Sigma} &\equiv - \ii \left( \frac{\partial^2 \cF}{\partial X^\Lambda \partial X^\Sigma} - \frac{\partial^2 \tilde{\cF}}{\partial \tilde{X}^\Lambda \partial \tilde{X}^\Sigma} \right) \, , \nn \\
	\cN_{\Lambda\Sigma} &= \frac{\partial^2 \tilde{\cF}}{\partial \tilde{X}^\Lambda \partial \tilde{X}^\Sigma} + \ii \frac{ N_{\Lambda P} X^P N_{\Sigma T}X^T }{N_{\Xi \Pi} X^\Xi X^\Pi} \, , \nn \\
	\tilde{\cN}_{\Lambda\Sigma} &= \frac{\partial^2 \cF}{\partial {X}^\Lambda \partial {X}^\Sigma} - \ii \frac{ N_{\Lambda P} \tilde{X}^P N_{\Sigma T} \tilde{X}^T }{N_{\Xi \Pi} \tilde{X}^\Xi \tilde{X}^\Pi} \, , \nn \\
	\cR_{\Lambda\Sigma} &\equiv \frac{1}{2} \left( \cN_{\Lambda\Sigma} + \tilde{\cN}_{\Lambda\Sigma} \right) \, , \qquad \cI_{\Lambda\Sigma} \equiv \frac{1}{2\ii} \left( \cN_{\Lambda\Sigma} - \tilde{\cN}_{\Lambda\Sigma} \right) \, .
\end{align}
Finally, we need the $n+2$ Fayet--Iliopoulos parameters for the gauging, $\zeta_\Lambda^{(4)}$, which we assume to be real, and enter the action in the scalar potential
\begin{align}
\label{eq:EuclideanScalarPotential_app}
	\cV_{(4)} &= \zeta^{(4)}_\Lambda \zeta^{(4)}_\Sigma \e^{\cK} \left( \cG^{I \tilde{J}} \nabla_I X^\Lambda \nabla_{\tilde{J}} \tilde{X}^\Sigma - 3 X^\Lambda \tilde{X}^\Sigma \right) \, ,
\end{align} 
where $\cG_{I\tilde{J}}\equiv \partial_I\partial_{\tilde J}\cK$, and
\begin{equation}
\label{eq:Cov_Derivates_Sections_app}
	\nabla_I X^\Lambda = \partial_I X^\Lambda + \partial_I \cK \, X^\Lambda \, , \qquad \nabla_{\tilde{I}} \tilde{X}^\Lambda = \partial_{\tilde{I}} \tilde{X}^\Lambda + \partial_{\tilde{I}} \cK \, \tilde{X}^\Lambda \, .
\end{equation}
The bosonic part of the action is written in terms of these quantities:
\begin{align}
\label{themainaction_app}
	I_{(4)}	&= - \frac{1}{16\pi G_{(4)}} \int \Big[ \left( R - 2 \cG_{I\tilde{J}} \partial^\mu z^I \partial_\mu \ztilde^{\tilde{J}} - \g^2\cV_{(4)} (z, \ztilde)  \right)\vol_4 \nn\\
	& \qquad \qquad \qquad \qquad + \frac{1}{2} {\cI}_{\Lambda \Sigma} * F^\Lambda \wedge F^{\Sigma} - \frac{\ii}{2} {\cR}_{\Lambda \Sigma} F^\Lambda \wedge F^\Sigma \Big] \, .
\end{align}

A solution to the equations of motion is supersymmetric if there exist two Dirac spinors $\epsilon$ and $\tepsilon$ satisfying the following equations
\begin{align}
\label{eq:Euclidean_KSE_epsilon_app}
	0 &= \nabla_\mu \epsilon + \frac{\ii}{2} \cA_\mu \gamma_5 \epsilon 
	-  \frac{\ii}{2} \g A_\mu^R  \epsilon +  \frac{1}{2\sqrt{2}} \g \gamma_\mu  \e^{\cK/2}\left( W \tinyspace \Pold_- + \tW\tinyspace \Pold_+ \right)\epsilon \nn \\
	& \ \ \  - \frac{\ii}{4\sqrt{2}} \mathcal{I}_{\Lambda\Sigma} F^{\Sigma}_{\nu\rho} \gamma^{\nu\rho}\gamma_\mu \left( L^\Lambda \tinyspace \Pold_- + \tL^\Lambda \tinyspace \Pold_+ \right) \epsilon \, ,\nn \\
	0 &= \frac{\ii}{2\sqrt{2}} \mathcal{I}_{\Lambda\Sigma} F^{\Sigma}_{\nu\rho} \gamma^{\nu\rho} 
	\left( \cG^{\tilde{I} J} \nabla_J L^\Lambda \tinyspace \Pold_- + \cG^{I\tilde{J}} \nabla_{\tilde{J}} \tL^\Lambda \tinyspace \Pold_+ \right) \epsilon + \gamma^\mu \left( \partial_\mu z^I \tinyspace \Pold_- + \partial_\mu \tz^{\tilde{I}} \tinyspace \Pold_+ \right) \epsilon \nn \\
	& \ \ \  - \frac{1}{\sqrt{2}} \g {\e^{\cK/2}} \left( \cG^{\tilde{I} J} \nabla_J W \tinyspace \Pold_- +  \cG^{I\tilde{J}} \nabla_{\tilde{J}} \tW \tinyspace \Pold_+ \right) \epsilon \, ,
\end{align}
and 
\begin{align}
\label{eq:Euclidean_KSE_tepsilon_app}
	0 &= \nabla_\mu \tepsilon + \frac{\ii}{2} \cA_\mu \gamma_5 \tepsilon 
	+  \frac{\ii}{2} \g A_\mu^R  \tepsilon + \frac{1}{2\sqrt{2}} \g \gamma_\mu \e^{\cK/2} \left( W \tinyspace \Pold_- + \tW\tinyspace \Pold_+ \right) \tepsilon \nn \\
	& \ \ \ + \frac{\ii}{4\sqrt{2}} \mathcal{I}_{\Lambda\Sigma} F^{\Sigma}_{\nu\rho} \gamma^{\nu\rho}\gamma_\mu \left( L^\Lambda \tinyspace \Pold_- + \tL^\Lambda \tinyspace \Pold_+ \right) \tepsilon \, ,\nn \\
	0 &= - \frac{\ii}{2\sqrt{2}} \mathcal{I}_{\Lambda\Sigma} F^{\Sigma}_{\nu\rho} \gamma^{\nu\rho} 
	\left( \cG^{\tilde{I} J} \nabla_J L^\Lambda \tinyspace \Pold_- + \cG^{I\tilde{J}}  \nabla_{\tilde J} \tL^\Lambda \tinyspace \Pold_+ \right) \tepsilon + \gamma^\mu \left( \partial_\mu z^I \tinyspace \Pold_- + \partial_\mu \tz^{\tilde{I}} \tinyspace \Pold_+ \right) \tepsilon \nn \\
	& \ \ \  -  \frac{1}{\sqrt{2}} \g {\e^{\cK/2}} \left( \cG^{\tilde{I} J} \nabla_J W \tinyspace \Pold_- + \cG^{I\tilde{J}} \nabla_{\tilde J} \tW \tinyspace \Pold_+ \right) \tepsilon \, .
\end{align}
Here we have used the same Euclidean conventions as in section \ref{app:4d_EuclideanSpinors}, and introduced the following objects
\begin{align}
\label{eq:Objects_app}
	\cA_\mu &\equiv - \frac{\ii}{2} \left( \partial_I \cK \, \partial_\mu z^I - \partial_{\tilde{I}} \cK \, \partial_\mu \tilde{z}^{\tilde{I}} \right) \, , \qquad A^R_\mu \equiv \frac{1}{2} \zeta_\Lambda^{(4)} A^\Lambda_\mu \, , \qquad \Pold_\pm \equiv \frac{1}{2} ( 1 \pm \gamma_5) \, , \nn \\
	W &\equiv \zeta^{(4)}_\Lambda X^\Lambda \, , \qquad \tW \equiv \zeta^{(4)}_\Lambda \tX^\Lambda \, , \qquad L^\Lambda \equiv \e^{\cK/2}X^\Lambda \, , \qquad \tL^\Lambda \equiv \e^{\cK/2} \tX^\Lambda \, , \nn \\
	\nabla_I L^\Lambda &\equiv \partial_I L^\Lambda + \frac{1}{2} \partial_I \cK \, L^\Lambda \, , \qquad \nabla_{\tilde{I}} \tL^\Lambda \equiv \partial_{\tilde{I}} \tL^\Lambda - \frac{1}{2} \partial_{\tilde{I}} \cK \, \tL^\Lambda \, .
\end{align}

Using these spinors, we construct the following $D=4$ bilinears
\begin{equation}
\label{eq:Bilinears_app}
	S \equiv \overline{\tepsilon} \epsilon \, , \quad P \equiv \overline{\tepsilon} \gamma_5 \epsilon \, , \quad 
	\xi^\flat \equiv - \ii \overline{\tepsilon} \gamma_{(1)} \gamma_5 \epsilon \, , \quad 
	K \equiv \overline{\tepsilon} \gamma_{(1)} \epsilon \, , \quad 
		U\equiv \ii \overline{\tepsilon} \gamma_{(2)} \epsilon \, .
\end{equation}
One then shows \cite{BenettiGenolini:2024lbj}, using the Killing spinor equations above, that $\xi$ generates a symmetry of the solution, and that each gauge field strength has an equivariant completion with respect to the action of $\xi$
\begin{equation}
\label{eq:Equivariant_PhiF_app}
	\Phi^\Lambda_{(F)} \equiv F^\Lambda + \sqrt{2} \left( C^\Lambda - \tilde{C}^\Lambda \right) \, ,
\end{equation}
where
\begin{align}
\label{Cdef_app}
	C^\Lambda \equiv L^\Lambda (S-P)\, , \qquad \widetilde{C}^\Lambda \equiv \tL^\Lambda (S+P)\, .
\end{align}
Moreover, the on-shell action is also the top-form of an equivariantly closed form:
\begin{equation}\label{phiecform_app}
    \Phi=\Phi_4+\Phi_2+\Phi_0\, ,
\end{equation}
where 
\begin{align}\label{Phidef_app}
\Phi_4&=-\frac{1}{2} \g^2 \mathcal{V} \smallspace \vol_{4}-\frac{1}{4} \mathcal{I}_{\Lambda \Sigma} *F^\Lambda \wedge F^\Sigma + \frac{\ii}{4} \mathcal{R}_{\Lambda\Sigma} F^\Lambda \wedge F^\Sigma \, ,\nn \\
\Phi_2
&=\frac{1}{\sqrt{2}} \g \xinew_\Lambda (L^\Lambda \tinyspace U_{[+]} + \tL^\Lambda \tinyspace U_{[-]} )- \frac{1}{\sqrt{2}}\mathcal{I}_{\Lambda\Sigma} \Big(C^\Lambda \tinyspace F^{\Sigma}_{[+]}+\tilde{C}^\Lambda \tinyspace F^{\Sigma}_{[-]}\big) + \frac{\ii}{\sqrt{2}}\mathcal{R}_{\Lambda\Sigma} F^\Sigma \big(C^\Lambda - \tilde{C}^\Lambda \big)\, ,\nn \\
\Phi_0 &= \ii \tinyspace \ex^{\mathcal{K}} \Big[{\cF}(X )(S-P)^2+\tilde{\cF}(\tX)(S+P)^2 \nonumber\\ 
& \qquad\quad - \frac{1}{2} (\partial_\Lambda \mathcal{F}(X)\tX^\Lambda +\partial_\Lambda \tilde{\cF}(\tX)X^\Lambda )(S^2-P^2)\Big]\,.
\end{align}
As already noticed in \cite{BenettiGenolini:2024lbj}, $\Phi_2$ can be expressed in a form that highlights the relation with Wald's formalism \cite{Wald:1993nt}, and is analogous to \eqref{eq:Wald_Phi3_5d}, namely
\begin{equation}
    \Phi_2 = - \frac{1}{2} *_{(4)} \rd \FourdSUSYVec^\flat - \frac{1}{2} \Phi_0^\Lambda \cJ_\Lambda \, ,
\end{equation}
where here $\cJ_\Lambda$ is the conserved electric current associated to the gauge equations of motion coming from \eqref{themainaction_app}:
\begin{equation}
    0 = \rd *_{(4)} \left( \cI_{\Lambda\Sigma}F^\Sigma - \ii \cR_{\Lambda\Sigma}*_{(4)} F^\Sigma \right) \equiv \rd *_{(4)} \cJ_\Lambda \, .
\end{equation}

In general, we have $\xi_\mu \xi^\mu = S^2 - P^2$. Thus,  on the fixed point set, where $\xi^\mu=0$, we have $P=\pm S$, and the Killing spinor is chiral. By definition, we find that on the fixed point set labelled by the $\pm$ chirality of the $D=4$ spinor
\begin{equation}
	\Phi^\Lambda_0 \rvert_+ = - 2\sqrt{2} S \tilde{L}^\Lambda \rvert_+ \, , \qquad \Phi^\Lambda_0 \rvert_- = 2\sqrt{2} S {L}^\Lambda \rvert_- \, .
\end{equation}
At an isolated fixed point, we can (skew-)diagonalize the Killing vector action and introduce the weights on the two orthogonal planes $(b_1, b_2)$. We find that at an isolated fixed point where the chirality is $\chi$
\begin{equation}
	b_1 - \chi b_2 = - \frac{\kappa}{2} \g \zeta^{(4)}_\Lambda \Phi^\Lambda_0 \, , 
\end{equation}
where $\kappa$ is a sign associated to the point, which can be fixed globally in a systematic way \cite{BenettiGenolini:2024hyd}. We then introduce the following combinations, evaluated at the fixed set
\begin{equation}
\label{eq:defn_u_pm_app}
	u_+^\Lambda = \left. \frac{\tilde{L}^\Lambda}{\zeta_\Sigma^{(4)} \tilde{L}^\Sigma} \right\vert_+ \,,\qquad
	u_-^\Lambda = \left. \frac{L^\Lambda}{\zeta_\Sigma^{(4)}L^\Sigma} \right\vert_- \, .
\end{equation}
It is then straightforward to check that at an isolated fixed point with chirality $\chi$
\begin{equation}
\label{eq:u_Phi0_nut_app}
	(b_1 - \chi b_2) u^\Lambda_\chi = - \frac{\kappa}{2} \g \Phi^\Lambda_0 \, .
\end{equation}
Similarly, at a fixed surface where the chirality is $\chi$, we write $b$ for the weight of the action in the normal plane, and we can write
\begin{equation}
	b = \frac{\kappa}{2} \chi\g \zeta_\Lambda^{(4)} \Phi^\Lambda_0 \, ,
\end{equation}
from which it follows that
\begin{equation}
\label{eq:u_Phi0_bolt_app}
	b u^\Lambda_\chi = \frac{\kappa}{2} \chi \g \Phi^\Lambda_0 \, .
\end{equation}

\section{Reduction of the \texorpdfstring{$D=5$}{D=5} Killing spinor equations}\label{appdimredkse}

As discussed in the text, in the orthonormal frame given in \eqref{Dequalsfiveframe} the $D=5$ Killing spinors can be taken to satisfy
\begin{equation}
	\cL_{\KKVec} {\chi} = \partial_{x_5}\chi= \ii Q^{(\ell)} \chi \,,\qquad
		\cL_{\KKVec} \tilde{\chi}= \partial_{x_5}\tilde \chi = \ii \tilde{Q}^{(\ell)} \tilde{\chi} \, .
\end{equation}
Using the dimensional reduction formulae of section \ref{sec:dimred}, we find that
along the base directions,
the gravitino variations for $\chi$ and $\tilde{\chi}$ in \eqref{eq:E5dKSEepsilon}, \eqref{eq:E5dKSEepsilontilde} reduce to
\begin{align}
\label{eq:Reduced_KSE_base}
	0 &= \Bigg[ \nabla_\mu - A^0_\mu \left( - \ii Q^{(\ell)} + \frac{\ii}{2} \g \zeta_I \nvxi^I \right) - \frac{\ii}{2} \g \zeta_I A^I_\mu - \frac{1}{2} \partial_\rho \lambda \gamma^\rho_{\ph{\rho}\mu} + \frac{1}{6} \g \e^{\lambda} W_{(5)} \gamma_\mu \nn\\
	& \qquad \ \ \  - \frac{1}{4} \e^{- 3 \lambda} F^0_{\mu\rho} \gamma^\rho \gamma_5 - \frac{\ii}{6} G_{IJ} z^J_2 \e^{4\lambda} \partial_\nu z^I_1 \left( \gamma_\mu^{\ph{\mu}\nu} - 2 \delta^\nu_\mu \right) \gamma_5  \nn \\
	& \qquad - \frac{\ii}{12} \e^{\lambda} G_{IJ} z^J_2 ( F^I_{\nu\rho} - z^I_1 F^0_{\nu\rho} )( \gamma^{\nu\rho}\gamma_\mu - 2 \delta_\mu^\nu \gamma^\rho ) \Bigg] \chi \, , \nn \\
	0 &= \Bigg[ \nabla_\mu {+} A^0_\mu \left( \ii \tilde Q^{(\ell)}+ \frac{\ii}{2} \g \zeta_I \nvxi^I \right) + \frac{\ii}{2} \g \zeta_I A^I_\mu - \frac{1}{2} \partial_\rho \lambda \gamma^\rho_{\ph{\rho}\mu} -\frac{1}{6} \g \e^{\lambda} W_{(5)} \gamma_\mu \nn \\
	& \qquad \ \ \  - \frac{1}{4} \e^{- 3 \lambda} F^0_{\mu\rho} \gamma^\rho \gamma_5 - \frac{\ii}{6} G_{IJ} z^J_2 \e^{4\lambda} \partial_\nu z^I_1 \left( \gamma_\mu^{\ph{\mu}\nu} - 2 \delta^\nu_\mu \right) \gamma_5 \nn  \\
	& \qquad - \frac{\ii}{12} \e^{\lambda} G_{IJ} z^J_2 ( F^I_{\nu\rho} - z^I_1 F^0_{\nu\rho} )( \gamma^{\nu\rho}\gamma_\mu - 2 \delta_\mu^\nu \gamma^\rho ) \Bigg] \tilde{\chi} \, .
\end{align}
Comparing with the Euclidean supersymmetry variations in $D=4$
(cf. \eqref{eq:Euclidean_KSE_epsilon_app}-\eqref{eq:Euclidean_KSE_tepsilon_app}),
where the FI parameters enter the gravitino variations as 
\begin{equation}
	\nabla_\mu \epsilon - \frac{\ii}{4} \g \zeta^{(4)}_\Lambda A_\mu^\Lambda \epsilon + \dots \, , \qquad \nabla_\mu \tilde{\epsilon} +  \frac{\ii}{4} \g \zeta^{(4)}_\Lambda A_\mu^\Lambda \tilde{\epsilon} + \dots \,,
\end{equation}
we identify the $D=4$ FI parameters as
\begin{equation}
\begin{split}
	 \g \zeta^{(4)}_0 & = -4Q^{(\ell)} + 2\g \zeta_I \nvxi^I \\
	& = 4 \tilde{Q}^{(\ell)} + 2\g \zeta_I \nvxi^I  \, ,  \\
	\g \zeta_I^{(4)} & = 2\g \zeta_I \, .
\end{split}
\end{equation}
To have a consistent reduction, leading to $D=4$ Killing spinor equations,
we thus require
\begin{equation}
	Q^{(\ell)} = - \tilde{Q}^{(\ell)} \, ,
\end{equation}
as discussed in the main text.

After some manipulations, the reduced gravitino variations for $\chi$ and $\tilde{\chi}$ along the $x^5$ fibre direction
can be written as
\begin{align}
\label{eq:Reduced_KSE_fibre}
	0
	&= \Bigg[ \frac{1}{4} \e^{ - 4\lambda} Y_J \left( F^J_{\nu\rho} - z^J_1 \, F^0_{\nu\rho} \right) \gamma^{\nu\rho}\gamma_5 - \frac{\ii}{4} \e^{- 6\lambda} F^0_{\nu\rho} \gamma^{\nu\rho} \nn \\
	& \qquad  + \e^{ - \lambda} \gamma^\mu \left(  Y_J \partial_\mu z^J_1 - 2 \ii e^{-2\lambda} \gamma_5 \partial_\mu \lambda \right) - \frac{1}{2} \g \zeta^{(4)}_0 - \frac{1}{2}\g \zeta_J^{(4)} z_1^J + \frac{\ii}{6} \g  \zeta_J^{(4)} z_2^J \gamma_5 \Bigg] \chi \,,\nn \\
	0
	&= \Bigg[  \frac{1}{4} \e^{ - 4\lambda} Y_J \left( F^J_{\nu\rho} - z^J_1 \, F^0_{\nu\rho} \right) \gamma^{\nu\rho}\gamma_5 -  \frac{\ii}{4} \e^{- 6\lambda} F^0_{\nu\rho} \gamma^{\nu\rho} \nn \\
	& \qquad + \e^{ - \lambda} \gamma^\mu \left(  Y_J \partial_\mu z^J_1 - 2 \ii e^{-2\lambda} \gamma_5 \partial_\mu \lambda \right)  + \frac{1}{2} \g \zeta^{(4)}_0 + \frac{1}{2} \g \zeta_J^{(4)} z^J_1 - \frac{\ii}{6} \g \zeta_J^{(4)} z_2^J \gamma_5 \Bigg] \tilde{\chi} \,.
\end{align}
To reduce the gaugino variations for $\chi$ and $\tilde{\chi}$, we use the identity for $G^{IJ}$ given in \eqref{eq:Useful_5d_Relations}, noting that $\delta^I_J - Y^I Y_J$ is a projector on the very special real manifold. This allows us to rewrite the gaugino equations in 
\eqref{eq:E5dKSEepsilon}, \eqref{eq:E5dKSEepsilontilde} as
\begin{align}
	0 & = \Bigg[ - \frac{\ii}{2} \partial_m Y^I \gamma^m + \frac{\ii}{2} \g \zeta_J (G^{IJ} - \frac{2}{3} Y^I Y^J) - \frac{1}{4} (\delta^I_J - Y^I Y_J) \cF^J_{mn} \gamma^{mn} \Bigg] \chi \, , \nn \\
	0 & = \Bigg[ - \frac{\ii}{2} \partial_m Y^I \gamma^m - \frac{\ii}{2} \g \zeta_J (G^{IJ} - \frac{2}{3} Y^I Y^J) - \frac{1}{4} (\delta^I_J - Y^I Y_J) \cF^J_{mn} \gamma^{mn} \Bigg] \tilde{\chi} \, .
\end{align}
Using the dimensional reduction ansatz, these reduce to
\begin{align}
\label{eq:Reduced_gaugino}
    0 & = \Bigg[ - \frac{1}{2} \e^{-3\lambda} \left( \delta^I_J - Y^I Y_J \right) (F^J_{\nu\rho}-z_1^J F^0_{\nu\rho}) \gamma^{\nu\rho} \gamma_5 \nn \\
    & \qquad + \gamma^\mu \Big( \partial_\mu ( z_1^I - \ii z_2^I \gamma_5)  - 2 \ii z_2^I \gamma_5 \partial_\mu \lambda - Y^IY_J \partial_\mu z_1^J \Big) \nn\\ 
& \qquad + \frac{\ii}{4} \e^{3\lambda} \mathfrak{g}_{(4)} \zeta_J^{(4)} \left(\mathcal{G}^{IJ} - \frac{4}{3} z_2^Iz_2^J \right) \gamma_5 \Bigg] \chi \, ,\nn\\
    0 & = \Bigg[ - \frac{1}{2} \e^{-3\lambda} \left( \delta^I_J - Y^I Y_J \right) (F^J_{\nu\rho}-z_1^J F^0_{\nu\rho}) \gamma^{\nu\rho} \gamma_5 \nn\\
    & \qquad + \gamma^\mu \Big( \partial_\mu ( z_1^I - \ii z_2^I \gamma_5)  - 2 \ii z_2^I \gamma_5 \partial_\mu \lambda - Y^IY_J \partial_\mu z_1^J \Big) \nn \\ 
& \qquad - \frac{\ii}{4} \e^{3\lambda} \mathfrak{g}_{(4)} \zeta_J^{(4)} \left(\mathcal{G}^{IJ} - \frac{4}{3} z_2^Iz_2^J \right) \gamma_5 \Bigg] \tilde{\chi} \,  .
\end{align}

Equations \eqref{eq:Reduced_KSE_fibre} and \eqref{eq:Reduced_gaugino} combine to give the $D=4$ gaugino variation. Conversely, the former is the $D=4$ equation contracted with $Y^I$, while the latter is that projected orthogonally to $Y^I$
(similar to \cite{Gutowski:2012yb}). To see this, it is useful to note the following identities that follow from \eqref{eq:Cov_Derivates_Sections_app}:
\begin{align}
\label{eq:holo_deriv}
	& \nabla_J X^0 = \frac{3\e^{2\lambda}}{2\ii} Y_J \, , \qquad \nabla_J X^I = \delta^I_J - \frac{3}{2} Y^I Y_J + \frac{3\e^{2\lambda}}{2\ii} Y_J z_1^I \, , \nn \\
	& \nabla_{\tilde{J}} \tilde{X}^0 = - \frac{3\e^{2\lambda}}{2\ii} Y_J \, , \qquad \nabla_{\tilde{J}} \tilde{X}^I =  \delta^I_J - \frac{3}{2} Y^I Y_J - \frac{3\e^{2\lambda}}{2\ii} Y_J z_1^I \, ,
\end{align}
as well as \eqref{eq:4d_ReducedMetric_ScalarManifold}, \eqref{thecalIs}.
Note we have made a choice of the phase in front of the sections here, by setting $X^0 = \tilde{X}^0 = 1$.
These results can be used to show 
\begin{align}
	& \frac{1}{2\sqrt{2}} \cI_{\Lambda \Sigma} F^\Sigma_{\nu\rho} \gamma^{\nu\rho} \, \e^{\cK/2} \left( \cG^{\tilde{I}J} \nabla_J X^\Lambda \mathbb{P}_- + \cG^{I\tilde{J}} \nabla_{\tilde{J}} \tilde{X}^\Lambda \mathbb{P}_+  \right) \nn \\
	&\qquad  \qquad \ \ \ =  - \frac{1}{2} \e^{-3\lambda} (\delta^I_J - Y^I Y_J) ( F^J_{\nu\rho} - z_1^J F^0_{\nu\rho} ) \gamma^{\nu\rho} \nn \\
	& \qquad \qquad \qquad +  \frac{1}{4} \e^{-3\lambda} Y^I Y_J ( F^J_{\nu\rho} - z_1^J F^0_{\nu\rho} ) \gamma^{\nu\rho} - \frac{\ii}{4} \e^{-5\lambda} Y^I F^0_{\nu\rho} \gamma^{\nu\rho} \gamma_5 \, .
\end{align}
Similarly,
\begin{align}
	& \frac{1}{\sqrt{2}} \g \e^{\cK/2} \left( \cG^{\tilde{I}J} \nabla_J W \, \mathbb{P}_- + \cG^{I\tilde{J}} \nabla_{\tilde{J}} \tilde{W} \, \mathbb{P}_+ \right) \nn\\
	= & \,\, \frac{1}{4} \g \e^{3\lambda} \zeta_J^{(4)} \left( \cG^{IJ} - \frac{4}{3} z_2^I z_2^J \right)
	+ \frac{\ii}{2} \g \e^{\lambda} Y^I \left( \zeta^{(4)}_0 + \zeta_J^{(4)} z_1^J - \frac{\ii}{3} \zeta_J^{(4)} z_2^J \gamma_5  \right) \gamma_5 \, .
\end{align}
It is also useful to write
\begin{align}
	\gamma^\mu \left( \partial_\mu z^I \mathbb{P}_- + \partial_\mu \tilde{z}^{\tilde{I}} \mathbb{P}_+ \right)
	& = \gamma^\mu \Big( \partial_\mu (z_1^I - \ii z_2^I \gamma_5) - 2 \ii z_2^I \gamma_5 \partial_\mu \lambda -  Y^I Y_J \partial_\mu z_1^J \Big) \nn \\
	& \quad + \gamma^\mu Y^I (Y_J \partial_\mu z_1^J - 2 \ii e^{-2\lambda} \gamma_5 \partial_\mu \lambda ) \, .
\end{align}
With the above identities, it is straightforward to see that \eqref{eq:Reduced_KSE_fibre} and \eqref{eq:Reduced_gaugino} combine to give
\begin{align}
\label{eq:Reduced_gaugino_4d_fin}
	0 & = \Bigg[ \frac{1}{2\sqrt{2}} \cI_{\Lambda \Sigma} F^\Sigma_{\nu\rho} \gamma^{\nu\rho} \gamma_5 \, \e^{\cK/2} \left( \cG^{\tilde{I}J} \nabla_J X^\Lambda \mathbb{P}_- + \cG^{I\tilde{J}} \nabla_{\tilde{J}} \tilde{X}^\Lambda \mathbb{P}_+  \right) \nn \\
	& + \gamma^\mu \left( \partial_\mu z^I \mathbb{P}_- + \partial_\mu \tilde{z}^{\tilde{I}} \mathbb{P}_+ \right) - \frac{\ii}{\sqrt{2}} \g \e^{\cK/2} \gamma_5 \left( \cG^{\tilde{I}J} \nabla_J W \, \mathbb{P}_- + \cG^{I\tilde{J}} \nabla_{\tilde{J}} \tilde{W} \, \mathbb{P}_+ \right) \Bigg] \chi \, , \nn \\
	0 & = \Bigg[ \frac{1}{2\sqrt{2}} \cI_{\Lambda \Sigma} F^\Sigma_{\nu\rho} \gamma^{\nu\rho} \gamma_5 \, \e^{\cK/2} \left( \cG^{\tilde{I}J} \nabla_J X^\Lambda \mathbb{P}_- + \cG^{I\tilde{J}} \nabla_{\tilde{J}} \tilde{X}^\Lambda \mathbb{P}_+  \right) \nn \\
	& + \gamma^\mu \left( \partial_\mu z^I \mathbb{P}_- + \partial_\mu \tilde{z}^{\tilde{I}} \mathbb{P}_+ \right) + \frac{\ii}{\sqrt{2}} \g \e^{\cK/2} \gamma_5 \left( \cG^{\tilde{I}J} \nabla_J W \, \mathbb{P}_- + \cG^{I\tilde{J}} \nabla_{\tilde{J}} \tilde{W} \, \mathbb{P}_+ \right) \Bigg] \tilde{\chi} \, .
\end{align}

Next, the reduced gravitino variations along the base directions \eqref{eq:Reduced_KSE_base} can be written as
\begin{align}
\label{eq:Reduced_KSE_base_4d}
	0 &= \Bigg[ \nabla_\mu - \frac{1}{2} \partial_\rho \lambda \gamma^\rho_{\ph{\rho}\mu} - \frac{\ii}{6} \cA_\nu \left( \gamma_\mu^{\ph{\mu}\nu} - 2 \delta^\nu_\mu \right) \gamma_5 - \frac{\ii}{2} \g A^R_\mu -\frac{1}{12} \g \e^{3\lambda} \zeta_I^{(4)} z_2^I \gamma_\mu \nn \\
	& \qquad \ \ \  - \frac{1}{4} \e^{- 3 \lambda} F^0_{\mu\rho} \gamma^\rho \gamma_5 - \frac{\ii}{6} \e^{-3\lambda} \cG_{IJ} z^J_2 ( F^I_{\nu\rho} - z^I_1 F^0_{\nu\rho} )( \gamma^{\nu\rho}\gamma_\mu - 2 \delta_\mu^\nu \gamma^\rho )\Bigg] \chi \, , \nn \\
	0 &= \Bigg[ \nabla_\mu - \frac{1}{2} \partial_\rho \lambda \gamma^\rho_{\ph{\rho}\mu} - \frac{\ii}{6} \cA_\nu \left( \gamma_\mu^{\ph{\mu}\nu} - 2 \delta^\nu_\mu \right) \gamma_5 + \frac{\ii}{2} \g A^R_\mu + \frac{1}{12} \g \e^{3\lambda} \zeta_I^{(4)} z_2^I \gamma_\mu \nn \\
	& \qquad \ \ \  - \frac{1}{4} \e^{- 3 \lambda} F^0_{\mu\rho} \gamma^\rho \gamma_5 - \frac{\ii}{6} \e^{-3\lambda} \cG_{IJ} z^J_2 ( F^I_{\nu\rho} - z^I_1 F^0_{\nu\rho} )( \gamma^{\nu\rho}\gamma_\mu - 2 \delta_\mu^\nu \gamma^\rho )\Bigg]\tilde{\chi} \, .
\end{align}
where we introduced the R-symmetry gauge field $A_\mu^R$ and the K\"ahler connection $\cA_\mu = 2  \cG_{IJ} z_2^J \partial_\mu z_1^I $, as defined in \eqref{eq:Objects_app}. To recover the $D=4$ gravitino variation in the conventions of \cite{BenettiGenolini:2024lbj}, it is useful to note that \eqref{eq:Reduced_KSE_fibre} can be written solely in terms of $D=4$ objects as
\begin{align}
\label{eq:Reduced_KSE_fibre_4d}
	0
	&= \Bigg[ \frac{1}{16} \e^{- 3\lambda} F^0_{\nu\rho} \gamma^{\nu\rho} \gamma_5 - \frac{\ii}{12} \e^{ - 3\lambda} \cG_{IJ} z_2^J \left( F^I_{\nu\rho} - z^I_1 \, F^0_{\nu\rho} \right) \gamma^{\nu\rho} \nn \\
	&  - \frac{1}{2} \gamma^\nu \partial_\nu \lambda + \frac{\ii}{6} \cA_\nu \gamma^\nu \gamma_5 - \frac{\ii}{8} \g \e^{3\lambda} \zeta^{(4)}_0 \gamma_5 - \frac{\ii}{8}\g \e^{3\lambda} \zeta_J^{(4)} z_1^J  \gamma_5 - \frac{1}{24} \g \e^{3\lambda}  \zeta_J^{(4)} z_2^J \Bigg] \chi \,, \nn \\
	0
	&= \Bigg[ \frac{1}{16} \e^{- 3\lambda} F^0_{\nu\rho} \gamma^{\nu\rho} \gamma_5 - \frac{\ii}{12} \e^{ - 3\lambda} \cG_{IJ} z_2^J \left( F^I_{\nu\rho} - z^I_1 \, F^0_{\nu\rho} \right) \gamma^{\nu\rho} \nn \\
	&  - \frac{1}{2} \gamma^\nu \partial_\nu \lambda + \frac{\ii}{6} \cA_\nu \gamma^\nu \gamma_5 + \frac{\ii}{8} \g \e^{3\lambda} \zeta^{(4)}_0 \gamma_5 + \frac{\ii}{8}\g \e^{3\lambda} \zeta_J^{(4)} z_1^J  \gamma_5 + \frac{1}{24} \g \e^{3\lambda}  \zeta_J^{(4)} z_2^J \Bigg]\tilde{\chi} \,.
\end{align}
Combining \eqref{eq:Reduced_KSE_base_4d} with $\gamma_\mu$ times \eqref{eq:Reduced_KSE_fibre_4d}, we get
\begin{align}
\label{eq:Reduced_gravitino_4d_fin}
	0 &= \Bigg[ \nabla_\mu - \frac{1}{2} \partial_\mu \lambda + \frac{\ii}{2} \cA_\mu \gamma_5 - \frac{\ii}{2} \g A^R_\mu + \frac{\ii}{2\sqrt{2}} \g \gamma_5 \gamma_\mu \e^{\cK/2} \left( W \mathbb{P}_- + \tilde{W} \mathbb{P}_+ \right) \nn \\
	& \qquad \ \ \ + \frac{1}{4\sqrt{2}} \cI_{\Lambda\Sigma} F^\Sigma_{\nu\rho} \gamma_5 \gamma^{\nu\rho} \gamma_\mu \e^{\cK/2} (X^\Lambda \mathbb{P}_- + \tilde{X}^\Lambda \mathbb{P}_+ ) \Bigg] \chi \, , \nn \\
	0 &= \Bigg[ \nabla_\mu - \frac{1}{2} \partial_\mu \lambda + \frac{\ii}{2} \cA_\mu \gamma_5 + \frac{\ii}{2} \g A^R_\mu - \frac{\ii}{2\sqrt{2}} \g \gamma_5 \gamma_\mu \e^{\cK/2} \left( W \mathbb{P}_- + \tilde{W} \mathbb{P}_+ \right) \nn \\
	& \qquad \ \ \ + \frac{1}{4\sqrt{2}} \cI_{\Lambda\Sigma} F^\Sigma_{\nu\rho} \gamma_5 \gamma^{\nu\rho} \gamma_\mu \e^{\cK/2} (X^\Lambda \mathbb{P}_- + \tilde{X}^\Lambda \mathbb{P}_+ ) \Bigg] \tilde{\chi} \, ,
\end{align}
where we made use of the following identities:
\begin{align}
	\frac{1}{2\sqrt{2}} \g \gamma_\mu \e^{\cK/2} \left( W \mathbb{P}_- + \tilde{W} \mathbb{P}_+ \right)  
	&= \frac{1}{8} \g \e^{3\lambda} \gamma_\mu \left( \zeta^{(4)}_0 + \zeta_I^{(4)} ( z_1^I - \ii z_2^I \gamma_5)  \right) \, , \nn\\
 \frac{1}{4\sqrt{2}} \cI_{\Lambda\Sigma} F^\Sigma_{\nu\rho} \gamma^{\nu\rho} \e^{\cK/2} (X^\Lambda \mathbb{P}_- + \tilde{X}^\Lambda \mathbb{P}_+ ) &= - \frac{1}{16} \e^{-3\lambda} F^0_{\nu\rho} \gamma^{\nu\rho}\nn \\
	& \quad  + \frac{\ii}{4} \e^{-3\lambda} \cG_{IJ} z_2^J ( F^I_{\nu\rho} - z_1^I F^0_{\nu\rho} ) \gamma^{\nu\rho} \gamma_5 \, .
\end{align}

If we now define $D=4$ spinors $\epsilon$, $\tilde\epsilon$ via 
\begin{equation}\label{epschirelsappb}
	\epsilon = {\ii} \e^{-\lambda/2} \e^{-\ii Q^{(\ell)}x^5} \e^{-\frac{\pi \ii}{4}\gamma_5} \chi \,,\qquad
		\tilde{\epsilon} = {\ii} \e^{-\lambda/2} \e^{-\ii \tilde{Q}^{(\ell)}x^5} \e^{\frac{\pi \ii}{4}\gamma_5} \tilde{\chi} \,,
\end{equation}
with $Q^{(\ell)} = - \tilde{Q}^{(\ell)}$, then \eqref{eq:Reduced_gaugino_4d_fin}, \eqref{eq:Reduced_gravitino_4d_fin} become the $D=4$ gaugino and gravitino variations in the conventions of \cite{BenettiGenolini:2024lbj}, i.e. we recover \eqref{eq:Euclidean_KSE_epsilon_app}--\eqref{eq:Euclidean_KSE_tepsilon_app}.

 We also record that in terms of the $D=4$ spinors $\epsilon$, $\tilde\epsilon$, the reduced $D=5$ gaugino variation \eqref{eq:Reduced_gaugino} reads:
 \begin{align}\label{appekse1}
    0 & = \Bigg[ - \frac{1}{2} \e^{-3\lambda} \left( \delta^I_J - \frac{4}{3} \cG_{JK} z_2^I z_2^K \right) (F^J_{\nu\rho}-z_1^J F^0_{\nu\rho}) \gamma^{\nu\rho} \gamma_5+ \frac{\ii}{4} \e^{3\lambda} \mathfrak{g}_{(4)} \zeta_J^{(4)} \left(\mathcal{G}^{IJ} - \frac{4}{3} z_2^Iz_2^J \right) \gamma_5 \nn \\
    & \qquad + \ii \gamma^\mu \gamma_5 \Big( \partial_\mu ( z_1^I - \ii z_2^I \gamma_5)  - 2 \ii z_2^I \gamma_5 \partial_\mu \lambda -  \frac{4}{3} \cG_{JK} z_2^I z_2^K \partial_\mu z_1^J \Big) \Bigg] \epsilon \, , \nn \\
    0 & = \Bigg[ - \frac{1}{2} \e^{-3\lambda} \left( \delta^I_J -  \frac{4}{3} \cG_{JK} z_2^I z_2^K \right) (F^J_{\nu\rho}-z_1^J F^0_{\nu\rho}) \gamma^{\nu\rho} \gamma_5 - \frac{\ii}{4} \e^{3\lambda} \mathfrak{g}_{(4)} \zeta_J^{(4)} \left(\mathcal{G}^{IJ} - \frac{4}{3} z_2^Iz_2^J \right) \gamma_5 \nn \\
    & \qquad - \ii \gamma^\mu \gamma_5 \Big( \partial_\mu ( z_1^I - \ii z_2^I \gamma_5)  - 2 \ii z_2^I \gamma_5 \partial_\mu \lambda -  \frac{4}{3} \cG_{JK} z_2^I z_2^K \partial_\mu z_1^J \Big) \Bigg] \tilde{\epsilon} \,  .
\end{align}
Similarly, the reduced $D=5$ gravitino variation along fibre \eqref{eq:Reduced_KSE_fibre} can be written
\begin{align}\label{appekse2}
	0
	&= \Bigg[ - \frac{1}{3} \e^{ - 6 \lambda} \cG_{IJ} z_2^I \left( F^J_{\nu\rho} - z^J_1 \, F^0_{\nu\rho} \right) \gamma^{\nu\rho}\gamma_5 - \frac{\ii}{4} \e^{- 6\lambda} F^0_{\nu\rho} \gamma^{\nu\rho}  - \frac{1}{2} \g \zeta_0^{(4)}\nn \\
	& \qquad  - 2 \ii \e^{ - 3 \lambda} \gamma^\mu \left( \frac{2}{3} \cG_{IJ} z_2^I \partial_\mu z^J_1 \gamma_5 + \ii \partial_\mu \lambda \right)  - \frac{1}{2}\g \zeta_J^{(4)} z_1^J + \frac{\ii}{6} \g  \zeta_J^{(4)} z_2^J \gamma_5 \Bigg] \epsilon \,, \nn\\
	0
	&= \Bigg[ - \frac{1}{3} \e^{ - 6 \lambda} \cG_{IJ} z_2^I \left( F^J_{\nu\rho} - z^J_1 \, F^0_{\nu\rho} \right) \gamma^{\nu\rho}\gamma_5 -  \frac{\ii}{4} \e^{- 6\lambda} F^0_{\nu\rho} \gamma^{\nu\rho} + \frac{1}{2} \g \zeta_0^{(4)}         \nn  \\
	& \qquad + 2 \ii \e^{ - 3 \lambda} \gamma^\mu \left( \frac{2}{3} \cG_{IJ} z_2^I \partial_\mu z^J_1 \gamma_5 + \ii \partial_\mu \lambda \right) + \frac{1}{2} \g \zeta_J^{(4)} z^J_1 - \frac{\ii}{6} \g \zeta_J^{(4)} z_2^J \gamma_5 \Bigg] \tilde{\epsilon} \,.
\end{align}
Finally, the reduced gravitino variation along base \eqref{eq:Reduced_KSE_base_4d} combined with 
$D=5$ gravitino variation along fibre \eqref{eq:Reduced_KSE_fibre} gives rise to the 
$D=4$ gravitino variation written in terms of $\epsilon$, $\tilde\epsilon$:
\begin{align}\label{appekse3}
	0 &= \Bigg[ \nabla_\mu + \ii  \cG_{IJ} z_2^J \partial_\mu z_1^I \gamma_5 - \frac{\ii}{4} \g \zeta_\Lambda^{(4)} A^\Lambda_\mu + \frac{1}{8} \g \e^{3\lambda} \gamma_\mu \left( \zeta_0^{(4)} + \zeta_I^{(4)} ( z_1^I - \ii z_2^I \gamma_5)  \right) \nn\\
	& \qquad \ \ \ +  \frac{\ii}{16} \e^{- 3\lambda} F^0_{\nu\rho} \gamma^{\nu\rho} \gamma_\mu + \frac{1}{4} \e^{ - 3\lambda} \cG_{IJ} z_2^J \left( F^I_{\nu\rho} - z^I_1 \, F^0_{\nu\rho} \right) \gamma^{\nu\rho} \gamma_\mu \gamma_5 \Bigg] \epsilon \, , \nn\\
	0 &= \Bigg[ \nabla_\mu + \ii  \cG_{IJ} z_2^J \partial_\mu z_1^I \gamma_5 +\frac{\ii}{4} \g \zeta_\Lambda^{(4)} A^\Lambda_\mu + \frac{1}{8} \g \e^{3\lambda} \gamma_\mu \left( \zeta_0^{(4)} + \zeta_I^{(4)} ( z_1^I - \ii z_2^I \gamma_5) \right) \nn\\
	& \qquad \ \ \ -  \frac{\ii}{16} \e^{- 3\lambda} F^0_{\nu\rho} \gamma^{\nu\rho} \gamma_\mu -  \frac{1}{4} \e^{ - 3\lambda} \cG_{IJ} z_2^J \left( F^I_{\nu\rho} - z^I_1 \, F^0_{\nu\rho} \right) \gamma^{\nu\rho} \gamma_\mu \gamma_5 \Bigg] \tilde{\epsilon} \, .
\end{align}
 
\bibliographystyle{utphys} 
\bibliography{biblio}{}

\end{document}